\documentclass[aps,nofootinbib,notitlepage,longbibliography,twocolumn]{revtex4-1}
\usepackage{amsmath,amssymb}
\usepackage{slashed}
\usepackage{graphicx}
\usepackage{bm}
\usepackage[T1]{fontenc}
\usepackage[utf8]{inputenc}
\usepackage{gauss} 
\usepackage[top=1.0in,bottom=1.0in,left=1.0in,right=1.0in]{geometry}
\usepackage[colorlinks,linkcolor=blue,citecolor=blue]{hyperref}
\newcommand{\be}{\begin{equation}}
\newcommand{\ee}{\end{equation}}
\newcommand{\bea}{\begin{eqnarray}}
\newcommand{\eea}{\end{eqnarray}}
\newcommand{\ba}{\begin{eqnarray}}
\newcommand{\ea}{\end{eqnarray}}
\usepackage{graphicx} 
\usepackage{xcolor}
\newcommand{\red}[1]{\textcolor{black}{#1}}

\usepackage[caption=false]{subfig}
\usepackage[toc,page,header]{appendix}
\makeatletter
\newcommand\appendix@section[1]{%
  \refstepcounter{section}%
  \orig@section*{Appendix \@Alph\c@section: #1}%
}
\let\orig@section\section
\g@addto@macro\appendix{\let\section\appendix@section}
\makeatother

\begin{document}

\title{The color force acting on a quark in  
the  pion and nucleon}

\author{Wei-Yang Liu}
\email{wei-yang.liu@stonybrook.edu}
\affiliation{Center for Nuclear Theory, Department of Physics and Astronomy, Stony Brook University, Stony Brook, New York 11794-3800, USA}

\author{Edward Shuryak}
\email{edward.shuryak@stonybrook.edu}
\affiliation{Center for Nuclear Theory, Department of Physics and Astronomy, Stony Brook University, Stony Brook, New York 11794-3800, USA}

\author{Ismail Zahed}
\email{ismail.zahed@stonybrook.edu}
\affiliation{Center for Nuclear Theory, Department of Physics and Astronomy, Stony Brook University, Stony Brook, New York 11794-3800, USA}

\begin{abstract} 
In the Operator Product Expansion (OPE) of hard scattering amplitudes, the twist-3 operators
describe local colored Lorentz forces acting on a quark, thereby providing a measure of the strength of the gluon fields. Its value is directly
accessible from the nucleon twist-3 polarized $g_2$-parton distribution function.
In the semiclassical (instanton-based) QCD  vacuum models, the leading non-perturbative contribution stems
from correlated instanton-anti-instanton pairs, or molecules.  We analyze the magnitude of the color force on a struck quark in light hadrons (pion and nucleon), in the context of the instanton liquid model (ILM). We derive explicitly the pertinent form factors associated with the color Lorentz force and show that they are intimately related to the pertinent hadronic gravitational and transversity form factors. Using the ILM enhanced by molecules, we detail the ensuing colored force distribution in the transverse plane for the luminal pions and nucleons. The results for the nucleons are in good agreement with those recently reported from a lattice collaboration.
\end{abstract}

\maketitle

\section{Introduction} \label{sec_intro}
Deep inelastic scattering (DIS) is the main process by which the quark
and gluon (parton) substructure of the nucleon can be quantitatively addressed. 
This is usually captured in terms of pertinent matrix elements of gauge
invariant and traceless operators, organized in the OPE as a {\it twist expansion},  
a series in inverse powers of the momentum transfer $1/Q^2$ of certain
light-cone current-current correlators. 
The leading {\it twist-2} operators are bilinear in the quark (or gluon) fields. When extracted from experimental data, they tell us about the parton 
distribution functions (PDFs), the densities  of gluons, as well as various quarks and antiquarks in the target.

Higher twist operators are of higher dimensions, including higher powers of quark and gluon fields that carry information
about the partonic correlations. The general theory for those operators has been carried out 
in the early 1980's in~\cite{Shuryak:1981kj} for unpolarized targets, and 
in~\cite{Shuryak:1981pi} for polarized ones. They provide the QCD corrections
to the partonic sum rules, such as the Bjorken and the Ellis-Jaffe sum rule. \red{A more recent study in~\cite{Hatta:2024otc} is also devoted to the twist-3 contributions for other (e.g. momentum) sum rules.}

The most interesting  correlation between partons are those 
 stemming from a 
polarized target. While the structure function $g_1(x,Q^2)$ 
starts with the usual twist-2 operators, the structure function
 $g_2(x,Q^2)$ starts with twist-3. Since in experiments the structure functions can be  separated {\it kinematically}, this fact  offers the most direct access to the higher twist  physics. For a transversely polarized nucleon, at $90^0$ to the incoming momentum, $g_1(x,Q^2)$ vanishes and the remaining DIS amplitude is purely twist-3. 
 The pertinent physics is related to the local operator\footnote{Although the operator carries three open indices $\rho,\mu,\nu$, the last pair is antisymmetrized, and the total spin is 2 and not 3.} 
\be 
\label{OFORCE}
O_{\bar q G q}= ig\big(\bar q \gamma^\rho  G^{\mu\nu} q \big)
\ee
given by the value of the gauge field strength on the struck quark. 
(The color indices  are not explicitly shown but assumed here and below.)

The magnitude of this effect has  been  discussed in~\cite{Burkardt:2008ps}. It was pointed out that in the large-$N_c$ limit, the forces on the $u$ and $d$ quarks in the proton should be equal in magnitude but  opposite in sign. The suggested magnitude was of the order of
\be \label{eqn_Burkardt} F_{u}=-F_{d}\sim 0.1\, \rm GeV/fm \ee
Note that it is an order of magnitude smaller than  
the confining "string tension"  force
\be F_{\bar Q Q}=\sigma\approx 1\,\rm  GeV/fm\approx 0.2\, GeV^2 \ee 

\subsection{Twist-3 force in the Instanton Liquid Model (ILM)}
In the late 1970s to early 1980s  studies of semiclassical pseudoparticles - instantons and anti-instantons - 
had led to describing the QCD vacuum as an "instanton liquid" \cite{Shuryak:1981ff}. The  vacuum is very inhomogeneous, with 4-quark operators of the type $LR+RL$ 
(here left-handed current means $L=\bar q (1+\gamma_5) q$, and right-handed with the opposite sine on front of $\gamma_5$)
strongly enhanced through fermionic zero modes, of a single pseudoparticle. Unfortunately, the operator (\ref{OFORCE}) is of different type, $LL+RR$, therefore it is not enhanced in a single pseudoparticle, and thus it does not appear in the
leading order in the pseudoparticle density.  The evaluation of matrix elements of such operators has been postponed in earlier literature, in favor of  discussion of the "most enhanced" effects.

A single instanton has a probability proportional to very small product of light quark (Higgs-induced)
masses $m_u m_d m_s/\Lambda_{QCD}^3\sim 10^{-4}$. Therefore, an ideal gas
of independent pseudoparticles would be extremely dilute and thus irrelevant.
Fortunately, 
the QCD vacuum is not an ideal gas but rather a "liquid" \cite{Shuryak:1981ff}, with strong correlations mediated by light quark exchanges. 
As a result, small Higgs-induced "Lagrangian" masses are substituted by much larger
effective quark masses. The
$SU(N_f)$ chiral symmetry is broken by a nonzero quark condensate. In the ILM, the density and typical size of the pseudoparticles are
\be \label{eqn_ILM}
{N\over V_4}=
n_{I+\bar{I}}\approx 1\rm  fm^{-4},\,\,\,\, \rho\sim {1\over 3 } fm  \ee
In spite of  smallness of the diluteness parameter,
\be {N \rho^4\over V_4 N_c}\approx \left({1 \over 3}\right)^5 \ll 1 \ee
the ensemble properties are not expandable in simple Taylor 
series in the packing fraction (diluteness). For example, (for two light quark flavors)
a constituent quark mass scales as a square root of the packing fraction,
\be
 M \rho\sim \sqrt {N \rho^4\over V_4}  
\ee
as it follows from Bethe-Salpeter equation summing infinite number of quark loop diagrams.

Derivation of effective action $S_{\rm eff}$ for constituent quarks in the mean field approach \cite{Diakonov:1985eg} 
complemented this mass by a specific form factor $\mathcal{F}(k)$
\be S_{\rm eff}=\int {d^4k \over (2\pi)^4} \psi^\dagger(k) [\slashed k - i M \mathcal{F}(k)] \psi(k)
\ee
related to quark zero mode. The form factor describes
dependence of the constituent quark mass on the 
momentum scale under investigation.
Further studies of quark
propagators 
were completed using numerical simulations of the instanton ensembles,
see review \cite{Schafer:1996wv}.

To our knowledge, the first attempts to theoretically quantify twist-3 and twist-4 matrix elements have been carried out in~\cite{Balla:1997hf}. 
\red{They have used a version of the ILM for the vacuum structure, and chiral soliton (large-$N_c$ realization of ILM) for the nucleon.} Their main conclusion is that the twist-3 matrix elements are non-zero but still suppressed by a power of the  diluteness parameter.

\subsection{Lattice studies}
We would not go into the technicalities and long history of lattice gauge field simulations, and many of the key achievements. It is sufficient to note that current lattice simulations are able to work with fermions as light as the quarks in the real world,  reproducing major parts of hadronic phenomenology quite well. This includes the nucleon mass, form factors, PDFs and even GPDs. 

Recently, there have been new attempts to  quantify the values of higher twist operators \cite{Vladimirov:2025qrh}. In particular,
the recent work by the Adelaide group \cite{Crawford:2024wzx}, where numerical evaluation of the twist-3 operators in the nucleon, was quoted well inside the (statistical) error bars. Remarkably, their results 
\be \label{eqn_lattice_Adelaide} 
F_{u} \approx 3\, \mathrm{GeV/fm},\,\, \,\, F_{d}= 0. \pm \rm 0.05\, GeV/fm
\ee
are  much larger than the one suggested earlier in (\ref{eqn_Burkardt}).
The forces on $u$ and $d$ are surprisingly different. Furthermore, as clearly shown in Fig.~5 of that paper, it is normal to both the directions of the nucleon's motion ($z$) and spin ($x$) being localized inside a small sphere of radius only $0.2-0.3\,\rm  fm$.
The vanishing force on the $d$ quark brings up a (perhaps simplistic) explanation:
if $d$ quark is always locked into the spin-zero $ud$ "good diquark", there would be no polarization-sensitive effects associated with it.

\red{
The present analysis is carried out within the instanton liquid model (ILM), which provides a semiclassical description of the QCD vacuum. As such, it does not incorporate confinement explicitly and relies on phenomenological input for the instanton ensemble, in particular the average instanton size $\rho$ and density $n_{I+\bar{I}}\equiv N/V_4$. While these parameters are constrained by phenomenology and lattice studies, the results retain a degree of model dependence.}

\red{Furthermore, the semiclassical treatment is not controlled by a strict small expansion parameter.  Although the packing fraction $n_{I+\bar{I}}\rho^4/ N_c$ is numerically small and allows for an systematic expansion for the many-body description, key observables are not analytic in this parameter. As a result, the present calculation should be viewed as providing robust qualitative trends and order-of-magnitude estimates, rather than precision predictions.}

\red{
In addition, observables sensitive to short-distance dynamics, such as the color Lorentz force, inherit a nontrivial dependence on the instanton size and ensemble characteristics, which propagates to the resulting matrix elements, as we will discuss below.}


\subsection{Outline of this paper}
We develop a theoretical framework for the analysis of pseudoparticles and their pairs (or instanton–anti-instanton molecules) connected by light quarks.  We first review  their emergence in the semiclassical treatment of the QCD vacuum and discuss how these correlations modify the gauge field topology compared to the uncorrelated instanton liquid. 

In section~\ref{SEC-IIBAR} we detail the structure of the molecular configurations using a ratio ansatz, and construct the explicit gauge field configurations, allowing for an estimate of the color force. The overlap of the quark zero modes in these correlated backgrounds is analyzed, leading to quantitative estimates for the overlap integral $T_{I\bar{I}}$ that encodes the effective range of the quark pairing between the instanton and anti-instanton. We then apply Monte Carlo techniques to evaluate the average color-electric field acting on a propagating quark in the molecular ensemble, and from this, determine the corresponding color Lorentz force. Our calculations show that the typical magnitude of this force is $F\sim2$–$3\,\mathrm{\rm GeV/fm}$, consistent with lattice extractions and well above the strength expected from earlier QCD estimates.

In section~\ref{SEC-PDF}, we show how these semiclassical configurations contribute  to the twist-3 parton distribution functions, focusing on the decomposition of $g_T(x)=g_1(x)+g_2(x)$ into its Wandzura–Wilczek and genuine twist-3 parts. Through this connection, the color Lorentz force is directly related to the expectation value of the local operator $\bar{q}gG^{+i}\gamma^+q$, thus establishing a bridge between non-perturbative vacuum dynamics and experimentally measurable structure functions.

In section~\ref{SEC-FF}, we show how the emergent momentum-dependent form factors provide explicit parameterizations of the quark–gluon interaction strength across a range of momentum transfers, revealing that the molecular configurations dominate certain non-perturbative effects. 

In the next sections~\ref{SEC-QUARK}-\ref{SEC-PION}-\ref{SEC-NUCLEON} we extend the analysis to a polarized quark, pion, and nucleon, respectively. While the pion —being spinless— exhibits a vanishing $d^\pi_2$ and no net transverse Lorentz force, the nucleon’s internal structure supports large color forces, consistent with the observed twist-3 moments.  We explicitly show that the color Lorentz forces in the light hadrons are tied to the hadronic gravitational and transversity form factors. 
Our conclusions are presented in section~\ref{SEC-CON}.  A number of appendices are added to complement the derivations in the main text.

\section{instanton-anti-instanton "molecules"}
\label{SEC-IIBAR}
\subsection{Generalities and estimates}
Instantons are tunneling events between topologically distinct gauge field configurations. In a theory with massless  quarks, an amplitude for an isolated tunneling event vanishes due to
fermionic zero modes. For light quarks beyond chiral limit, it is proportional to the product of the light quark masses $m_u m_d m_s\rho^3$ which is nonzero but numerically negligibly small. 

Therefore, there are only two ways in which QCD instantons can produce significant observable effects. 
One of them - the dominant one on which the ILM is based - is in which the instanton ensemble  $collectively$ breaks  chiral symmetry, and trade  the small "Lagrangian" quark masses of order  few MeV by "constituent" ones of order $\sim {\cal O}(400\, \rm {MeV})$.
 
The other is that few instantons and anti-instantons together can produce a  cluster with zero topological charge. If so, the amplitude is nonzero even for vanishing quark masses. The simplest cluster of this kind is an instanton-anti-instanton ($I\bar I$) pair, or "molecule". It can be considered as an "incomplete tunneling" event, in which the quantum path wanders
 into the classically forbidden area and then returns back. The light quarks must propagate in loops
 between $I$ and $\bar I$, which provide the correct
 chirality flips, see the sketch in Fig.\ref{fig_sketch}.

Before going through the history of this mechanism and its technical details, let us present simple (but naive) estimates of
the fields and forces involved, to make connection to
the numbers mentioned in the preceding section.
The instanton field squared at distance $r$ from its center is
\be 
\big(gG^a_{\mu\nu}(r)\big)^2={192 \rho^4 \over (\rho^2+r^2)^4}
\ee
so at the origin ($r=0$) and $\rho=1/3\, \rm fm$ the r.m.s.
field is numerically $G_{r.m.s.}\sim 4.4\, \rm GeV^2$. The lattice estimates \cite{Athenodorou:2018jwu} of the
density of $I\bar I$ molecules is an order of magnitude larger than that of ILM instantons (\ref{eqn_ILM})
$ n_{\rm mol}\rho^4\sim 0.1 $. A very crude estimate of a field acting on a quark thus produce a value
\be 
F\sim (n_{\rm mol}\rho^4)G_{\rm r.m.s.} \sim 0.44 \rm GeV^2\sim 2.2\, GeV/fm
\ee 
 comparable to the lattice results (\ref{eqn_lattice_Adelaide}), even twice $larger$ than the $\bar QQ$ force in quarkonia $\sigma\sim 1 \,\rm  GeV/fm $. Note that this is a crude estimate, but more detailed 
calculations will be reported below. Here we just remind the reader that
in our paper \cite{Shuryak:2021fsu} the confining quark-antiquark potential was also derived from the same  "molecular" contribution.

\subsection{Historical comments}
 The overlapping $I\bar I$ configurations were first considered in
the framework of the famous quantum mechanical  problem of a   double-well potential.  In \cite{Shuryak:1987tr}, by keeping certain points on the path fixed and minimizing the action, one
finds that the "gradient flow" produces a certain set of configurations, which are {\it conditional} minima of the action (in all variables except along the flow gradient).  This construction was developed in parallel to previously known concept in complex analysis known as  "\red{Lefschetz} thimbles",  particular lines connecting all extrema of a function or functional on a complex plane. The $I\bar I$ streamline connects well-separated $I$ and $\bar I$ (the extremum at $R_{I\bar I}\rightarrow \infty $) to the zero field "perturbative vacuum".

In gauge theory, the derivation of the gradient flow equation were pioneered in~\cite{Balitsky:1986qn}. The solution 
was first found for large distances (small overlap) of $I\bar I$, and then at all distances in \cite{Verbaarschot:1991sq}, via  the special conformal transformation of $I$ and $\bar I$ into a co-centering configuration.

The first applications of the "molecular configurations" were in finite temperature QCD
\cite{Ilgenfritz:1988dh}. Indeed, in the QGP phase above the chiral phase transition there is no quark condensate, and "molecules" are the only \red{leftover} of the instanton ensemble.

The fact that the $I\bar I$ thimble goes all the way to zero
(perturbative) fields was the main difficulty in the  studies of instanton ensembles. Numerical simulations
\cite{Schafer:1996wv} 
treated it "by brute force": sufficiently close instanton–anti-instanton ($I\bar I$) pairs were excluded through the introduction of an artificial repulsive core, thereby removing configurations deemed insufficiently semiclassical. 
It is fair to say that quantitative theory of $I\bar I$ thimbles still remains undeveloped.

The issue has been studied on the lattice, with the advent of the gradient flow method a decade ago. The density
of $I\bar I$ correlated pairs was quantified by \cite{Athenodorou:2018jwu}. With increasing \red{flow} time $\tau$, their study shows how $I\bar I$ pairs get annihilated, leading to a dilute instanton ensemble and once more confirming the ILM parameters (\ref{eqn_ILM}). The important new observation was the density of correlated pairs or "molecules" at finite $\tau$. The extrapolation 
to zero \red{flow} time ($\tau\rightarrow 0$) is $n_{mol}\sim 10 \, \mathrm{fm}^{-4}$, about an order of magnitude larger than the instanton density in ILM.

The contribution of the correlated  $I\bar I$ pairs also resurfaced a few years ago in several applications, where the
nonperturbative gauge fields (rather than their fermionic zero modes) are directly involved. The first was our study of the pion form factors in the semi-hard regime~\cite{Shuryak:2020ktq}. The second was our study of the heavy quark static and spin-dependent potentials~\cite{Shuryak:2021fsu}. In both analyses, the results were in agreement with phenomenology, after the contribution of the $I\bar I$ "molecules" was included.

The application of the $I\bar I$ molecules to the quarkonium central and spin-dependent potentials was worked out in our recent paper \cite{Miesch:2024fhv}. It bridges the gap between the vacuum and hadronic structure by focusing on the interquark potentials in heavy quarkonia. In this case, the basic  theory is well known. The central potential
is related to the expectation value of the Wilson loop, while the spin-dependent potentials follow by dressing the loop with magnetic (or electric) field strengths. These calculations are methodologically quite close to those to be reported below. For the spin-dependent potentials, one has {\it two} gauge field strength insertions on the temporal Wilson line, while the matrix element of the operator (\ref{OFORCE}) contains {\it one} gauge field strength on a light-like Wilson line.

\subsection{Field configurations}
We now proceed to the space-time shape of the $I\bar I$
configurations. Since the conformal mapping leads to rather complicated expressions to implement,  we will use a simpler (but we think still qualitatively representative) ansatz to describe them, a variant of the so called "ratio ansatz"
\be \label{eqn_ansatz}
A^{\mu a}(x)={ \bar\eta^{a\mu\nu} y_I^\nu \rho^2/ Y_I^2 + \eta^{a\mu\nu}y_A^\nu*\rho^2 /Y_A^2 \over
1+\rho^2/Y_A +\rho^2/ Y_I }
\ee
where $I,A$ stand for instanton and anti-instanton.
Their centers are  located at $y_{I,A}^i=x^i,i=1,2,3$ and $y^4_I=x^4-R/2, y^4_A=x^4+R/2$, with $Y_{I,A}$  referring to their squared lengths, i.e.
$Y_{I,A}=(y_{I,A}^\mu y_{I,A}^\mu)$. Near one of the centers (e.g. $Y_I$ small and $Y_A$ large),
the dominant contribution in the numerator is the first term, and  with the leading terms in  the denominator, yield the familiar field of an instanton in singular gauge.

We obtain lengthy but manageable expressions for the field strength and its squares, though they remain too long to present here. The distributions of the action density $(G_{\mu\nu}^a)^2$ and the corresponding fields were shown in \cite{Miesch:2024fhv}. However, in small separations $R<0.5$, it does not go to zero but displays a
small repulsive core, in contrast to streamline configurations.


The qualitative shape of the field $G^a_{\mu\nu}$ depends on the direction of the  vector $\vec R_{I \bar I}$ 
between the two centers. If it is directed along the time axes, then the electric fields of $I$ and $\bar I$
are of opposite sign and tend to cancel, while the magnetic fields are
of the same sign and are doubled. However, in the calculations, one has to
take $\vec R_{I \bar I}$ to be homogeneously distributed over the 4-sphere.  
This produces nontrivial relation between electric-induced and magnetic-induced
forces.

\subsection{Quark zero modes and propagators}
\begin{figure}[h!]
    \centering
    \includegraphics[width=0.85\linewidth]{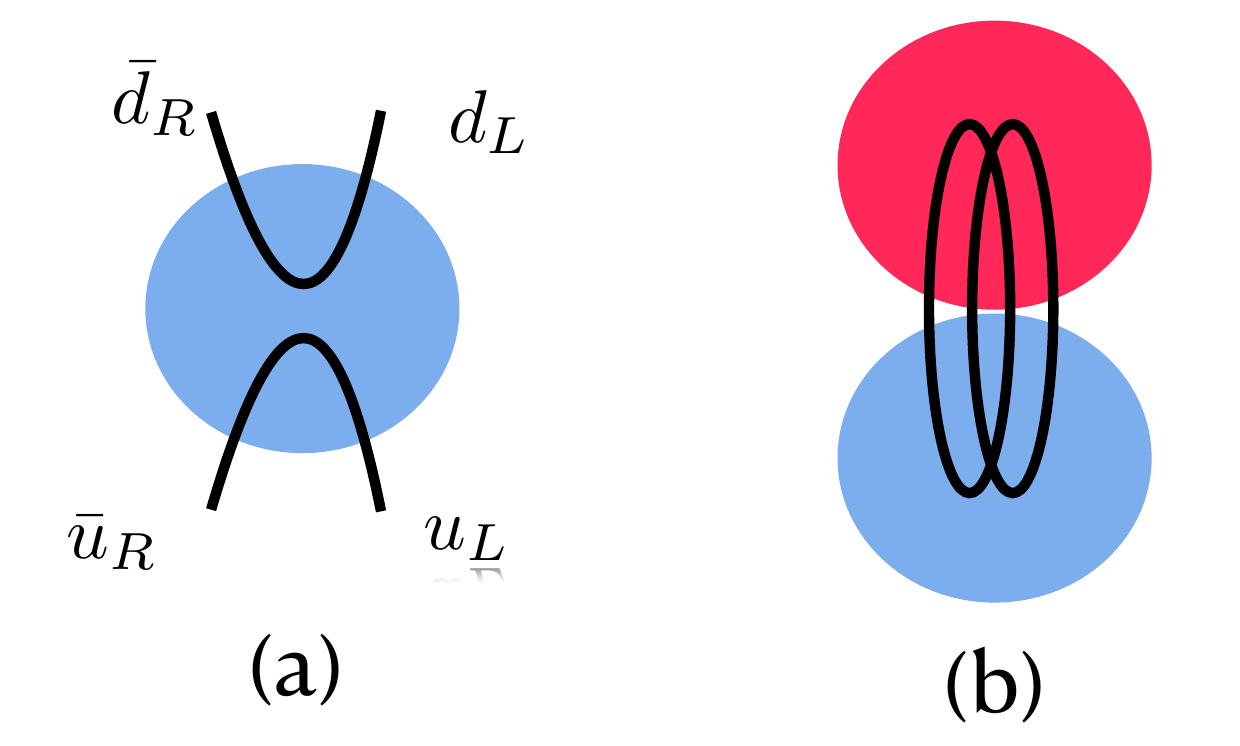}
    \caption{(a) An instanton as an effective 't Hooft operator with 4 external lines; (b) Quark propagation in the $I \bar I$ molecule. Both 
    figures assume
    two light quark flavors, $N_f=2$}
    \label{fig_quark_lines}
\end{figure}

As shown by 't Hooft \cite{tHooft:1976snw},  in the instanton case
the chirality flip is described via
 the fermionic zero modes. As a result, an instanton should be considered as an effective operator with $2N_f$ external quark lines, see Fig.\ref{fig_quark_lines}(a).

\begin{figure}[h!]
    \centering
    \includegraphics[width=0.75\linewidth]{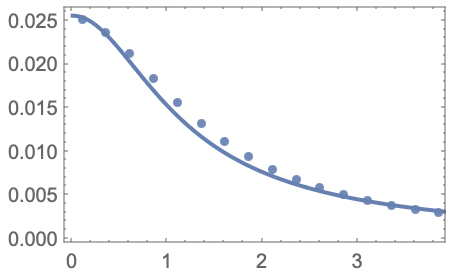}
    \caption{The points show the numerical results for the  dimensionless combination $\rho^2 T(r)$ versus $r/\rho$. The line is our approximate parameterization $0.0257/(r^2 + 1)^{0.75}$}
    \label{fig_M_of_r}
\end{figure}

The $I\bar I$ molecule is a configuration with zero topological charge,  therefore  its amplitude is 
always nonzero. The quark part of the functional determinant includes in this case loop diagrams proportional
to the power of the so called "zero mode overlap" $\sim |T_{I\bar I}|^{2N_f}$, 
representing quarks travelling between an instanton and anti-instanton, as illustrated in Fig.\ref{fig_quark_lines}(b).
The analytic expression for this "hopping amplitude" is
\ba
\label{eq:hop}
T_{I\bar I} &=& \int d^4 x \phi^\dagger(x-z_{\bar I}) i \slashed{\partial} \phi(x-z_{I})\nonumber \\
&=& \mathrm{Tr}(U_I \tau^-_\mu U^\dagger_{\bar I}) {R_\mu \over R}{d T(R)\over dR}\ea
where $R=z_I-z_{\bar I}$ and
\be T(R)={1 \over 2\pi^2R} \int_0^\infty dp p^2 \frac{|\phi(p)|^2}{4\pi^2\rho^2} J_1(pR)\ee
and expressed via the Fourier transform
\be 
\label{eq:zm}
\phi(k)=\pi \rho^2{d \over dx}\bigg(I_0(x)K_0(x)-I_1(x)K_1(x)\bigg)_{x=\frac 12 k\rho}
\ee
of the fermion zero mode 
\be \phi(x)={\rho \over \pi} {1 \over \sqrt{x^2}(x^2+\rho^2)^{3/2} } \ee
Since $M(R)$ is an integral including many Bessel functions, we evaluated it numerically and found
simple parameterization of the result, see Fig.\ref{fig_M_of_r}

The amplitude of the $I\bar I$ molecule contains the overlap integral in the power $2N_f$. At large distances 
$R\gg \rho$ is decreasing very strongly 
As a result, the distance between the centers
is strongly restricted.
As shown in the lower plot, after the overlap $|T_{I\bar I}|^2$
is multiplied by the 4d volume element $R^3$,  we  find a rather sharp peak
at $R/\rho\approx 1$. Therefore, we would focus on
the $I\bar I$ molecules with a distance between the centers $R=\rho$.

\begin{figure}[b!]
    \centering
    \includegraphics[width=0.75\linewidth]{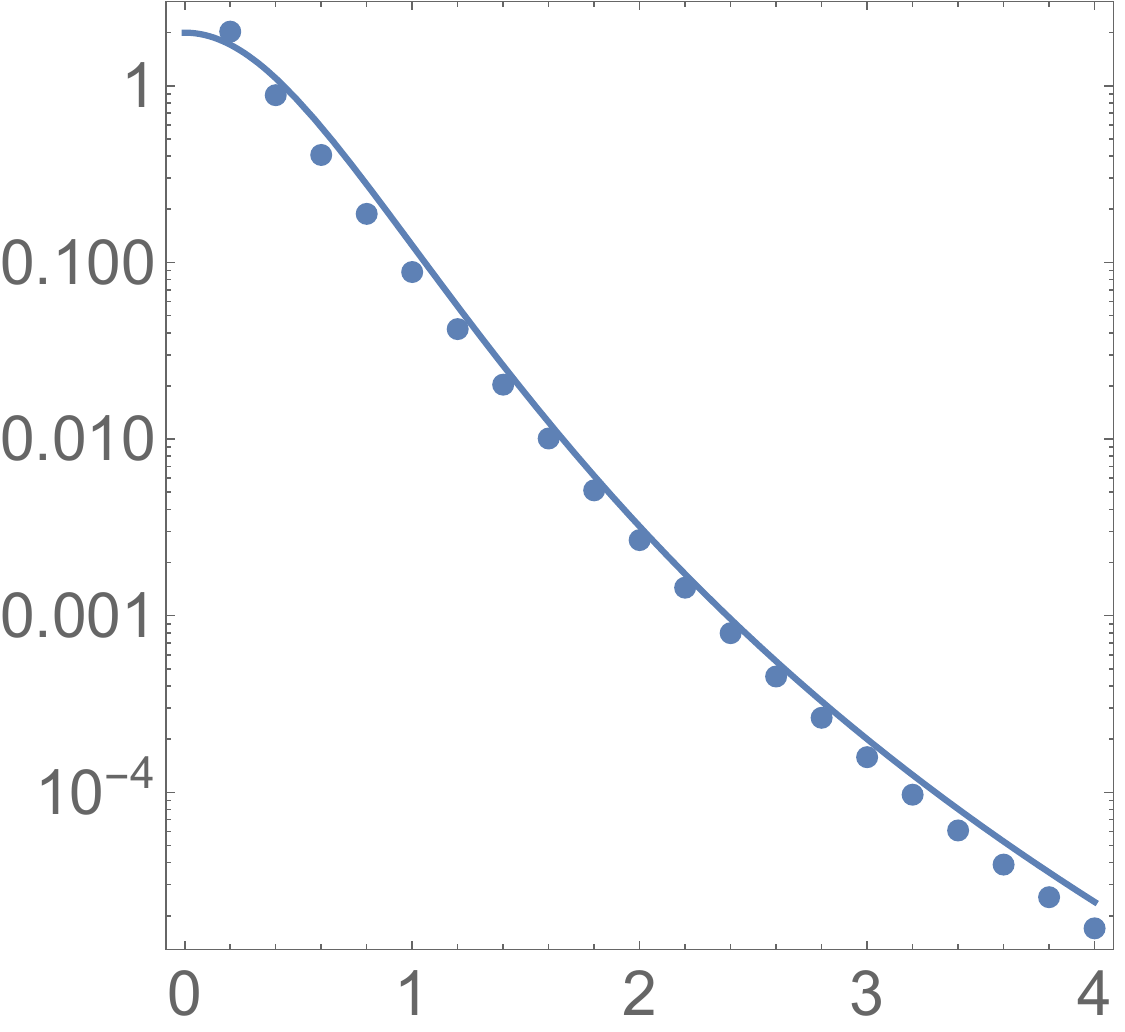}
     \includegraphics[width=0.75\linewidth]{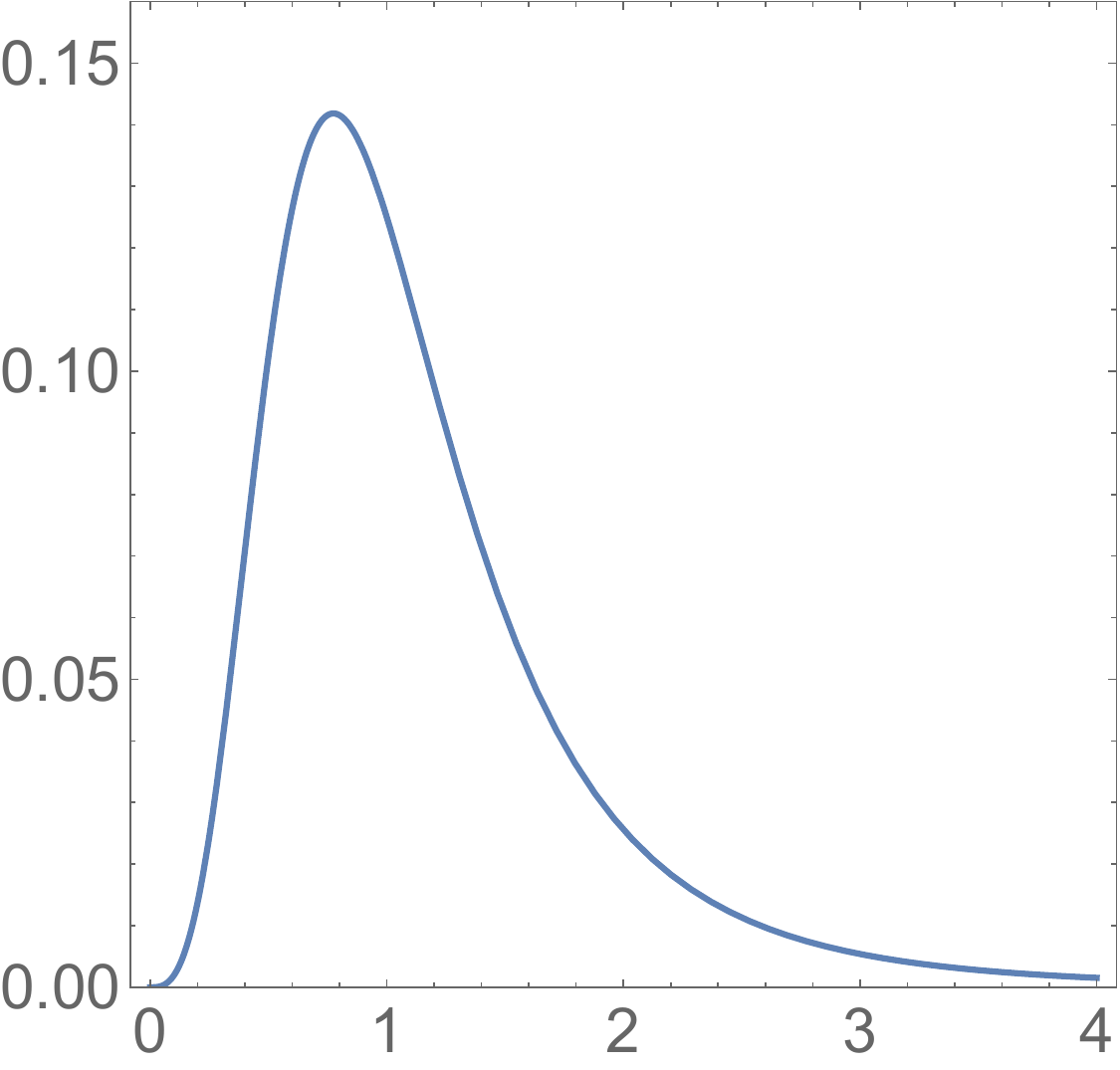}
    \caption{The points in  the upper plot show the numerical results for the zero mode overlap integral 
    $|T_{I\bar I}(R)   |^2$
    versus $R/\rho$.  The line is our  approximate parameterization $2/(R^2 + 1)^4$. The lower
    plot shows the same amplitude times $R^3$ from  the 4d
    radial integral.  It displays a sharp peak at $R/\rho\approx 1$.}
    \label{fig_TIA2}
\end{figure}

\begin{figure}
    \centering
    \includegraphics[width=0.4\linewidth]{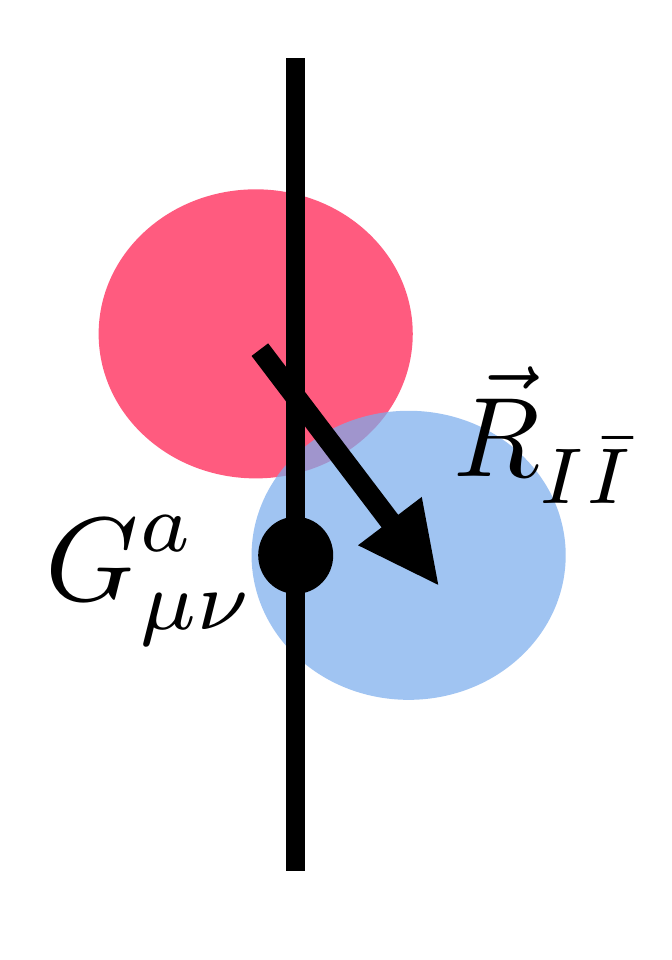}
    \includegraphics[width=0.75\linewidth]{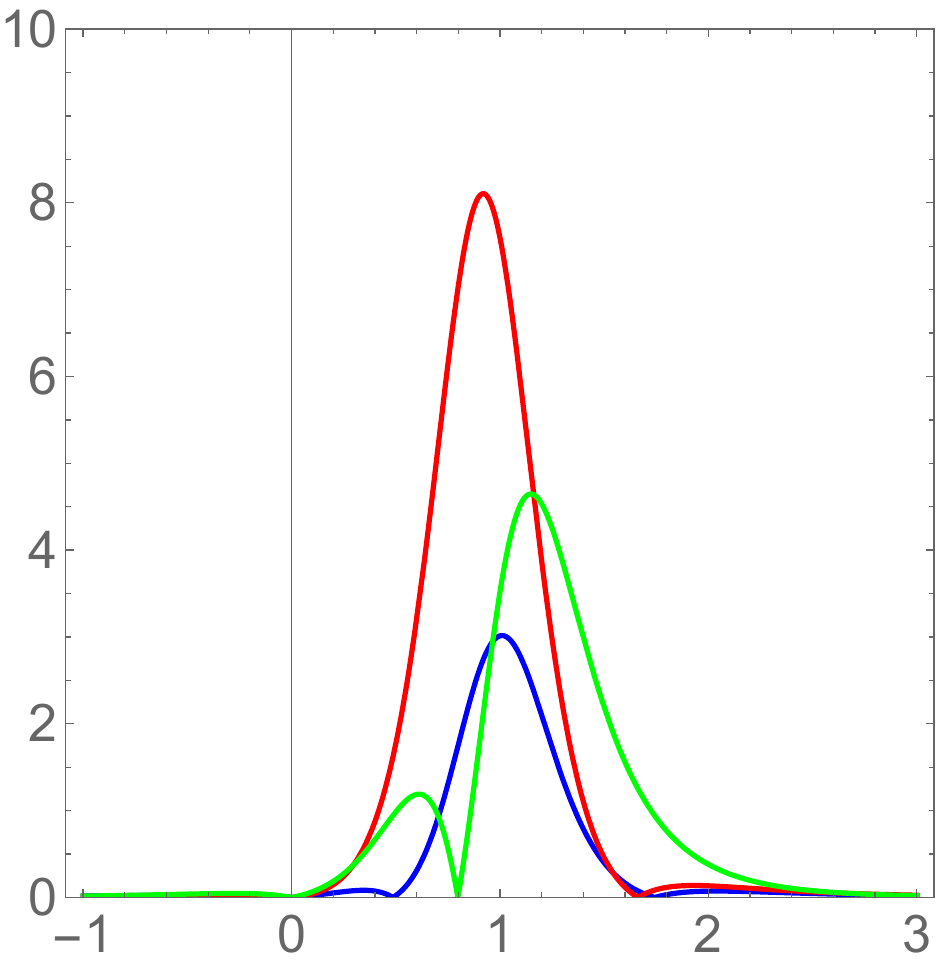}
    \caption{The upper figure is a sketch of a molecular configuration
    pierced by a quark path in the time direction. The lower figure
    is a random example of how three components of the field (m=1,2,3, blue red,green respectively) depend on time along the line. }
    \label{fig_sketch}
\end{figure}

The average of the operator in question on the fields of the $I\bar I$ molecule is schematically shown in Fig.~\ref{fig_sketch} (upper). The black dot corresponds to the location of the operator (\ref{OFORCE}), where it picks up both the local field strength (shown) and the corresponding  quark zero modes (not shown). It should then be integrated along the quark path (the vertical line). The orientation of the 
molecule needs to be averaged over the 4d orientation $ \vec R_{I\bar I}$ (shown) and the location (not shown) of the molecule relative to the quark. An example of the (absolute values) of the electric field components
$E_m=(gG^a_{4m} \hat r^a)$ 
along the line (4th coordinate) are shown in Fig.\ref{fig_sketch}(lower).

Using a Monte-Carlo method with 4d-random orientation of the vector $R_{I\bar I}$ and random shift of the molecule center, we generated thousands of such configurations of the fields and evaluated the integrated kick on a quark
\begin{equation}
\begin{aligned}
\langle \int E d\tau \rangle^2 &= (\int d\tau E_1)^2+(\int d\tau E_2)^2+(\int d\tau E_3)^2 \\
&\approx 
\left[(4.15\pm 0.21)/\rho\right]^2
\end{aligned}
\end{equation}
\red{The spatial shift and temporal integration are selected accordingly with the ensemble measure normalized by the 4-volume. 
The distances between the molecules scale as the inverse 4th root of their dimenionless 4d density $n_{mol}\rho^4$.} The force, defined by the integrated kick per time, thus reads 
\be F= (n_{\rm mol}\rho^4)^{\frac14} \langle \int E d\tau \rho \rangle {1\over \rho^2}\ee


Using $n_{\rm mol}\rho^4\approx 0.10-0.15$ from \cite{Athenodorou:2018jwu} and $\rho=0.3$ fm \cite{Shuryak:1981ff}, we obtain an estimate for the  force 
\be F\approx (2-3)~\mathrm{\rm GeV/fm} \ee
The unstruck virtual quark loop connecting $I$ and $\bar I$ contributes an additional factor $|T_{I\bar I}(R)|^2$, which is not shown here.
We now proceed to a more refined analysis of this force in the low lying hadrons, e.g. pion and nucleon.


\section{Twist-3 PDF}
\label{SEC-PDF}
As mentioned earlier, in DIS the twist-3 contribution describes the average Lorentz force acting on a quark in the nucleon. This contribution is accessible in DIS process on a polarized nucleon target.
In contrast to the  structure function $g_1(x, Q^2)$ which is leading twist-2 and polarization independent, the structure function $g_2(x, Q^2)$ is a twist-3 and polarization dependent~\cite{Shuryak:1981kj,Shuryak:1981pi}. We now proceed to its quantitative description in the QCD instanton vacuum, using the general framework developed in~\cite{Liu:2024rdm} that includes 
individual pseudoparticles (instantons, anti-instantons) and their molecular configurations (instanton-anti-instanton). We will focus on this average force acting on the emerging quark, the pion and the nucleon. 

DIS process on a polarized target, splits into longitudinal and transverse contributions
\begin{equation}
\begin{aligned}
    &g_1(x)s_L=\int_{-\infty}^{\infty} \frac{d\xi^-}{4\pi} e^{i \xi^-p^+ x} \\
    &\times 
\langle ps | \bar{q}(0) \gamma^+ \gamma_5 W(0,\xi^-) q(\xi^-) | ps \rangle
\end{aligned}
\end{equation}
\begin{equation}
\begin{aligned}
    &\frac{m_N}{p^+} g_T(x)s_\perp=\int_{-\infty}^{\infty} \frac{d\xi^-}{4\pi}e^{i\xi^-p^+x} \\
    &\times 
\langle ps | \bar{q}(0) \gamma^\perp \gamma_5 W(0,\xi^-)q(\xi^-) | ps \rangle
\end{aligned}
\end{equation}
The transverse distribution is the sum of $g_2$ and $g_1$ 
\begin{equation}
    g_T(x)=g_1(x)+g_2(x)
\end{equation}
a measure of the transverse spin. We now recall  that $g_2$ satisfies the 
Burkhardt–Cottingham (BC) sum rule  \cite{Bhattacharya:2021boh}
\begin{equation}
    \int dx\, g_2(x,Q^2) = 0 
\end{equation}
connecting the twist-2 and twist-3 parton distributions. 
Since the structure functions $g_1, g_2$ can be separated kinematically in experiments, this observation  allows for the description of higher twist effects initially developed in~\cite{Shuryak:1981pi,Shuryak:1981dg,Jaffe:1981td,Jaffe:1982pm}.

For a transversely polarized nucleon  $g_1(x, Q^2)$ vanishes and the remaining DIS amplitude is purely $g_2$ of twist-3. By analogy with the twist-2 PDFs $f_1(x), g_1(x), h_1(x)$, the twist-3 PDFs are referred to as 
$e(x), g_T(x), h_L(x)$. In particular,  the twist-3 PDF $g_T(x)$ can be expressed as a sum of the Wandzura-Wilczek (WW) term, a piece that is determined entirely in terms of twist-2 helicity PDF $g_1(x)$, and an interaction dependent dynamical twist-3 term $\bar g_2(x)$, which involves quark-gluon correlations. \cite{Aslan:2019jis}
\begin{equation}
    g_T(x, Q^2)=g_T^{WW}(x, Q^2)+\bar{g}_2(x, Q^2)
\end{equation}
The first contribution is the twist-2  Wandzura-Wilczek fixed by~\cite{Wandzura:1977qf}
\begin{equation}
g_T^{WW}(x, Q^2) \equiv \int_x^1 dy \frac{g_1(y, Q^2)}{y}
\end{equation}
%
After subtraction of this twist-2 contribution,
the second Mellin moment of $\bar g_2(x)$  is related to the Lorentz color force matrix element ~\cite{Aslan:2019jis,Vladimirov:2025qrh}
\bea
\frac {d_2}3=\int_0^1dx\,x^2 \bar g_2(x)
\eea

More specifically, the Lorentz force is defined by the matrix element 
\bea
\label{FY}
d_2\epsilon_{\perp ij}S^j=&&-\frac {\langle PS|\bar q (0)\gamma^+gG^{+i}(0)q(0)|PS\rangle}
{2m_N(P^{+})^2 }\nonumber\\
\eea
This result can be seen to follow from the identity~\cite{Aslan:2019jis}
\begin{align}
&2S^xm_N(P^+)^2\int_0^1dx\,x^2 g_T(x) \nonumber\\
=&-\langle ps | \bar{q}(0) \gamma^x \gamma^5 (\overleftrightarrow{D}^+)^2 q(0) | ps\rangle\nonumber\\
=& 2 S^x m_N (P^+)^2 \int_{-1}^{1} \mathrm{d}x \, x^2 g_T^{WW}(x) \nonumber \\
& - \frac{1}{3} \langle ps| \bar{q}(0) \gamma^+ g G^{+y}(0) q(0) | ps \rangle
\end{align}

%
%
For a polarized nucleon along the $x$-direction  with light-cone momentum $P^+$ along the $z$-direction and struck by a transverse probe $q_\perp$, the average color Lorentz force density on a transverse plane is \cite{Crawford:2024wzx}
\begin{equation}
\begin{aligned}
    &F_{q/h}^i(b)=i\int \frac{d^2q_\perp}{(2\pi)^2}e^{-iq_\perp\cdot b}\frac {\langle h(p')|\bar q \gamma^+igG^{+i}q|h(p)\rangle}{\sqrt{2}\bar{p}^+ }
\end{aligned}
\end{equation}
The presence of $\bar q\gamma^+ q$ in the operator means that this force is proportional to the quark density~\cite{Aslan:2020eqo}
\begin{equation}
    \rho_{q/h}(b)=\int \frac{d^2q_\perp}{(2\pi)^2}e^{-iq\cdot b_\perp}\frac{\langle h(p')|\bar q \gamma^+q|h(p)\rangle}{\sqrt{2}\bar{p}^+}
\end{equation}
which is a measure of the electromagnetic form factor in the transverse plane.


We note that the force involves the quark current
 with a single $\gamma_\mu$ which is chiral
even ($LL+RR$). Since the instanton-induced operators are
chiral-odd ($LR+RL$), their contribution to the force is a priori suppressed,
except for their non-zero mode contributions which are chiral even. This
brings about the role and contribution of the molecular instanton-anti-instanton  configurations to the force as we discussed earlier, although formally suppressed by an extra power of the packing fraction.

\begin{figure}
\centering
\subfloat[\label{fig:ONEv1}]{\includegraphics[width=.35\linewidth]{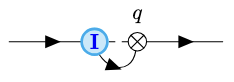}}
\hfill
\subfloat[\label{fig:ONEv2}]
{\includegraphics[width=.35\linewidth]{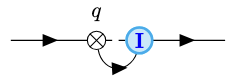}}
\hfill
\subfloat[\label{fig:ONEv3}]{\includegraphics[width=.28\linewidth]{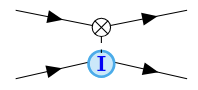}}
    \caption{The single instanton/anti-instanton  vertices with the insertion of the color Lorentz operator (crossed-circle): (a), (b) along a fermion line and (c) with a pair of \red{fermion} lines. The latter is surppessed by $1/N_c$ compared to the former.}
    \label{fig:ONEvertices}
\end{figure}

\begin{figure}
\centering
\subfloat[\label{fig:v1}]{\includegraphics[width=.35\linewidth]{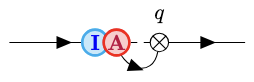}}
\hfill
\subfloat[\label{fig:v2}]{\includegraphics[width=.35\linewidth]{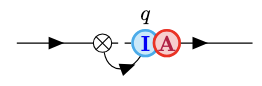}}
\hfill
\subfloat[\label{fig:v3}]{\includegraphics[width=.28\linewidth]{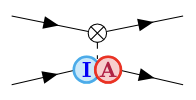}}
    \caption{A molecular pair of instanton-anti-instanton with the insertion of the color Lorentz operator (crossed-circle): (a), (b) along a fermion line and (c) with a pair of \red{fermion} lines. The latter is suppressed by $1/N_c$ compared to the former.}
    \label{fig:vertices}
\end{figure}

\section{Emergent form factors and the Color Lorentz Force  }
\label{SEC-FF}
The Lorentz force contributions from
individual instantons or anti-instantons are illustrated in Fig.~\ref{fig:ONEvertices}, and the molecular contributions illustrated in Fig.~\ref{fig:vertices}. The details of their analytic forms are given in Appendix~\ref{App:mol}. More specifically, the  non-forward amplitude of the twist-3 operator can be deduced from the  4-fermion contribution, as illustrated in Fig.~\ref{fig:ONEvertices}, with the result
\begin{widetext}
\begin{equation}
\begin{aligned}
\label{eq:force_op_0}
    \langle h'|ig\bar \psi G_{\mu\nu}\gamma_\sigma\psi|h\rangle=&\frac{1}{4(N_c^2-1)}\frac{n_{I+\bar{I}}}2\left(\frac{4\pi^2\rho^2}{m^*}\right)i\int \frac{d^4k}{(2\pi)^4}8G(\rho k)\left(\frac{k_\lambda k_\nu}{k^2}-\frac14g_{\lambda\nu}\right)\\
    &~\times\int d^4xe^{-i(q+k)x}\langle h'|\bar\psi(x)\gamma_{\sigma}\lambda^A\psi(x)\bar\psi(0) \sigma_{\mu\lambda}\lambda^A\psi(0)|h\rangle\\
    &-\frac{n_{mol}}{4(N_c^2-1)^2}\gamma_{I\bar{I}}\, i\,t_{\mu\nu\rho\lambda\alpha\beta}\int \frac{d^4k}{(2\pi)^4}\rho^2\frac{k_{\rho}k_{\lambda}}{k^2}G(\rho k)\\
    &~\times\int d^4xe^{-i(q+k)x}\langle h'|\bar\psi(x)\gamma_\sigma\lambda^A\psi(x)\bar\psi(0)i\gamma_{(\alpha}\gamma^5\overleftrightarrow{\partial}_{\beta)}\lambda^A\psi(0)|h\rangle
\end{aligned}
\end{equation}
\end{widetext}
Here, each emergent quark carries  a non-local form factor reflecting on its origin
as a quark zero mode. In momentum space, this form factor is related to the  Fourier transform  of the zero mode in \eqref{eq:zm},
\begin{equation}
\label{instanton-quark}
    \mathcal{F}(k)=\left|\frac{\phi(k)}{2\pi\rho}\right|^2
\end{equation}
 The Fourier transform  of the instanton field profile gives 
\begin{equation}
\begin{aligned}
    &G(k)
    =\frac{4\pi^2}{k^2}\left( 1-\frac{16}{k^{2}} +\frac{k^{2}}2 \, K_{2}(k) + 2 k \, K_{3}(k) \right)
\end{aligned}
\end{equation}
with $K_{2,3}$ are  modified Bessel functions of the second kind. More details are given in Appendix~\ref{App:force}.
The emerging instanton-anti-instanton  coefficient is estimated as  
\begin{widetext}
\bea
    \gamma_{I\bar{I}}=\frac{2\pi^2}{n_{\rm mol}}\left(\frac{n_{I+\bar{I}}}2\right)^2\int_0^\infty dRR^3\, \left|\frac{T_{I\bar{I}}}{m^*}\right|^{2N_f}
   \left(\frac{4\pi^2\rho^2}{|T_{I\bar{I}}|}\right)^{2}\left[\frac{-1}4R\frac{dT(R)}{dR}\right]
\eea
\end{widetext}
where the molecule density $n_{\rm mol}=7.248$ fm$^{-4}$ is obtained after summing over all the 3-flavors in the molecular pairing, as detailed in Appendix~\ref{App:mol}.

\red{
We note that the derivation above relies on a local approximation for the quark-gluon operator. This approximation is justified by the strong localization of the instanton profile and the restriction on the quark separation, $|x|\lesssim \rho$, as follows from the structure of the zero modes and the instanton-induced interaction. However, this simplification neglects nonlocal contributions inherent in the full expression (30), and therefore contains an additional source of systematic uncertainty in the evaluation of the color-force operator.}

For the quark-gluon operator inside a hadron, the resulting effective quark operators are usually related to a four point correlation (two quark current and two hadronic source).
Since the instanton profile is highly localized, the separation between the two-quark source in \eqref{eq:force_op_0} is controlled by $|x|\lesssim\rho\ll \sqrt[4]{V_4/N}$. Thus, we can further approximate the hopping quark propagator in Fig.~\ref{fig:ONEv1} and \ref{fig:ONEv2} for a single instanton, and \ref{fig:v1} and \ref{fig:v2} for an instanton-anti-instanton pair. The vertices in Fig. \ref{fig:ONEv3} and \ref{fig:v3} will be neglected as they contribute to higher power of $1/N_c$.




With this in mind, the contribution to the (non-forward) matrix element of Fig.~\ref{fig:ONEvertices}  is
\begin{widetext}

\begin{equation}
\begin{aligned}
\label{eq:CFOp2}
    &\langle h'|ig\bar\psi\gamma_\sigma G_{\mu\nu}\psi|h\rangle_{I+\bar{I}}=-\left(\frac{n_{I+\bar{I}}}2\right)\frac{1}{2N_c}\left(\frac{4\pi^2\rho^2}{m^*}\right)\beta^{(+)}_{\bar{q}Gq,1}(\rho q)\left(g_{\mu\sigma}q_\nu-g_{\nu\sigma}q_\mu\right)\langle h'|\bar\psi\psi|h\rangle\\
    &+\left(\frac{n_{I+\bar{I}}}2\right)\frac{1}{2N_c}\left(\frac{4\pi^2\rho^2}{m^*}\right)\beta^{(+)}_{\bar{q}Gq,1}(\rho q)\epsilon_{\mu\nu\sigma\alpha}q_\alpha\langle h'|\bar\psi i\gamma^5\psi|h\rangle\\
    &-\left(\frac{n_{I+\bar{I}}}2\right)\frac{1}{2N_c}\left(\frac{4\pi^2\rho^2}{m^*}\right)\rho^2\beta^{(+)}_{\bar{q}Gq,2}(\rho q)2i\epsilon_{[\mu\lambda\sigma\rho}\left(q_\lambda q_{\nu]}-\frac 1{4}g_{\lambda\nu]}q^2\right)m^*\langle h'|\bar\psi\gamma_\rho\gamma^5\psi|h\rangle\\
\end{aligned}
\end{equation}
\end{widetext}
where two new instanton form factors are defined as 
\begin{align}
    \beta^{(+)}_{\bar q Gq,1}(q)=&\frac1q\int_0^\infty dx \frac{16}{(x^2+1)^2}\frac{J_2(qx)}{qx}K_D(x) \\
    \beta^{(+)}_{\bar q Gq,2}(q)=&\frac1q\int_0^\infty dx \frac{16x^2}{(x^2+1)^2}\frac{J_3(qx)}{q^2x^2}K_m(x)
\end{align}
with the quark zero mode induced modification during the quark propagation
\begin{equation}
    K_D(x)=
    \frac{x^3}{(1 + x^2)^{3/2}}
\end{equation}
and
\begin{equation}
    K_m(x)=\int_0^\infty dk x J_1(kx)\sqrt{\mathcal{F}(k)}
\end{equation}
At zero momentum transfer, the values of those two instanton form factors are
$\beta^{(+)}_{\bar q Gq,1}(0)=\frac25$ and
$\beta^{(+)}_{\bar q Gq,2}(q\rightarrow0)=-\frac13\ln q$. 
The "molecular" contribution from Fig.~\ref{fig:vertices} is
\begin{widetext}
\begin{equation}
\begin{aligned}
\label{eq:CFOp3}
    &\langle h'|ig\bar\psi\gamma_\sigma G_{\mu\nu}\psi|h\rangle_{I\bar{I}}=
    \\
    &-\frac{n_{mol}\gamma_{I\bar{I}}}{2N_c(N_c^2-1)}\beta^{(+-)}_{\bar{q}Gq,1}(\rho q)\left(g_{\mu\sigma}g_{\nu\alpha}-g_{\nu\sigma}g_{\mu\alpha}\right)q_\beta\langle h'|\bar{\psi}\left(\gamma_{(\alpha} i\overleftrightarrow{\partial}_{\beta)}-\frac{1}{4}g_{\alpha\beta}i\overleftrightarrow{\slashed{\partial}}\right)\psi|h\rangle\\
    &-\frac{n_{mol}\gamma_{I\bar{I}}}{2N_c(N_c^2-1)}\beta^{(+-)}_{\bar{q}Gq,2}(\rho q)\left(g_{\mu\alpha}q_\nu-g_{\nu\alpha}q_\mu\right)\langle h'|\bar{\psi}\left(\gamma_{(\alpha} i\overleftrightarrow{\partial}_{\sigma)}-\frac{1}{4}g_{\alpha\sigma}i\overleftrightarrow{\slashed{\partial}}\right)\psi|h\rangle\\
    &+\frac{n_{mol}\gamma_{I\bar{I}}}{2N_c(N_c^2-1)}\frac1{\rho^2}\left[\frac14\rho^2Q^2\beta^{(+-)}_{\bar{q}Gq,2}(\rho q)-\beta^{(+-)}_{\bar{q}Gq,3}(\rho q)\right]i\epsilon_{\mu\nu\sigma\rho}\langle h'|\bar\psi\gamma_\rho\gamma^5\psi|h\rangle\\
    &-\frac{n_{mol}\gamma_{I\bar{I}}}{2N_c(N_c^2-1)}\left[\frac14\beta^{(+-)}_{\bar{q}Gq,2}(\rho q)-\beta^{(+-)}_{\bar{q}Gq,4}(\rho q)\right]i\epsilon_{\mu\nu\sigma\lambda}q_\lambda q_\rho\langle h'|\bar\psi\gamma_\rho\gamma^5\psi|h\rangle\\
    &-\frac{n_{mol}\gamma_{I\bar{I}}}{2N_c(N_c^2-1)}\left[\frac34\beta^{(+-)}_{\bar{q}Gq,2}(\rho q)-\beta^{(+-)}_{\bar{q}Gq,4}(\rho q)\right]i\epsilon_{\mu\nu\lambda\rho}q_\lambda q_\sigma\langle h'|\bar\psi\gamma_\rho\gamma^5\psi|h\rangle\\
   &-\frac{n_{mol}\gamma_{I\bar{I}}}{2N_c(N_c^2-1)}\beta^{(+-)}_{\bar{q}Gq,5}(\rho q)m^*\epsilon_{\mu\nu\sigma\lambda}q_\lambda \langle h'|\bar\psi i\gamma^5\psi|h\rangle\\
    &-\frac{n_{mol}\gamma_{I\bar{I}}}{2N_c(N_c^2-1)}\rho^2\beta^{(+-)}_{\bar{q}Gq,6}(\rho q)im^*\epsilon_{\mu\nu\alpha\lambda}q_\lambda q_\beta\langle h'|\bar\psi\sigma_{\sigma(\alpha}\gamma^5\overleftrightarrow{\partial}_{\beta)}\psi|h\rangle\\
   &+\frac{n_{mol}\gamma_{I\bar{I}}}{2N_c(N_c^2-1)}\beta^{(+-)}_{\bar{q}Gq,7}(\rho q)im^*\epsilon_{\mu\nu\alpha\lambda}\langle h'|\bar\psi\sigma_{\sigma\alpha}\gamma^5\overleftrightarrow{\partial}_{\lambda}\psi|h\rangle
\end{aligned}
\end{equation}

where the new "molecular form factors" $\beta^{(+-)}$ are defined in Appendix~\ref{MOL-FF}.
\end{widetext}

\section {DIS on a single quark: color force}
\label{SEC-QUARK}
To analyze the effect in a nucleon we start with the simplest model, assuming
that the $ud$ quark pair form a spin-zero ("good scalar") diquark and therefore do not contribute to the color force. The color force is then only active on the unpaired $u$-quark. For (non-forward) quark states,  with incoming momentum $p$ and spin $s$, and outgoing momentum $p'$ and spin $s$, covariance
implies
\begin{equation}
\begin{aligned}
\langle p's|\bar \psi igG_{\mu\nu} \gamma_\sigma \psi|ps\rangle=\bar u_s(p')\Gamma^{(q)}_{\mu\nu\sigma}(p',p) u_s(p)
\end{aligned}
\end{equation}
where $u_s(p)$ stands for the in and out-going (constituent) quark 4-spinor. In the forward limit, the contribution from the single instanton vanishes as those contributions are proportional to the momentum transfer $Q$. The single instanton only contributes to the matrix element when the quark is kicked by non-zero $Q$. Therefore, near $Q=0$ the molecular contribution is dominant.
We now start the evaluation of the vertex function for a single quark using the fields of the instanton-anti-instanton molecules in QCD vacuum
\begin{widetext}
\bea
\Gamma^{(q)}_{\mu\nu\sigma}(p',p)\approx-\frac{n_{mol}}{2N_c(N_c^2-1)}\gamma_{I\bar{I}}\rho^2
 t_{\mu\nu\rho\lambda(\alpha\beta)}\left(\Gamma_{\rho\lambda\alpha\beta\sigma}(p')+\bar\Gamma_{\rho\lambda\alpha\beta\sigma}(p)\right)\nonumber\\
\eea
with
\begin{equation}
\begin{aligned}
    &\Gamma_{\rho\lambda\alpha\beta\sigma}(p)=i\int \frac{d^4k}{(2\pi)^4}\frac{(2p-k)_{\beta}k_{\rho}k_{\lambda}}{2k^2}G(k)\sqrt{\mathcal{F}(p)\mathcal{F}(p-k)}\gamma_{\alpha}\gamma^5S(p-k)\gamma_\sigma\\
    =&-\left(g_{\alpha\delta}\gamma_\sigma\gamma^5+g_{\sigma\delta}\gamma_\alpha\gamma^5 -g_{\alpha\sigma}\gamma_\delta\gamma^5-i\epsilon_{\sigma\alpha\gamma\delta}\gamma_\gamma\right)  \Bigg[\int \frac{d^4k}{(2\pi)^4}\frac{(2p-k)_{\beta}k_{\rho}k_{\lambda}(p-k)_\delta}{2k^2}\frac{G( k)\sqrt{\mathcal{F}(p)\mathcal{F}(p-k)}}{(p-k)^2-M^2}\Bigg]\\
    &+\left(g_{\sigma\alpha}+i\sigma_{\sigma\alpha}\right)\gamma^5M\Bigg[\int \frac{d^4k}{(2\pi)^4}\frac{(2p-k)_{\beta}k_{\rho}k_{\lambda}}{2k^2}\frac{G( k)\sqrt{\mathcal{F}(p)\mathcal{F}(p-k)}}{(p-k)^2-M^2}\Bigg]
\end{aligned}
\end{equation}
and
\begin{equation}
\begin{aligned}
    &\bar\Gamma_{\rho\lambda\alpha\beta\sigma}(p)=i\int \frac{d^4k}{(2\pi)^4}\frac{(2p-k)_{\beta}k_{\rho}k_{\lambda}}{2k^2}G(k)\sqrt{\mathcal{F}(p)\mathcal{F}(p-k)}\gamma_\sigma S(p-k)\gamma_{\alpha}\gamma^5\\
    =&-\left(g_{\alpha\delta}\gamma_\sigma\gamma^5+g_{\sigma\delta}\gamma_\alpha\gamma^5 -g_{\alpha\sigma}\gamma_\delta\gamma^5+i\epsilon_{\sigma\alpha\gamma\delta}\gamma_\gamma\right)  \Bigg[\int \frac{d^4k}{(2\pi)^4}\frac{(2p-k)_{\beta}k_{\rho}k_{\lambda}(p-k)_\delta}{2k^2}\frac{G(k)\sqrt{\mathcal{F}(p)\mathcal{F}(p-k)}}{(p-k)^2-M^2}\Bigg]\\
    &-\left(g_{\sigma\alpha}-i\sigma_{\sigma\alpha}\right)\gamma^5M\Bigg[\int \frac{d^4k}{(2\pi)^4}\frac{(2p-k)_{\beta}k_{\rho}k_{\lambda}}{2k^2}\frac{G( k)\sqrt{\mathcal{F}(p)\mathcal{F}(p-k)}}{(p-k)^2-M^2}\Bigg]
\end{aligned}
\end{equation}
\end{widetext}
We used the effective quark propagator  $S(k)=\frac{i(\slashed{k}+M)}{k^2-M^2}$.
\subsection{Forward limit}
Since these expressions are involved, let us evaluate them first 
at zero momentum transfer, following the decomposition~\cite{Ji:1995mg}, 
\begin{equation}
\begin{aligned}
    \langle ps|\bar \psi igG^{\mu\nu} \gamma^\sigma \psi|ps\rangle&=2id_2^qM\epsilon^{\mu\nu\alpha\beta}s_\alpha p_\beta p^\sigma \\
    &+\frac{2i}3\left(f^q_2+d^q_2\right)M^3\epsilon^{\mu\nu\sigma\alpha}s_\alpha 
\end{aligned}
\end{equation}
Here $d^q_2$ is related to the "twist-3 moment
on the light cone~\cite{Ji:1995mg}. It represents a measure of the quark-gluon correlations and higher-order effects with spin structure.
The parameter $f^q_2$ is a twist-4 and spin-1 matrix element. This matrix element is a higher-twist correction to the spin-dependent deep inelastic scattering structure function, which describes the nucleon's spin structure, and is related to the twist-4 moment

After averaging over quark the spin ("spin independent"), the  contribution vanishes 
\begin{equation}
\begin{aligned}
\langle ps|\bar \psi igG_{\mu\nu} \gamma_\sigma \psi|ps\rangle_{\rm spin-indep}=0
\end{aligned}
\end{equation}
leaving only the spin contribution (proportional to quark the spin four vector $s_\mu$)
\begin{widetext}


\begin{equation}
\begin{aligned}
\label{eq:integral}
&\langle ps|\bar \psi igG_{\mu\nu} \gamma_\sigma \psi|ps\rangle\approx\frac{n_{mol}}{2N_c(N_c^2-1)}\gamma_{I\bar{I}}\, \rho^2t_{\mu\nu\rho\lambda\alpha\beta}4M \\
    &\times\Bigg[\int \frac{d^4k}{(2\pi)^4}\frac{(2p-k)_{\beta}k_{\rho}k_{\lambda}}{2k^2}\frac{G( k)\sqrt{\mathcal{F}(p-k)}}{(p-k)^2-M^2}\left(2p_\sigma s_\alpha+s\cdot kg_{\sigma\alpha}-k_\sigma s_\alpha-k_\alpha s_\sigma\right)\Bigg]\\
    &=\frac{n_{mol}}{N_c(N_c^2-1)}\gamma_{I\bar{I}}\, \left(M\mathcal{I}_1  2i\epsilon_{\mu\nu\alpha\beta}s^\alpha p^\beta  p_\sigma+M^3\mathcal{I}_22i\epsilon_{\mu\nu\sigma\alpha}s^\alpha \right)
\end{aligned}
\end{equation}
The  integrals defined in \eqref{eq:integral} simplify after the Lorentz symmetry reduction.
\begin{equation}
\begin{aligned}
    \mathcal{I}_1\approx&~2\rho^2\int \frac{d^4k}{(2\pi)^4}\frac{G(k)\sqrt{\mathcal{F}(k)}}{(p-k)^2+M^2}\left(\frac{16 \, (k \cdot p)^3}{3 \, k^2 p^4}
- \frac{32 \, (k \cdot p)^2}{3 \, k^2 p^2}
- \frac{8 \, (k \cdot p)^2}{3 \, p^4}
+ \frac{2 k^2}{3 p^2}
+ \frac{8 (k \cdot p)}{3 p^2}
+ \frac{8}{3}\right)\\
    \mathcal{I}_2\approx&~2\rho^2\int \frac{d^4k}{(2\pi)^4}\frac{G( k)\sqrt{\mathcal{F}(k)}}{(p-k)^2+M^2}\left( \frac{8 \, (k \cdot p)^3}{3 \, k^2 p^4}
- \frac{4 \, (k \cdot p)^2}{3 \, p^4}
+ \frac{4 k^2}{3 p^2}
- \frac{8 (k \cdot p)}{3 p^2}\right)
\end{aligned}
\end{equation}

\end{widetext}
with the values  
$\mathcal{I}_1\sim 0.1306$ and ${\mathcal I}_2\sim \,-3.90$.
Using the canonical constituent mass $M=395$ MeV and the enhanced ILM with molecules with $n_{\rm mol}=7.248$ fm$^{-4}$ 
at the low resolution $\mu\sim1/\rho$, 
and  assuming that the unpaired quark carries the whole momentum, we obtain
\bea
    \label{I1I2}
    d_2^q&=&\frac{n_{\rm mol}}{N_c(N_c^2-1)}\gamma_{I\bar{I}}\mathcal{I}_1=0.96\nonumber\\
    \frac13\left(f^q_2+d^q_2\right)&=&\frac{n_{mol}}{N_c(N_c^2-1)}\gamma_{I\bar{I}}\mathcal{I}_2=-28.52 \nonumber\\
\eea
which are to be compared to, 
\bea
    d_2^q&=&\frac{n_{mol}\gamma_{I\bar{I}}}{2N_c(N_c^2-1)}\beta^{(+-)}_{\bar{q}Gq,7}(0) =0.83
    \nonumber\\
    \frac13\left(f^q_2+d^q_2\right)&=&\frac{-n_{\rm mol}\gamma_{I\bar{I}}}{2N_c(N_c^2-1)}\frac{\beta^{(+-)}_{\bar{q}Gq,3}(0)}{\rho^2M^2}=-28.39
    \nonumber\\
\eea
using the short distance approximation as detailed in Appendix~\ref{App:force}. In this case, the results are similar, 
a measure of the accuracy of the approximation. Using the results (\ref{I1I2}), the ensuing average Lorentz force on a constituent quark   
produced by the instanton molecules, is
\bea
F^y= -2M^2d^q_2=-\frac{1.54\,\rm GeV}{\rm fm}d^q_2
\sim 1.47\,\frac{\rm GeV}{\rm fm}\nonumber\\
\eea
which is still larger than the value of the string tension!

\begin{figure}[t!]
    \centering
\subfloat[\label{fig:WF_m}]{\includegraphics[width=\linewidth]{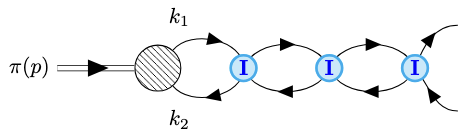}}
    \caption{The pion at low resolution $\mu\lesssim 1/\rho$, traveling  through the QCD instanton vacuum.} 
    \label{fig:Had}
\end{figure}

\begin{figure}
    \centering
    \includegraphics[width=0.9\linewidth]{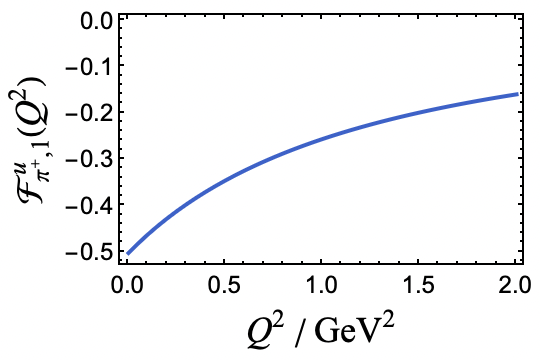}
    \caption{Emergent form factor  $\mathcal{F}^q_{\pi,1}(Q^2)$ induced by a color Lorentz operator in the pion in the ILM enhanced by "molecular" $I\bar I$ pairs. }
    \label{fig:pi_force}
\end{figure}

\begin{figure}
    \centering
    \includegraphics[width=\linewidth]{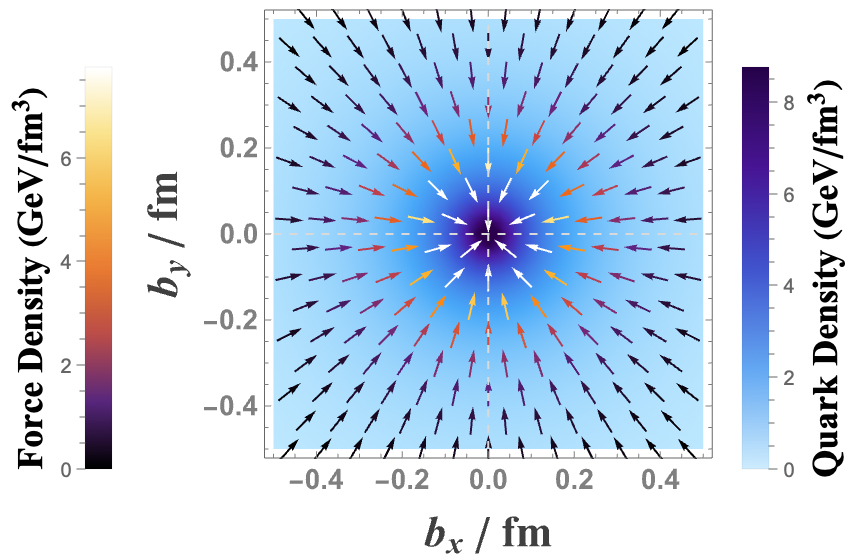}
    \caption{Transverse field distribution of the color Lorentz force
in an unpolarized up quark (arrows), along with the  up quark density distribution (heat map), in impact parameter space for a pion in the ILM enhanced by molecular pairs.}
    \label{fig:pi_den}
\end{figure}

\section{DIS on a pion: color force}
\label{SEC-PION}
The estimate of the effect on a nucleon in the previous section
was done in a schematic quark-diquark model. For  more accurate
estimates, one would need some realistic LF nucleon wave functions.
Fortunately, we will show below that this can be by-passed by using
pertinent form factors. 

To illustrate these ideas, let us start with a simpler case, that of 
deep inelastic scattering on a pion, as its description in the instanton vacuum is well established.  The idea of how the pion emerges 
solely from the short distance 't Hooft $\bar qq$ attraction is diagrammatically illustrated  in Fig.~\ref{fig:Had}.


The non-forward amplitude for the twist-3 operator in a pion target,
is constrained by intrinsic parity and hermiticity. Its Lorentz covariant form
is characterized by two invariant form factors,

\begin{equation}
\label{TWIST3FFPION}
\begin{aligned}
    &\langle \pi(p') |\bar \psi igG^{\mu\nu} \gamma^\sigma \psi| \pi(p)\rangle\\
=&2\Big(\bar{p}^\mu q^{\nu} - \bar{p}^\nu q^\mu \Big)\bar{p}^\sigma \Phi^q_{\pi,1}(Q^2)\\
&+2m_\pi^2\Big(q^\mu g^{\nu\sigma} - q^\nu g^{\mu\sigma} \Big)\Phi^q_{\pi,2}(Q^2)
\end{aligned}
\end{equation}
where
\begin{widetext}

\begin{align}
    \Phi^q_{\pi,1}(Q^2)=&-\frac{n_{mol}}{2N_c(N_c^2-1)}\gamma_{I\bar{I}}\beta^{(+-)}_{\bar{q}Gq,2}(\rho q)A^q_\pi(Q^2)\\
    \Phi^q_{\pi,2}(Q^2)=&\left(\frac{n_{I+\bar{I}}}2\right)\frac{1}{2N_c}\left(\frac{4\pi^2\rho^2}{m^*}\right)\beta^{(+)}_{\bar{q}Gq,1}(\rho q)\frac{\sigma^q_{\pi}(Q^2)}{m}\nonumber\\
    &-\frac{n_{mol}}{8N_c(N_c^2-1)}\gamma_{I\bar{I}}\left[\beta^{(+-)}_{\bar{q}Gq,1}(\rho q)+\beta^{(+-)}_{\bar{q}Gq,2}(\rho q)\right]\left(1+\frac{Q^2}{4m_\pi^2}\right)A^q_\pi(Q^2)\nonumber\\
    &-\frac{n_{mol}}{8N_c(N_c^2-1)}\gamma_{I\bar{I}}\left[\beta^{(+-)}_{\bar{q}Gq,1}(\rho q)-\frac13\beta^{(+-)}_{\bar{q}Gq,2}(\rho q)\right]\frac{3Q^2}{4m_\pi^2}D^q_\pi(Q^2)
\end{align}
\end{widetext}
Here $\sigma_\pi^q$ is the quark scalar form factor for each flavor $q$ in the pion, and $A_\pi^q$ and $D_\pi^q$ are the quark gravitational form factors for each flavor $q$ in the pion. Their detailed definitions are given in Appendix~\ref{App:had_FF}.

At zero momentum transfer, the forward matrix element of the color Lorentz force vanishes 
\begin{equation}
\begin{aligned}
    &\langle \pi(p) |\bar \psi igG^{\mu\nu} \gamma^\sigma \psi| \pi(p)\rangle=0 
\end{aligned}
\end{equation}
This is expected, since the pion does not carry spin.
However, the off-forward matrix element - the form factor of the color Lorenttz force - is non-vanishing. Indeed, 
in the Drell-Yan frames with $q^+=0$,  a collection of frames related by light-front boosts, we specialize to the Breit frame with $p_\perp=0$, where the pion momenta have light-front components
\bea
p^\mu=&&\bigg(p^+, -\frac 12 q_\perp, \frac{m_\pi^2+\frac 14 q_\perp^2}{2p^+}\bigg)
\nonumber\\
p'^\mu=&&\bigg(p^+, +\frac 12 q_\perp, \frac{m_\pi^2+\frac 14 q_\perp^2}{2p^+}\bigg)
\eea
The color force form factor in momentum space, is then
\begin{equation}
\begin{aligned}
\label{eq:pi_f}
    &\langle \pi (p')|\bar \psi \gamma^+igG^{+i}\psi|\pi(p)\rangle=2(p^+)^2q^i\mathcal{F}^q_{\pi,1}(Q^2)
\end{aligned}
\end{equation}
where
\begin{equation}
    \mathcal{F}^q_{\pi,1}(Q^2)=\Phi^q_{\pi,1}(Q^2)
\end{equation}
In transverse coordinate, it is given by
\eqref{eq:pi_f}
\begin{equation}
\begin{aligned}
    F_{q/\pi}^i(b)
    =&\frac{2b^i p^+}{2\pi b}\int_0^\infty dQ Q^2 \mathcal{F}^q_{\pi,1}(Q^2)J_1(Q b)
\end{aligned}
\end{equation}

The behavior of the color force form factor in the pion \eqref{eq:pi_f} is shown in Fig.~\ref{fig:pi_force} for a range of $Q^2$. 
We used the ILM  parameters $n_{\rm mol}=7.248$ \rm fm$^{-4}$ and the physical pion mass \red{$m_\pi=140$ MeV}. The final result is evolved from $\mu=1/\rho\approx650$ MeV to 2 GeV for possible comparison to future lattice simulations.
Note that the vanishing at $Q=0$ of (\ref{eq:pi_f}) is caused by the extra factor of $Q^i$.
In Fig.~\ref{fig:pi_den} we show the distribution of the color-Lorentz force
acting on an unpolarized up quark in the transverse plane (indicated by the vector field), superimposed on the up quark density distribution in
impact parameter space in a  pion. The quark density we get from the pion electromagnetic form factor, using a  vector meson dominance model with the vector meson mass $m_\rho=791$ MeV \cite{Liu:2023yuj,Liu:2023fpj}.

\begin{table*}
    \centering
    \begin{tabular}{|c|c|c|c|c|c|}
    \hline
         & ILM & Göckeler et. al \cite{Gockeler:2005vw} & QCDSF \cite{Crawford:2024wzx} & RQCD \cite{Burger:2021knd} & E143 \cite{E143:1998hbs} \\
         \hline
        $d_{2,N}^u$ & 0.0255 & $0.010(12)$ & 0.079(18) &  0.025(4)(12) & 0.02(4)\\
        \hline
        $d_{2,N}^d$ & $-0.00496$ &  $-0.0056(50)$ & $-0.007(6)$ & $-0.0081(25)(138)$ & 0.02(4) \\
        \hline
        $d_2^p$ & 0.0108& 0.004(5) & 0.046(7)(16) & 0.0105(68) & 0.0122(106) \\
        \hline
        $d_2^n$ & 0.000636& $-0.001(3)$ & 0.023(5)(8) & $-0.0009(70)$ & 0.0106(443) \\
        \hline
    \end{tabular}
    \caption{ILM calculation with $n_{mol}=7.248$ fm$^{-4}$ and a physical pion mass $m_\pi=140$ MeV, evolved to 2 GeV by \eqref{eq:d2_evol}. Our results are compared to the lattice calculation in~\cite{Gockeler:2005vw} at $\mu=2.24$ GeV, the recent lattice calculation from the QCDSF collaboration in~\cite{Crawford:2024wzx} at a renormalization scale of 2 GeV and a heavy pion mass =450 MeV, and the RQCD lattice collaboration in \cite{Burger:2021knd} at a renormalization scale of 2.24 GeV. Our  results are also compared to the experimental fit from the E143 collaboration \cite{E143:1998hbs} at $\mu=2.24$ GeV.}
    \label{tab:d_2}
\end{table*}

\begin{figure}
    \centering
    \includegraphics[width=0.89\linewidth]{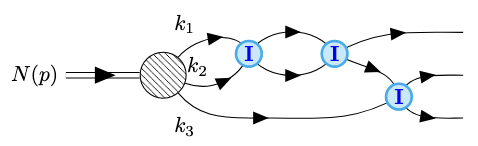}
    \caption{The nucleon at low resolution $\mu\lesssim 1/\rho$, traveling  through the QCD instanton vacuum.}
    \label{fig:placeholder}
\end{figure}

\section{DIS on the nucleon: color force}
\label{SEC-NUCLEON}
The color Lorentz force on a quark in a nucleon can be carried out in the same spirit. Parity invariance, time reversal symmetry and Lorentz covariance, imply that the  matrix element of the Lorentz force in the nucleon is characterized by 8 form factors,
\begin{widetext}
\begin{equation}
\begin{aligned}
    &\langle N(p',s') |\bar \psi igG^{\mu\nu} \gamma^\sigma \psi| N(p,s)\rangle = \bar{u}_{s'}(p') \Bigg\{\left(\bar{p}^{\mu} q^{\nu}-\bar{p}^{\nu} q^{\mu}\right)\frac{\bar{p}^{\sigma}}{m_N}\Phi^q_{N,1}(Q^2)+m_N\left(q^{\mu}g^{\sigma\nu}-q^{\nu}g^{\sigma\mu}\right)\Phi^q_{N,2}(Q^2)\\
    &+m_N i\sigma^{\mu\nu}\bar{p}^\sigma \Phi^q_{N,3}(Q^2) +m^2_Ni\epsilon^{\mu\nu\sigma\lambda}\gamma_\lambda\gamma^5\Phi^q_{N,4}(Q^2)+m_Ni \epsilon^{\mu\nu\sigma\lambda}q_\lambda\gamma^5\Phi^q_{N,5}(Q^2)\\
    &+\frac{i\sigma^{\mu\alpha}\bar{p}^{\nu} -i\sigma^{\nu\alpha}\bar{p}^{\mu}}{2m_N}  q_\alpha q^{\sigma}\Phi^q_{N,6}(Q^2)+ \left(\bar{p}^{\mu} q^{\nu}-\bar{p}^{\nu} q^{\mu}\right)\frac{i\sigma^{\sigma\lambda}q_\lambda}{2m_N}\Phi^q_{N,7}(Q^2)+ \frac{i\sigma^{\mu\alpha} q^{\nu}-i\sigma^{\nu\alpha}q^{\mu}}{2m_N}q_\alpha \bar{p}^{\sigma}\Phi^q_{N,8}(Q^2)\Bigg\} u_s(p)
\end{aligned}
\end{equation}
With the help of the short distance approximation from \eqref{eq:CFOp2} and \eqref{eq:CFOp3}, each form factor can be expressed in terms of more standard nucleon form factors, as detailed in Appendix \ref{App:force}, 
\begin{align}
    \Phi^q_{N,1}(Q^2)=&
    -\frac{n_{mol}\gamma_{I\bar{I}}}{2N_c(N_c^2-1)}\beta^{(+-)}_{\bar{q}Gq,2}(\rho q)A^q_N(Q^2)+\frac{n_{mol}\gamma_{I\bar{I}}}{2N_c(N_c^2-1)}\beta^{(+-)}_{\bar{q}Gq,7}(\rho q) \frac{m}{m_N} \left[\tilde{A}^q_{T}(Q^2)+\frac12B^q_{T}(Q^2)\right]
\end{align}

\begin{align}
    \Phi^q_{N,2}(Q^2)=&\left(\frac{n_{I+\bar{I}}}2\right)\frac{1}{2N_c}\left(\frac{4\pi^2\rho^2}{m^*}\right)\beta^{(+)}_{\bar{q}Gq,1}(\rho q)\frac{\sigma^q_N(Q^2)}{m}\nonumber\\
    &-\frac{n_{mol}\gamma_{I\bar{I}}}{8N_c(N_c^2-1)}\left[\beta^{(+-)}_{\bar{q}Gq,1}(\rho q)+\beta^{(+-)}_{\bar{q}Gq,2}(\rho q)\right]\left[\left(1+\frac{Q^2}{4m_N^2}\right)A^q_N(Q^2)-\frac{Q^2}{2m_N^2}J^q_N(Q^2)\right]\nonumber\\
    &-\frac{n_{mol}\gamma_{I\bar{I}}}{8N_c(N_c^2-1)}\left[\beta^{(+-)}_{\bar{q}Gq,1}(\rho q)-\frac{1}3\beta^{(+-)}_{\bar{q}Gq,2}(\rho q)\right]\frac{3Q^2}{4m_N^2}D_N^q(Q^2)\nonumber\\
    &+\frac{n_{mol}\gamma_{I\bar{I}}}{2N_c(N_c^2-1)}\beta^{(+-)}_{\bar{q}Gq,7}(\rho q) \frac{m}{m_N} \left[\frac12A^q_{T}(Q^2)+\frac12B^q_{T}(Q^2)+\left(1+\frac{Q^2}{4m_N^2}\right)\tilde{A}^q_{T}(Q^2)+2C^q_T(Q^2)\right]\\[7pt]
    \Phi^q_{N,3}(Q^2)=&\frac{n_{mol}\gamma_{I\bar{I}}}{2N_c(N_c^2-1)}\frac{m}{m_N}\left[\beta^{(+-)}_{\bar{q}Gq,7}(\rho q)-\frac12\rho^2Q^2\beta^{(+-)}_{\bar{q}Gq,6}(\rho q)\right]A^q_{T}(Q^2)\\
    \Phi^q_{N,4}(Q^2)=&-\left(\frac{n_{I+\bar{I}}}2\right)\frac{1}{2N_c}\left(\frac{4\pi^2\rho^2}{m^*}\right)\frac{\rho^2Q^2}{2}\beta^{(+)}_{\bar{q}Gq,2}(\rho q)\frac{m}{m_N^2}G^q_A(Q^2)\nonumber\\
    &-\frac{n_{mol}\gamma_{I\bar{I}}}{2N_c(N_c^2-1)}\frac{1}{\rho^2m^2_N}\left[\beta^{(+-)}_{\bar{q}Gq,3}(\rho q)-\frac14\rho^2Q^2\beta^{(+-)}_{\bar{q}Gq,2}(\rho q)\right]G^q_{A}(Q^2)
\end{align}

\begin{align}
    \Phi^q_{N,5}(Q^2)=&\left(\frac{n_{I+\bar{I}}}2\right)\frac{1}{2N_c}\left(\frac{4\pi^2\rho^2}{m^*}\right)\left[\beta^{(+)}_{\bar{q}Gq,1}(\rho q)\frac{\tilde{G}^q_P(Q^2)}{m}-\frac14\rho^2Q^2\beta^{(+)}_{\bar{q}Gq,2}(\rho q)\frac{m}{m_N^2}G^q_P(Q^2)\right]\nonumber\\
    &+\frac{n_{mol}\gamma_{I\bar{I}}}{N_c(N_c^2-1)}\left[\beta^{(+-)}_{\bar{q}Gq,4}(\rho q)-\frac14\beta^{(+-)}_{\bar{q}Gq,2}(\rho q)\right]G^q_{A}(Q^2)\nonumber\\
    &+\frac{n_{mol}\gamma_{I\bar{I}}}{2N_c(N_c^2-1)}\frac{1}{2\rho^2m^2_N}\left[\frac12\rho^2Q^2\beta^{(+-)}_{\bar{q}Gq,2}(\rho q)-\beta^{(+-)}_{\bar{q}Gq,3}(\rho q)-\rho^2Q^2\beta^{(+-)}_{\bar{q}Gq,4}(\rho q)\right]G^q_{P}(Q^2)\nonumber\\
    &-\frac{n_{mol}\gamma_{I\bar{I}}}{2N_c(N_c^2-1)}\beta^{(+-)}_{\bar{q}Gq,5}(\rho q)\tilde{G}^q_P(Q^2)\\
    &+\frac{n_{mol}\gamma_{I\bar{I}}}{2N_c(N_c^2-1)}\beta^{(+-)}_{\bar{q}Gq,6}(\rho q)\frac{m}{m_N}\left[\rho^2m_N^2D_T^q(Q^2)-\frac12\rho^2Q^2\tilde{D}_T^q(Q^2)\right]\\
    \Phi^q_{N,6}(Q^2)=
    &-\frac{n_{mol}\gamma_{I\bar{I}}}{N_c(N_c^2-1)}\left[\frac34\beta^{(+-)}_{\bar{q}Gq,2}(\rho q)-\beta^{(+-)}_{\bar{q}Gq,4}(\rho q)\right]G^q_A(Q^2)\nonumber\\
    &+\frac{n_{mol}\gamma_{I\bar{I}}}{2N_c(N_c^2-1)}\left[\beta^{(+-)}_{\bar{q}Gq,7}(\rho q)+\frac12\rho^2Q^2\beta^{(+-)}_{\bar{q}Gq,6}(\rho q)\right]\frac{m}{m_N} \tilde{B}^q_{T}(Q^2)\\
    &-\frac{n_{mol}\gamma_{I\bar{I}}}{2N_c(N_c^2-1)}\beta^{(+-)}_{\bar{q}Gq,6}(\rho q)\rho^2m m_N D^q_{T}(Q^2)
\end{align}

\begin{align}
    \Phi^q_{N,7}(Q^2)=&-\left(\frac{n_{I+\bar{I}}}2\right)\frac{1}{N_c}\left(\frac{4\pi^2\rho^2}{m^*}\right)\beta^{(+)}_{\bar{q}Gq,2}(\rho q)\rho^2mG^q_A(Q^2)\nonumber\\
    &-\frac{n_{mol}\gamma_{I\bar{I}}}{2N_c(N_c^2-1)}\beta^{(+-)}_{\bar{q}Gq,2}(\rho q)J^q_{N}(Q^2)\\
    \Phi^q_{N,8}(Q^2)=&\left(\frac{n_{I+\bar{I}}}2\right)\frac{1}{N_c}\left(\frac{4\pi^2\rho^2}{m^*}\right)\beta^{(+)}_{\bar{q}Gq,2}(\rho q)\rho^2mG^q_A(Q^2)\nonumber\\
    &-\frac{n_{mol}\gamma_{I\bar{I}}}{2N_c(N_c^2-1)}\beta^{(+-)}_{\bar{q}Gq,2}(\rho q)J^q_{N}(Q^2)\nonumber\\
    &+\frac{n_{mol}\gamma_{I\bar{I}}}{2N_c(N_c^2-1)}\beta^{(+-)}_{\bar{q}Gq,7}(\rho q) \frac{m}{2m_N} B^q_{T}(Q^2)
\end{align} 

\end{widetext}

\subsection{ Parameter $d_2$ of the nucleons}
At zero momentum transfer, the single instanton contribution is suppressed. The zero momentum transfer receives contributions from molecular configurations. $\Phi^q_{3,N}(0)$ is equal to $d^q_{2,N}$, 
\begin{equation}
\begin{aligned}
    d^q_{2,N}=&\frac{n_{mol}\gamma_{I\bar{I}}}{2N_c(N_c^2-1)}\beta^{(+-)}_{\bar{q}Gq,7}(0)\frac{m}{m_N}A^{q}_{T}(0)
\end{aligned}
\end{equation}

If we assume the evolution does not mix flavors at one-loop, the one-loop evolution equation is \cite{Burger:2021knd} 
\begin{equation}
\label{eq:d2_evol}
    d_{2}(\mu) = 
\left( \frac{\alpha_{s}(\mu')}{\alpha_{s}(\mu)} \right)^{-\frac{\gamma_{d_2}}{\beta_0}}
\, d_{2}(\mu')
\end{equation}
where $\beta_0=\frac{11}{3}N_c - \frac{2}{3}N_f$ and the one-loop anomalous dimension is defined by
\begin{equation}
  \gamma_{d_2}= 3N_c - \frac{1}{6}\left( N_c - \tfrac{1}{N_c}\right) 
\end{equation}
The total proton and neutron $d_2$ can be constructed from $d^q_{2,N}$ using~\cite{Burger:2021knd}
\begin{align}
\label{eq:d2N}
d_{2}^{p} &= \left( \tfrac{2}{3} \right)^{2} d_{2,N}^{u} 
             + \left( -\tfrac{1}{3} \right)^{2} d_{2,N}^{d} \\[6pt]
d_{2}^{n} &= \left( -\tfrac{1}{3} \right)^{2} d_{2,N}^{u} 
             + \left( \tfrac{2}{3} \right)^{2} d_{2,N}^{d}
\end{align}

To compare our results with the lattice and global analyses, we evolve our ILM result from $\mu=1/\rho\approx650$ MeV to $\mu=2$ GeV. The one-loop perturbative correction gives around $$d_2(\mu=2~\mathrm{GeV})\approx0.48\,d_2(\mu=1/\rho)$$ 
The final result is shown in Table~\ref{tab:d_2} with our result using $n_{\rm mol}=7.248$ fm$^{-4}$ and the physical pion mass $m_\pi=140$ MeV, corresponding to current quark mass $m(\mu=1/\rho)=10$ MeV, which has been suggested in many low energy vacuum-based models \cite{Liu:2025ldh,Liu:2023yuj,Schafer:1996wv,Hatsuda:1994pi,Liu:2025kuc,Diakonov:2002fq}. 
Note that the result of QCDSF collaboration in Table~\ref{tab:d_2}, there is a mismatch between the $d_2^p$ and $d_2^n$ obtained from $d_2^u$ and $d_2^d$ by \eqref{eq:d2N} presented in their plot, and the direct value of $d_2^p$ and $d_2^n$  they present in \cite{Crawford:2024wzx}.

\red{We note that the one-loop evolution from the low renormalization point $\mu \sim 1/\rho \approx 650$ MeV to higher scales introduces sizable uncertainties and scheme dependence, particularly below 1 GeV where perturbation theory is marginal. Therefore, the evolution presented here should be regarded as qualitative and illustrative. A quantitatively reliable matching to higher scales would require nonperturbative renormalization or lattice input as outlined in~\cite{Shuryak:2026pqt}, which is still under development and lies beyond the scope of the present work.}


\subsection{Color force FFs of the nucleons}
To evaluate the color force form factors (FFs) in the nucleon, we specialize 
again to the Breit frames with $q^+=0$ and $p_\perp=0$ for the nucleon. 
The form factors defined on the light front are
\begin{widetext}
\begin{equation}
\begin{aligned}
    &\langle N(p's)|\bar \psi \gamma^+igG^{+i}\psi|N(ps)\rangle=\\
    &\bar u_s(p')\Bigg[\gamma^+q^i\mathcal{F}^q_{N,1}(Q^2)+m_Ni\sigma^{+i}\mathcal{F}^q_{N,2}(Q^2)-\frac{i\sigma^{+j}q_jq^i}{m_N}\mathcal{F}^q_{N,3}(Q^2)\Bigg]p^+u_s(p)
\end{aligned}
\end{equation}
\end{widetext}
with 
\bea
    \mathcal{F}^q_{N,1}(Q^2)&=&\Phi^q_{N,1}(Q^2)\nonumber\\
    \mathcal{F}^q_{N,2}(Q^2)&=&\Phi^q_{N,3}(Q^2)
\eea    
and
\bea
    &&\mathcal{F}^q_{N,3}(Q^2)=\nonumber\\
    &&-\frac12\left[\Phi^q_{N,7}(Q^2)+\Phi^q_{N,8}(Q^2)-\Phi^q_{N,1}(Q^2)\right]
    \nonumber\\
\eea
Note that all  $\mathcal{F}^q_{N,1,2,3}(Q^2)$ are proportional to $n_{\rm mol}$. The single instanton contribution in $\Phi_{N,7}$ and $\Phi_{N,8}$ cancel each other. Thus, even though in general the off-foward matrix element of $\bar\psi\gamma_\sigma G_{\mu\nu}\psi$ receives contribution from single instantons, the leading twist-3 contribution only receives contribution from $I\bar I$ molecules.

\begin{figure}
    \centering
\includegraphics[width=\linewidth]{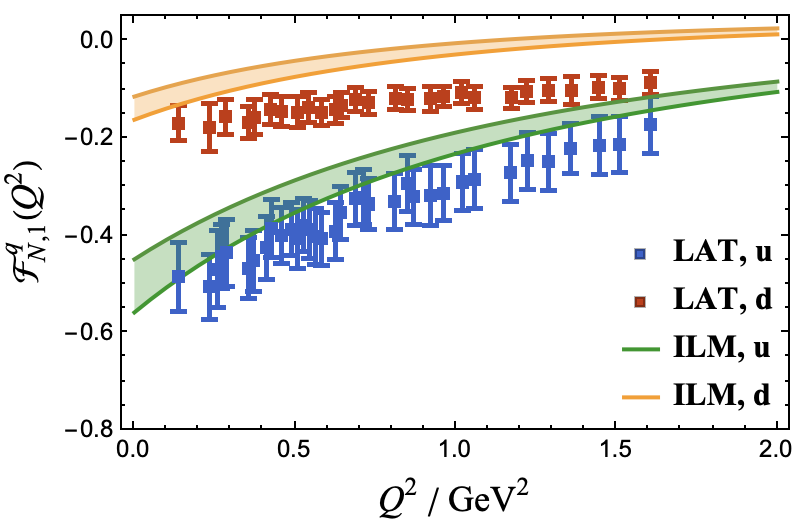}
\caption{Color force form factors $\mathcal{F}^q_{N,1}(Q^2)$ from ILM with parameters $n_{mol}=7.248$ fm$^{-4}$ and \red{ current quark mass $m=10-58$ MeV ($m_\pi=140-338$ MeV) showing with bands}, evolved to 2 GeV compared to the lattice calculation with pion mass $450$ MeV in \cite{Crawford:2024wzx}.}
\label{fig:nuc_force1}
\end{figure}

The color force form factor $\mathcal{F}^q_{N,1}(Q^2)$ is related to the {\it unpolarized nucleon gravitational form factor} $A^q_{N}$ \red{and {\it transversity gravitational form factor} $2\tilde{A}^q_{T}+B^q_{T}$}. Its behavior is shown in Fig.~\ref{fig:nuc_force1}, 
using the ILM with molecule density   $n_{mol}=7.248$ fm$^{-4}$.

\red{In order to facilitate comparison with lattice results obtained at heavier pion masses ($m_\pi = 450$ MeV), we adjust the quark mass $m$ in the ILM using the chiral relation $m_\pi \propto \sqrt{m}$. 
The explicit numerical relation can be found in Table~\ref{tab:mass}. Under the assumption that the other ILM parameters ($n_{mol}$, $\gamma_{I\bar I}$) and the nucleon mass $m_N$ are only weakly dependent on the quark mass $m$, this procedure provides an effective matching between the model parameters and the lattice setup. For completeness, we also present results over a range of the quark mass $m=10$ MeV ($m_\pi=140$ MeV) and $m=58$ MeV ($m_\pi=338$ MeV), which allows for a comparison of the quark-mass sensitivity. The plots suggest strong quark mass dependence in the second color force fomr factor $\mathcal{F}^q_{N,2}(Q^2)$, which is opposite to the lattice observation in \cite{Burger:2021knd}.}


The results are evolved  to 2 GeV, and compared to the lattice calculation with pion mass $450$ MeV in~\cite{Crawford:2024wzx}. Both the $u$- and $d$-flavor components of the Lorentz force are in good agreement with the reported lattice results, over a relatively broad range of $Q^2$.

\begin{figure}
    \centering
\includegraphics[width=\linewidth]{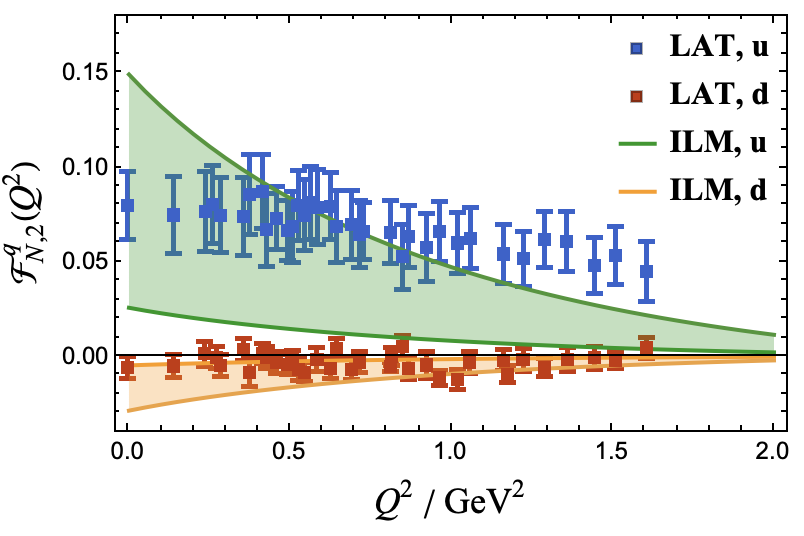}
\caption{Color  Lorentz force form factors $\mathcal{F}^q_{N,2}(Q^2)$ using the ILM with parameters $n_{\rm mol}=7.248$ fm$^{-4}$ and \red{ current quark mass $m=10-58$ MeV ($m_\pi=140-338$ MeV) showing with bands}, evolved to 2 GeV and compared to the lattice calculation with pion mass $450$ MeV in~\cite{Crawford:2024wzx}.}
\label{fig:nuc_force2}
\end{figure}

The color force form factor $\mathcal{F}^q_{N,2}(Q^2)$ is fully related to the {\it transversity gravitational form factor}  $A^q_{T}$. Its behavior is shown in Fig.~\ref{fig:nuc_force2} using the ILM parameters. The comparison to the recently reported lattice results in~\cite{Crawford:2024wzx} is fair. Note that  the smallness of this form factor originates from its  suppression by the quark mass $m$ and in the ILM enhanced by molecules.  

\begin{figure}
    \centering
\includegraphics[width=\linewidth]{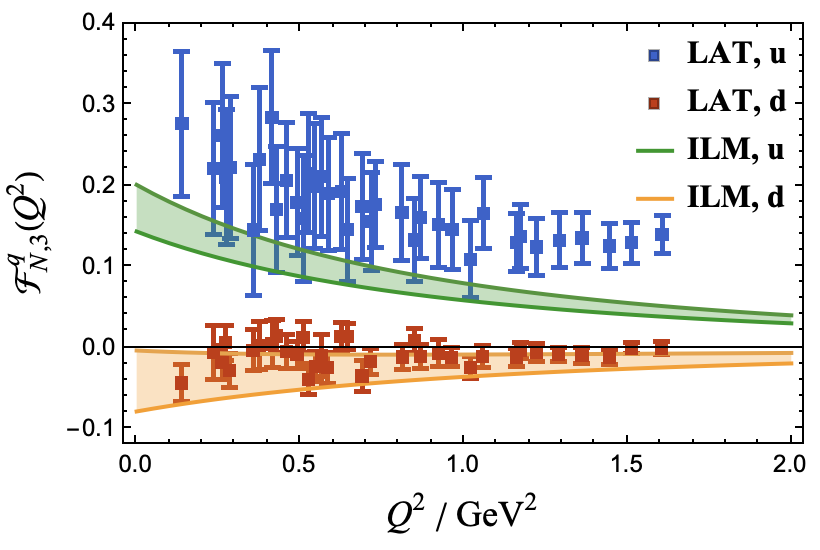}
\caption{Color Lorentz force form factors $\mathcal{F}^q_{N,3}(Q^2)$ from the ILM with parameters $n_{\rm mol}=7.248$ fm$^{-4}$ and \red{ current quark mass $m=10-58$ MeV ($m_\pi=140-338$ MeV) showing with bands}, evolved to 2 GeV and compared to the lattice calculation with pion mass $450$ MeV in~\cite{Crawford:2024wzx}.}
\label{fig:nuc_force3}
\end{figure}

The color force form factor $\mathcal{F}^q_{N,3}(Q^2)$ is a combination of  {\it gravitational form factor} $B^q_N$ and \red{{\it transversity gravitational form factor} $\tilde{A}^q_{T}$}. This form factor is displayed in Fig.~\ref{fig:nuc_force3} using the ILM enhanced by molecules, and compared 
to the recent lattice results~\cite{Crawford:2024wzx}. Again the comparison is satisfactory.  Since $B^u_N+B^d_N\approx0$, the asymmetry between $u$ and $d$ arises from \red{the non-vanishing isoscalar transversity GPD moment $\tilde{A}^{u+d}_{T}(0)$.} 

\begin{figure}
    \centering
\subfloat[]{\includegraphics[width=1\linewidth]{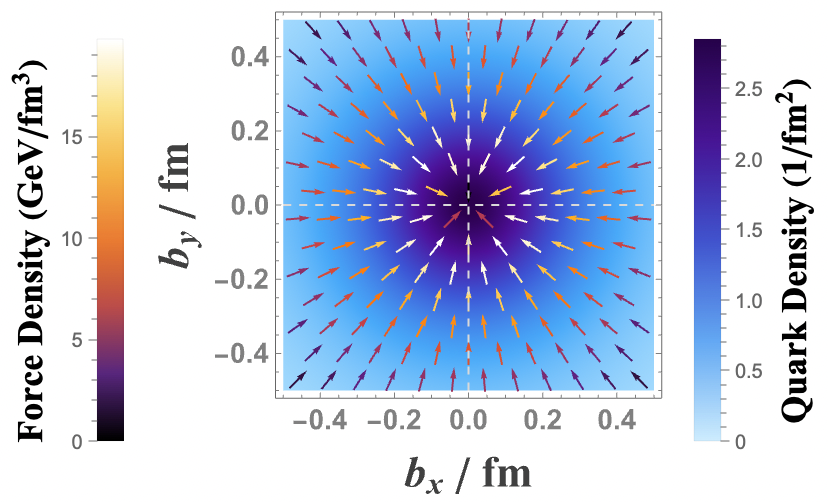}}
\hfill
\subfloat[]{\includegraphics[width=1\linewidth]{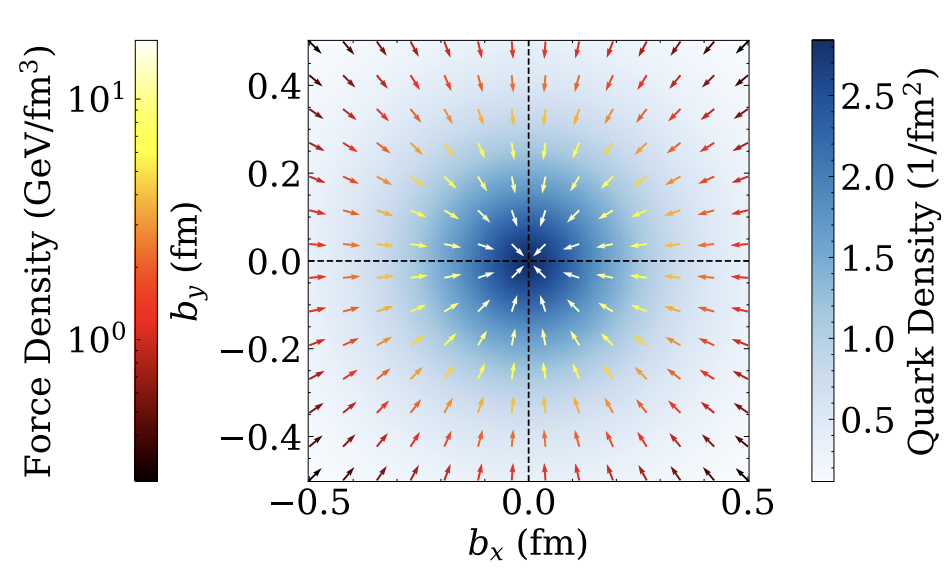}}
    \caption{Transverse field distribution of the color Lorentz
force in an  unpolarized up quark  (arrows), along with the up quark density distribution (heat map), 
 for an unpolarized  proton  in the enhanced ILM. The up quark density follows from  the proton Dirac and Pauli form factors from the phenomenological fit in \cite{Alberico:2008sz}. The results are compared to the lattice results in~\cite{Crawford:2024wzx}.}
    \label{fig:nuc_den}
\end{figure}

\begin{figure}
    \centering
\subfloat[]{\includegraphics[width=1\linewidth]{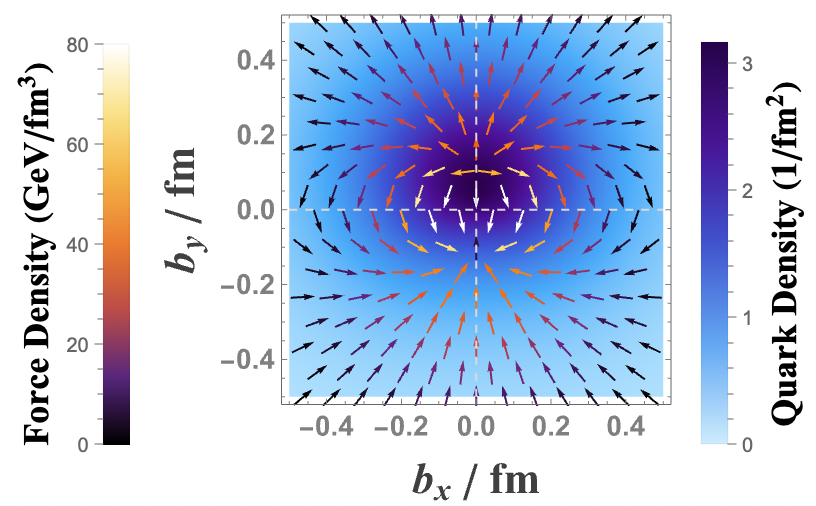}}
\hfill
\subfloat[]{\includegraphics[width=1\linewidth]{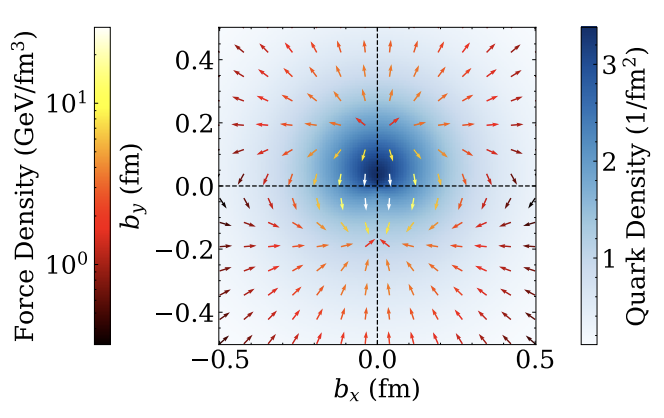}}
    \caption{Transverse field distribution of the color Lorentz
force in an  unpolarized up quark  (arrows), along with the up quark density distribution (heat map), 
 for a proton transversely polarized in $\hat{x}$ direction in the enhanced ILM. The up quark density follows from  the proton Dirac and Pauli form factors from the phenomenological fit in \cite{Alberico:2008sz}. The results are compared to the lattice results in~\cite{Crawford:2024wzx}.}
    \label{fig:nuc_den_2}
\end{figure}

\red{In Fig.~\ref{fig:nuc_den} and \ref{fig:nuc_den_2}, we plot the color force density on a transverse plane at physical pion mass ($m_\pi=140$ MeV). The result is also compared to the lattice calculation in \cite{Crawford:2024wzx} at a heavy pion mass ($m_\pi=450$ MeV).}

\section{Summary}
\label{SEC-CON}


We have developed a comprehensive and unified treatment of the twist-3 color Lorentz force in pions and nucleons, within the framework of the instanton liquid model and its "molecular" $I \bar I$ extension. 

The color Lorentz force arises naturally from the non-perturbative QCD vacuum fluctuations of the gauge field, including  correlated instanton–anti-instanton  pairs.
By combining semiclassical strength and spatial structure of these fields  with the operator structure of higher-twist quark–gluon correlators, we have shown  that   molecular correlations provide an important mechanism behind the large higher-twist effects observed in polarized DIS, and confirmed in recent lattice simulations.
This agrees with our previous derivation of the confining potential
in quarkonia, from the same setting.

One new result of this study is the explicit construction and analysis of the gauge field structure of the molecular ensemble,
and its role in the quark-colored force. By employing the "ratio ansatz" for the correlated instanton–anti-instanton pair, we derived explicit expressions for the color-electric and color-magnetic field components and computed their spatial profiles. The overlap of the quark zero modes across the molecular pair was analyzed in detail, revealing a strong delocalization effect that enhances the color force.  Using Monte Carlo sampling over molecular orientations and separations, we obtained a robust estimate of the average color Lorentz force, $F\approx2$–$3\,\mathrm{GeV/fm}$, acting on a single quark. This result indicates that the nonperturbative color forces are comparable to, or can even exceed the magnitude of the confining string tension.

Our  analysis shows a  direct connection between these microscopic nonperturbative fields and the experimentally accessible twist-3 observables. The color Lorentz force operator $\bar{q}gG^{+i}\gamma^+q$—whose expectation value defines the twist-3 matrix element $d_2$—is shown to acquire  contributions from the molecular component of the instanton liquid. The induced  form factors  encode the nonlocal structure of this coupling and provide a unified description of the quark–gluon interaction over a broad range of momentum transfers. In particular, the molecular contributions dominate at low $Q^2$, leading to enhanced color forces consistent with phenomenological extractions from polarized structure functions.  These findings confirm that the twist-3 sector is a direct manifestation of the underlying topological fields in the QCD vacuum.
Remarkably, the   emergent color Lorentz force  form factors are shown to be intimately related to the hadronic gravitational and transversity form factors, offering additional insights to the nature of the mass and force distributions in hadrons. \red{However, one has to keep in mind that the relations shown in \eqref{eq:CFOp2} and \eqref{eq:CFOp3} are built on the local approximation of the original relation in \eqref{eq:force_op_0}, which involves a more complicated nonlocal two-current structure (see Appendix~\ref{App:force}).}

From a phenomenological perspective, the implications of our results are as follows. For a constituent quark, the transverse color Lorentz force reaches $F_y\sim1.5\,\mathrm{GeV/fm}$, reproducing the scale of the force inferred from lattice evaluations of $d_2$ and phenomenological analysis of $g_2(x,Q^2)$. In contrast, the pion, as a spinless bound state, exhibits a vanishing $d^\pi_2$, consistent with the absence of a net transverse Lorentz force. This clear dichotomy between spin-1/2 and spin-0 systems demonstrates that the twist-3 color force is intrinsically spin-dependent and directly tied to the internal color-magnetic structure of the hadron. Moreover, the ability of the instanton–molecule framework to simultaneously describe both systems underscores its universality as a source of nonperturbative QCD dynamics.

Conceptually, this study shows how topological excitations of the QCD vacuum—once considered peripheral to the hadron structure—directly govern measurable high-energy observables. The emergence of strong, localized color forces from correlated instanton configurations provides a compelling microscopic picture that complements and extends conventional models based on confinement and gluon exchange. The instanton–molecule framework offers a natural bridge between the semiclassical field picture and the partonic description of hadrons, thereby unifying aspects of nonperturbative vacuum structure and experimental phenomenology.

The results also suggest new directions for future research. Extending the present formalism to generalized parton distributions (GPDs) and transverse-momentum-dependent distributions (TMDs) would enable mapping of the spatial and dynamical structure of color forces in both position and momentum space. Lattice simulations that isolate instanton and molecular contributions could provide stringent tests of the predicted force magnitudes and their dependence on quark flavor and spin. 


\red{
The present analysis is subject to several limitations inherent to the ILM framework. First, confinement is not explicitly incorporated, and the results depend on phenomenological input for the instanton ensemble, including the typical instanton size and density. Second, the semiclassical treatment itself is not systematically controlled by a small parameter although the many-body expansion in our current framework is, and therefore the quantitative predictions should be interpreted at the level of order-of-magnitude estimates. Third, the use of a local approximation for the quark-gluon operator neglects nonlocal contributions that may become relevant at larger distances or lower momenta. Finally, the perturbative evolution from low scales introduces additional uncertainties. 
These notwithstanding, the framework captures essential nonperturbative features of the QCD vacuum and provides a coherent and physically transparent mechanism for the emergence of sizable color Lorentz forces, in qualitative agreement with recent lattice observations.}

In summary, the instanton–anti-instanton molecular component of the QCD vacuum is an important  source of the nonperturbative color Lorentz forces responsible for twist-3 phenomena in light hadrons. By  relating the semiclassical field theory to measurable quantities, we have provided both qualitative understanding and quantitative predictions that connect the topology of the QCD vacuum to hadronic observables. The framework presented here not only enhances our comprehension of the QCD vacuum but also lays the foundation for a broader, unified picture of nonperturbative dynamics in strong interaction physics, ranging from hadronic spectroscopy  to 
partonic physics on the light front.

\begin{acknowledgments}\noindent
This work is supported by the U.S. Department of Energy, Office of Science, Office of Nuclear Physics under Contract No. DE-FG-88ER40388. This work is also supported in part by the Quark-Gluon Tomography (QGT) Topical Collaboration, with Award DE-SC0023646.
\end{acknowledgments}

\appendix

\section{Emergent instanton induced interactions}
\label{App:mol}
To help organize the many-body calculus in the ILM enhanced by molecules, we introduce the effective vertices induced by single pseudoparticles and molecule, and we use the  $1/N_c$ counting rules to organize the contributions. For that, consider the ILM vacuum partition function 

\begin{widetext}
\begin{equation}
\begin{aligned}
\label{eq:Z_N2}
   Z_{\rm ILM}=&\int \mathcal{D}\psi \mathcal{D}\psi^\dagger \frac1{N_+!N_-!}\prod_{I=1}^{N_++N_-}\left(\int d^4z_IdU_I\int d\rho_I n(\rho_I)\rho_I^{N_f}\Theta_I(z_I) \right)\exp\left(\int d^4x\psi^\dagger i\slashed{\partial}\psi\right)\\
   \rightarrow&\int \mathcal{D}\psi \mathcal{D}\psi^\dagger\exp\left[-\int d^4x\left(-\psi^\dagger i\slashed{\partial}\psi+\mathcal{L}_{\rm inst}+\mathcal{L}_{\rm mol}+\cdots\right)\right]
\end{aligned} 
\end{equation}
\end{widetext}
The molecular contributions follow from $\mathcal{L}_{\rm mol}$, while the single pseudoparticle contributions from  $\mathcal{L}_{\rm inst}$, which of the detailed expression can be found in \cite{Liu:2025ldh}. 

\begin{figure}
    \centering
\subfloat[\label{mol_1}]{\includegraphics[width=0.33\linewidth]{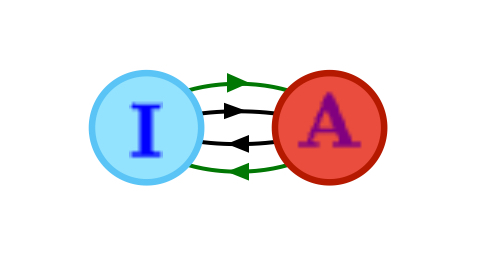}}
\hfill
\subfloat[\label{mol_2}]{\includegraphics[width=0.33\linewidth]{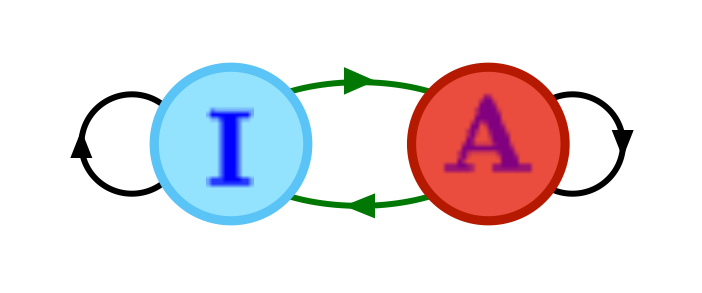}}
\hfill
\subfloat[\label{mol_3}]{\includegraphics[width=0.33\linewidth]{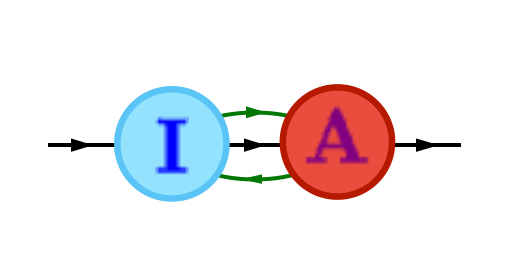}}
    \caption{Feynman diagrams for the vertices induced by a close pair of instanton ($I$) and anti-instanton ($A$): (a) the vacuum tunneling rate of a fully connected molecule, (b) its flavor reduced vacuum tunneling rate, and (c) the one-body (two-Fermi) vertex induced by an instanton-anti-instanton pair.}
    \label{fig:mole_0}
\end{figure}


The molecular  interactions  between an instanton-anti-instanton pair mediated by light quarks,  follow from 
\begin{equation}
\begin{aligned}
\label{eq:L_mol}
    \mathcal{L}_{\rm mol}=&\int d\rho_Id\rho_{\bar{I}}n(\rho_I)n(\rho_{\bar{I}})\rho_I^{N_f}\rho_{\bar{I}}^{N_f}\\
    &\times\int dud^4R\Theta_{I\bar{I}}(z_I,z_{\bar{I}})
\end{aligned}
\end{equation}
where each instanton-anti-instanton pair induced vertex is defined by 
\begin{equation}
    \Theta_{I\bar{I}}(z_I,z_{\bar{I}})=[\Theta_{I}(z_I)\Theta_{\bar{I}}(z_{\bar{I}})]_{\rm conn},
\end{equation}
with all possible contractions. Some contractions are illustrated in Fig.~\ref{fig:mole_0} where the instanton vertex $\Theta_{I}$ is defined as
\begin{equation}
    \begin{aligned}
        \label{eq:tHooft}
    &\Theta_{I}=(4\pi^2\rho^2)^{N_f}\\
    &\times\prod^{N_f}_{q=1}\left(\frac{m_q}{4\pi^2\rho^2}-\frac{1}{8}\bar{q}_{R}U_I\tau^-_\mu\tau^+_\nu\gamma_\mu\gamma_\nu U_I^\dagger q_{L}\right)
    \end{aligned}
\end{equation}
Here  $q$ denotes the quark flavor with current mass $m_q$, and $L,R$ represents the chirality of the quark fields. A similar vertex holds for the anti-instantons, differing only by the interchange between $\tau^-_\mu\leftrightarrow\tau^+_\mu$ and $q_{L}\leftrightarrow q_{R}$.

The molecular tunneling rate can be obtained by resumming all closed diagrams such as in Fig.~\ref{mol_1} and \ref{mol_2}. The dominant contribution is from Fig.~\ref{mol_1}, due to the strong attraction induced by fermion exchange. The molecular  tunneling rate is approximately of the form 
\begin{equation}
\begin{aligned}
\label{pair_density}
    n_{mol}=&\int d\rho_Id\rho_{\bar{I}}n(\rho_I)n(\rho_{\bar{I}})\rho_I^{N_f}\rho_{\bar{I}}^{N_f}\\
    &\times\int d^4Rdu|T_{I\bar{I}}(u,R)|^{2N_f}\\    =&\left[\frac{n_{I+\bar{I}}}2\left(\frac{4\pi^2\rho^2}{m^*}\right)^{N_f}\right]^2\left(\frac{|T_{I\bar{I}}|}{4\pi^2\rho^2}\right)^{2N_f}
\end{aligned}
\end{equation}
where the quark hopping across the instanton-anti-instanton pair yields the hopping integral $T_{I\bar{I}}$ (the quark zero mode overlap between the pair) defined in \eqref{eq:hop}.

\red{The resulting determinantal mass is determined by}

\begin{equation}
\label{eq:det_gap}
    m^*=m+\frac{2\pi^2\rho^2}{N_c}\langle\bar{q}q\rangle
\end{equation}
\red{which characterizes the leading contribution in the ensemble average of the fermionic determinant with a measure of the unquenched the gauge configurations~\cite{Liu:2025kuc,Liu:2025ldh}.}
\begin{equation}
    \left\langle\rho^{NN_f} \prod_f\mathrm{Det}(\slashed{D}+m)_{\mathrm{ZM}}\right\rangle=(\rho m^*)^{NN_f}
\end{equation}

For completeness, we present the numerical relation among the quark mass $m$, pion mass $m_\pi$, and determinantal mass $m^*$ in  Table~\ref{tab:mass}.
\begin{table}[h!]
\centering
\begin{tabular}{cccc}
\hline
$m(\mu=2~\mathrm{GeV})$ & $m(\mu=1/\rho)$ & $m^*$ (MeV) & $m_\pi$ (MeV)  
\\
\hline
3 MeV & 10 MeV & 142  & 140  
\\
7.8 MeV & 26 MeV & 164  & 226  
\\
9.6 MeV & 32 MeV & 158  & 250  
\\
17.4 MeV & 58 MeV & 190  & 337  
\\
\hline
\end{tabular}
\caption{The numerical chiral relation among pion mass $m_\pi$ and quark mass $m$ at different energy scale with pion decay constant $f_\pi=91$ MeV and quark condensate $\langle\bar{\psi}\psi\rangle=\langle\bar{u}u\rangle+\langle\bar{d}d\rangle=(300~\mathrm{MeV})^3$.}
\label{tab:mass}
\end{table}






As shown in Fig.~\ref{mol_3}, the leading $1/N_c$ contribution to the color-singlet molecular vertex, is given by
\begin{widetext}
\begin{equation}
\begin{aligned}
\label{eq:sin_mol}
    &\Theta^{\rm sin}_{I\bar{I}}(z_I,z_{\bar{I}})=\frac 1{2N_c}\left|T_{I\bar{I}}\right|^{2N_f}\left(\frac{1}{\left|T_{I\bar{I}}\right|^{2}}\frac{iR_\mu}{R}\frac{dT(R)}{dR}\right)\\
    &\times(4\pi^2\rho^2)^2\bigg[\mathrm{tr}_c(\tau_\mu^+u)\mathrm{tr}_c(\tau^-_\nu u^\dagger)\bar\psi_R(z_I)\gamma_\nu\psi_R(z_{\bar{I}})-\mathrm{tr}_c(\tau_\mu^-u^\dagger)\mathrm{tr}_c(\tau^+_\nu u)\bar\psi_L(z_{\bar{I}})\gamma_\nu\psi_L(z_I)\bigg]\\
    &+\cdots
\end{aligned}
\end{equation}
and the contribution to the color-octet vertex, is given by
\begin{equation}
\begin{aligned}
\label{eq:oct_mol}
    &\Theta^{\rm oct}_{I\bar{I}}(z_I,z_{\bar{I}})=\frac 1{4}\left|T_{I\bar{I}}\right|^{2N_f}\left(\frac{1}{\left|T_{I\bar{I}}\right|^{2}}\frac{iR_\mu}{R}\frac{dT(R)}{dR}\right)\\
    &\times(4\pi^2\rho^2)^2\bigg[\mathrm{tr}_c(\tau_\mu^+u)\mathrm{tr}_c(\tau^-_\nu u^\dagger\lambda^A)\bar\psi_R(z_I)\lambda^A\gamma_\nu\psi_R(z_{\bar{I}})-\mathrm{tr}_c(\tau_\mu^-u^\dagger)\mathrm{tr}_c(\tau^+_\nu \lambda^Au)\bar\psi_L(z_{\bar{I}})\lambda^A\gamma_\nu\psi_L(z_I)\bigg]\\
    &+\cdots
\end{aligned}
\end{equation}
\end{widetext}
where the Gell-Mann matrices are normalized to $\mathrm{tr}_c(\lambda^A\lambda^B)=2\delta^{AB}$. If we insert the vertices in \eqref{eq:sin_mol} and \eqref{eq:oct_mol} into \eqref{eq:L_mol} and evaluate the integral in a similar way to \eqref{pair_density}, we can simplify the integral to a typical coupling strength $\gamma_{I\bar{I}}$ for those one-body molecular vertices, which is defined by
\begin{widetext}
\begin{equation}
    \int d\rho_Id\rho_An(\rho_I)n(\rho_{A})\rho_I^{N_f}\rho_A^{N_f}\\
    \int dud^4R\left[\left|T_{I\bar{I}}\right|^{2N_f}\frac{(4\pi^2\rho^2)^2}{|T_{I\bar{I}}|^2}\left(\frac{-1}4R\frac{dT(R)}{dR}\right)\right] \rightarrow n_{\rm mol}\gamma_{I\bar{I}}
\end{equation}
\end{widetext}
with the molecular density $n_{\rm mol}$ is defined in \eqref{pair_density}. The best-fit value of $\gamma_{I\bar{I}}$ is obtained by matching to the RQCD estimate of $d_2^u=0.025$ \cite{Burger:2021knd}, yielding 
$311.13$ fm$^4$.

\section{Lorentz force operator}
\label{App:force}
In the ILM enhanced by molecules, the Lorentz force operator
$    g\bar \psi G^{\mu\nu} \gamma^\sigma \psi$
involves color-octet vertices, and follows by averaging over the individual pseudoparticles and their molecular  pairs. 
To evaluate the Lorentz force produced by single instantons and $I\bar{I}$ molecules on a light quark, we decompose the field strength for a $I+\bar{I}$ pseudo-particle pair as the sum of the self-dual single field strengths plus a non-self-dual interaction term,
\begin{equation}
G^a_{\mu\nu}[I\bar{I}]=G^a_{\mu\nu}[A_{\bar{I}}]+G^a_{\mu\nu}[A_{\bar{I}}]+G^a_{\mu\nu}[A_I,A_{\bar{I}}]
\end{equation}

\subsection{Single instanton}
The single instanton contribution to the Lorentz force operator reads
\begin{widetext}
\begin{equation}
\begin{aligned}
\label{eq:CFOp}
    &g\bar \psi(x)G_{\mu\nu}(x) \gamma_\sigma \psi(x)=\int d\rho n(\rho)\rho^{N_f}\int d^4z\int dU\bar\psi(x) \frac{\lambda^A}2G^A_{\mu\nu}(x)\gamma_\alpha \psi(x)\left[\Theta^{}_{I}(z)+\Theta^{}_{\bar{I}}(z)\right]\\
    =&-\frac{1}{4(N_c^2-1)}\frac{n_{I+\bar{I}}}2\left(\frac{4\pi^2\rho^2}{m^*}\right)\int d^4z\frac{8\rho^2}{[(x-z)^2+\rho^2]^2(x-z)^2}\left[\frac{(x-z)_\lambda (x-z)_\nu}{(x-z)^2}-\frac14 \delta_{\lambda\nu}\right]\\
    &\times\bar\psi(z)\sigma_{\mu\lambda}\lambda^A\psi(z)\bar\psi(x)\gamma_\sigma\lambda^A\psi(x)-(\mu\leftrightarrow\nu)
\end{aligned}
\end{equation}
\end{widetext}
where the Gell-Mann matrices here are normalized to $\mathrm{tr}_c(\lambda^A\lambda^B)=2\delta^{AB}$.
Carrying the color average and the size integration, yield 
\begin{widetext}
\begin{equation}
\begin{aligned}
\label{eq:CFOpINS}
    g\bar \psi G_{\mu\nu}\gamma_\sigma\psi(x)=&-\frac{1}{4(N_c^2-1)}\frac{n_{I+\bar{I}}}2\left(\frac{4\pi^2\rho^2}{m^*}\right)\int d^4z\frac{8\rho^2}{[(x-z)^2+\rho^2]^2(x-z)^2}\\
    &\times\left(\frac{(x-z)_\lambda (x-z)_\nu}{(x-z)^2}-\frac14 g_{\lambda\nu}\right)\bar\psi(x)\gamma_{\sigma}\lambda^A\psi(x)\bar\psi(z) \sigma_{\mu\lambda}\lambda^A\psi(z)-(\mu\leftrightarrow\nu)
\end{aligned}
\end{equation}
\end{widetext}


\subsection{Instanton pairs}
Since we are interested in the Lorentz color force, the color-octet vertices ae relevant. The operator $g\bar \psi G_{\mu\nu}\gamma_\sigma\psi$ in the instanton vacuum molecules, receives a dominant contribution from  Fig~\ref{mol_3} in the form 
\begin{widetext}
\begin{equation}
\begin{aligned}
\label{eq:CFOp}
    &\int d\rho_Id\rho_An(\rho_I)n(\rho_{\bar{I}})\rho_I^{N_f}\rho_A^{N_f}\int d^4zdud^4R\bar\psi(x)G_{\mu\nu}(x)\gamma_\alpha \psi(x)\Theta^{\rm oct}_{I\bar{I}}(z_I,z_{\bar{I}})\\
    =&\int d^4zd^4R\left[\frac{n_{I+\bar{I}}}2\left(\frac{4\pi^2\rho^2}{m^*}\right)^{N_f}\right]^2
    \left(\frac{|T_{I\bar{I}}|}{4\pi^2\rho^2}\right)^{2N_f}
    \left[\frac{(4\pi^2\rho^2)^2}{|T_{I\bar{I}}|^2}\frac{iR_{\beta}}{R}\frac{dT(R)}{dR}\right]\\
    &\times\frac{1}{4(N_c^2-1)^2}\frac{\epsilon^{abc}\left(\bar\eta^b_{\mu\rho}\eta^d_{\nu\lambda}-\bar\eta^b_{\nu\rho}\eta^d_{\mu\lambda}\right)\rho^4(x-z_I)_\rho(x-z_{\bar{I}})_\lambda}{[(x-z_I)^2+\rho^2](x-z_I)^2[(x-z_{\bar{I}})^2+\rho^2](x-z_{\bar{I}})^2}\bar\psi(x)\gamma_\sigma\lambda^A\psi(x)\\
    &\times\bigg[\mathrm{tr}_c(\tau_{\alpha}^-\tau^d\tau^+_{\beta}\tau^c\tau^a)\bar\psi(z_I)\gamma_{\alpha}\frac{1+\gamma^5}{2}\lambda^A\psi(z_{\bar{I}})-\mathrm{tr}_c(\tau_{\alpha}^+\tau^a\tau^c\tau^-_{\beta}\tau^d)\bar\psi(z_{\bar{I}})\gamma_{\alpha}\frac{1-\gamma^5}{2}\lambda^A\psi(z_I)\bigg]\\
\end{aligned}
\end{equation}

\end{widetext}
Here  $\bar{\eta}^a_{\mu\nu}$ and $\eta^a_{\mu\nu}$ are the 't Hooft symbols for the instanton and anti-instanton in singular gauge,  and $\tau^a$ an $N_c\times N_c$ matrix with $2\times2$ Pauli matrices embedded in the upper left corner. 

In general, the effective quark operators induced by the higher order instanton clusters are non-local in the quark fields $\bar{\psi}(z_I)\psi(z_{\bar{I}})$. However, as a result of the highly localized nature of the instanton profiles, the effective range for the instanton vertex is contrained within the instanton size $R=|z_I-z_{\bar{I}}|\sim\rho\ll \sqrt[4]{V_4/N}$. Hence  we can approximate the non-local quark operators with local quark operators by localizing the instanton cluster profiles. This reduction will be referred to as the local approximation, as originally suggested in~\cite{Liu:2024rdm}. The contribution to those matrix elements in leading  order, comes from those clusters. Therefore,  in the local approximation
$$
\bar{\psi}(z_I)\psi(z_{\bar{I}})\simeq\bar{\psi}(z)\psi(z)-R_\mu\bar{\psi}(z)\overleftrightarrow{\partial_\mu}\psi(z)+\cdots
$$
where $z$ is the center of the pseudoparticle pair, and $R$ is their relative distance. 

With this in mind, \eqref{eq:CFOp} can be written as
\begin{widetext}
\begin{equation}
\begin{aligned}
\label{eq:CFOpAPP}
    g\bar \psi(x)G_{\mu\nu}(x)\gamma_\sigma\psi(x)=&\frac{1}{4(N_c^2-1)^2}\int d^4z d^4R\,\left(\frac{n_{I+\bar{I}}}2\right)^2\left(\frac{|T_{IA}|^2}{(m^*)^2}\right)^{N_f}\left(\frac{4\pi^2\rho^2}{|T_{IA}|}\right)^{2}\left[\frac{-1}4R\frac{dT(R)}{dR}\right]\\
    &\times t_{\mu\nu\rho\lambda\alpha\beta}\frac{\rho^4(x-z)_\rho (x-z)_\lambda}{[(x-z)^2+\rho^2]^2(x-z)^4}\bar\psi(x)\gamma_\sigma\lambda^A\psi(x)\bar\psi(z) i\gamma_{(\alpha}\gamma^5\overleftrightarrow{\partial}_{\beta)}\lambda^A\psi(z)
\end{aligned}
\end{equation}
\end{widetext}
where $(\cdots)$ bracketing the Lorentz indices refers to full symmetrization, 
and
$$
\bar\psi i\overleftrightarrow{\partial}_{\mu}\psi=\frac12\bar\psi \left(i\overrightarrow{\partial}_{\mu}-i\overleftarrow{\partial}_{\mu}\right)\psi
$$
The color factor is defined as
\begin{equation}
t_{\mu\nu\rho\lambda\alpha\beta}=2i\left(\bar\eta^b_{\mu\rho}\eta^d_{\nu\lambda}-\bar\eta^b_{\nu\rho}\eta^d_{\mu\lambda}\right)
\mathrm{tr}(\tau^-_{\alpha}\tau^d\tau^+_{\beta}\tau^b)
\end{equation}
where the last two indices are symmetric $t_{\mu\nu\rho\lambda(\alpha\beta)}=t_{\mu\nu\rho\lambda\alpha\beta}$, and the middle two indices are also symmetric but traceless. Finally, the Fourier transform is then
\begin{widetext}
\begin{equation}
\begin{aligned}
\label{eq:force_op}
    \frac1V\int d^4xe^{iqx}g\bar \psi(x)G_{\mu\nu}(x)\gamma_\sigma\psi(x)=&\frac{1}{4(N_c^2-1)}\frac{n_{I+\bar{I}}}2\left(\frac{4\pi^2\rho^2}{m^*}\right)\int \frac{d^4k}{(2\pi)^4}8G(k)\left(\frac{k_\lambda k_\nu}{k^2}-\frac14g_{\lambda\nu}\right)\\
    &~\times\int d^4xe^{i(q-k)x}\bar\psi(x)\gamma_{\sigma}\lambda^A\psi(x)\int d^4z e^{ikz}\bar\psi(z) \sigma_{\mu\lambda}\lambda^A\psi(z)\\
    &-\frac{n_{\rm mol}}{4(N_c^2-1)^2}\gamma_{I\bar{I}}\, t_{\mu\nu\rho\lambda\alpha\beta}\int \frac{d^4k}{(2\pi)^4}\rho^2\frac{k_{\rho}k_{\lambda}}{k^2}G(k)\\
    &~\times\int d^4xe^{i(q-k)x}\bar\psi(x)\gamma_\sigma\lambda^A\psi(x)\int d^4ze^{ikz}\bar\psi(z) i\gamma_{(\alpha}\gamma^5\overleftrightarrow{\partial}_{\beta)}\lambda^A\psi(z)
\end{aligned}
\end{equation}
\end{widetext}
where the instanton field profile $G(k)$ in momentum space is defined by
\begin{equation}
\begin{aligned}
    &G(k)=\frac{4\pi^2}{t}\int_0^\infty dx\frac{1}{(x^2+1)^2}J_3(tx)\bigg|_{t=\rho k}\\
    &=\frac{4\pi^2}{t^2}\left( 1-\frac{16}{t^{2}} +\frac{t^{2}}2 \, K_{2}(t) + 2 t \, K_{3}(t) \right)\bigg|_{t=\rho k}
\end{aligned}
\end{equation}

\begin{figure}
    \centering
    \includegraphics[width=\linewidth]{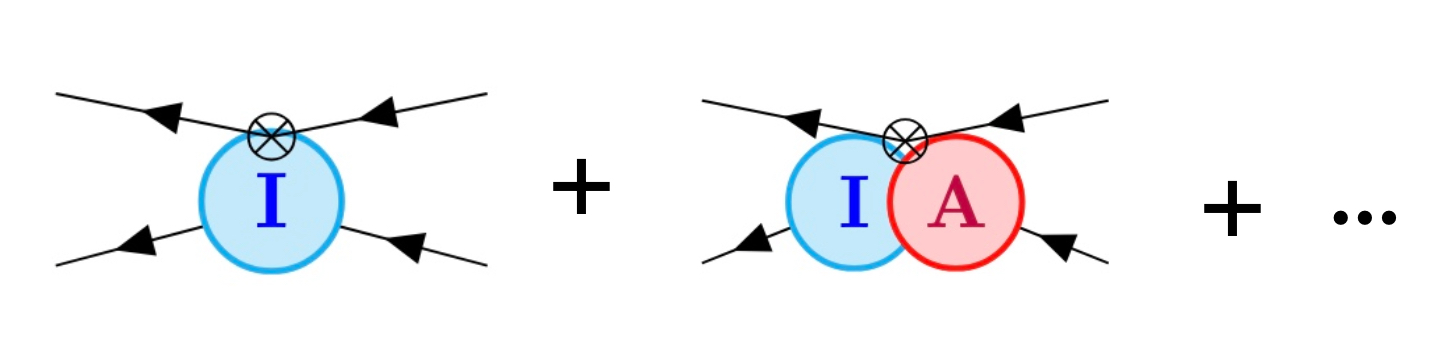}
    \caption{Feynman diagrams for the Lorentz force operator (crossed-dot)  in the ILM enhanced by molecules. The quark lines connected to the cross dot at the top of each diagrams are located at $x$ while the quark lines at the bottom of each diagrams are connected to the blob of the classical pseudoparticle profile centered at $z$.}
    \label{fig:Oqg}
\end{figure}

The final result \eqref{eq:force_op} is illustrated in Fig.~\ref{fig:Oqg}. The Lorentz operator (cross-dot) takes the form of a product of two color–octet currents, analogous to a one–gluon exchange between quarks. The  function $G(k)$ plays the role of an effective gluon propagator. The evaluation of its nucleon matrix element is therefore closely related to the calculation of gluon–exchange corrections to the nucleon mass \cite{Balla:1997hf,Diakonov:1991}

\subsubsection{Short distance approximation}
As we noted earlier, the pseudo-particle profile is highly localized. The  separation between two quark sources in \eqref{eq:force_op}, is controlled by $|x-z|\lesssim\rho\ll \sqrt[4]{V_4/N}$, hence the approximation
\begin{equation}
\begin{aligned}
&\sqrt{\mathcal{F}(i\rho\partial)}S(x)\equiv\langle\psi(x)\sqrt{\mathcal{F}(i\rho\partial)}\bar\psi(0)\rangle\\
\simeq&\frac{i\slashed{x}}{2\pi^2x^4}K_D(x/\rho)+\frac{m}{4\pi^2x^2}K_m(x/\rho)+\mathcal{O}(m^2)
\end{aligned}
\end{equation}
In the ILM the  quark propagator $S(x)$ is defined as 
\begin{equation}
    S(x)=\frac{i\slashed{x}}{2\pi^2x^4}+\frac{m}{4\pi^2x^2}+\mathcal{O}(m^2)
\end{equation}
with the distorsions induced by the instanton zero mode profiles
\bea
    K_D(x)&=&\frac12\int_0^\infty dk kx^2 J_2(kx)\sqrt{\mathcal{F}(k)}
    =\frac{x^3}{(1 + x^2)^{3/2}}\nonumber\\
    K_m(x)&=&\int_0^\infty dk x J_1(kx)\sqrt{\mathcal{F}(k)}
\eea

In this short distance approximation, the effective Lorentz force operator can be further reduced to a sum of local fermionic operators induced by the zero modes in the ensemble,
\begin{widetext}
\begin{equation}
\begin{aligned}
    &g\bar\psi\gamma_\sigma G_{\mu\nu}\psi(x)=\\
    &-\left(\frac{n_{I+\bar{I}}}2\right)\frac{1}{2N_c}\left(\frac{4\pi^2\rho^2}{m^*}\right)\int d^4z\left[F^{(+)}_{\bar{q}Gq,1,\mu\nu\sigma}(x-z)\bar\psi(z)\frac{1-\gamma^5}2\psi(z)+F^{(+)}_{\bar{q}Gq,2,\mu\nu\sigma\rho}(x-z)m\bar\psi(z)\gamma_\rho\gamma^5\psi(z)\right]\\
    &-\left(\frac{n_{I+\bar{I}}}2\right)\frac{1}{2N_c}\left(\frac{4\pi^2\rho^2}{m^*}\right)\int d^4z\left[F^{(-)}_{\bar{q}Gq,1,\mu\nu\sigma}(x-z)\bar\psi(z)\frac{1+\gamma^5}2\psi(z)+F^{(-)}_{\bar{q}Gq,2,\mu\nu\sigma\rho}(x-z)m\bar\psi(z)\gamma_\rho\gamma^5\psi(z)\right]\\
    &+\left(\frac{n_{I+\bar{I}}}2\right)^2\frac{1}{2N_c(N_c^2-1)}\int d^4R\,\left(\frac{|T_{IA}|^2}{(m^*)^2}\right)^{N_f}\frac{(4\pi^2\rho^2)^2}{|T_{IA}|^2}\left[\frac{-1}4R\frac{dT(R)}{dR}\right]\\
    &\times\int d^4z\Bigg[F^{(+-)}_{\bar{q}Gq,1,\mu\nu\sigma\alpha\beta}(x-z)\bar\psi(z)\gamma_{\alpha}i\overleftrightarrow{\partial}_{\beta}\psi(z)+F^{(+-)}_{\bar{q}Gq,2,\mu\nu\sigma\rho}(x-z)\bar\psi(z)\gamma_\rho\gamma^5\psi(z)\\
    &\qquad\qquad+F^{(+-)}_{\bar{q}Gq,3,\mu\nu\alpha\beta}(x-z)m\bar\psi(z)\sigma_{\sigma(\alpha}\gamma^5\overleftrightarrow{\partial}_{\beta)}\psi(z)+F^{(+-)}_{\bar{q}Gq,4,\mu\nu\sigma}(x-z)m\bar\psi(z)i\gamma^5\psi(z)\Bigg]
\end{aligned}
\end{equation}
with
\begin{equation}
    F^{(\pm)}_{\bar{q}Gq,1,\mu\nu\sigma}(x)=\frac{8\rho^2}{(x^2+\rho^2)^2}K_D(x/\rho)\frac{g_{\mu\sigma}x_\nu-g_{\nu\sigma}x_{\mu}\pm i\epsilon_{\mu\nu\sigma \alpha}x_\alpha}{2\pi^2x^4}
\end{equation}
\begin{equation}
    F^{(\pm)}_{\bar{q}Gq,2,\mu\nu\sigma\rho}(x)=\pm\frac{4\rho^2}{(x^2+\rho^2)^2}K_m(x/\rho)\left[\frac{i\left(g_{\mu\rho} g_{\lambda\sigma}-g_{\lambda\rho} g_{\mu\sigma}\right)\pm\epsilon_{\mu\lambda\sigma\rho}}{2\pi^2x^2}\left(\frac{x_\lambda x_\nu}{x^2}-\frac14 g_{\lambda\nu}\right)-(\mu\leftrightarrow\nu)\right]
\end{equation}
\begin{equation}
\begin{aligned}
    F^{(+-)}_{\bar{q}Gq,1,\mu\nu\sigma\alpha\beta}(x)=&\frac{\rho^4}{(x^2+\rho^2)^2x^2}K_D(x/\rho)\frac{x_\rho x_\lambda}{x^2}t_{\mu\nu\rho\lambda(\gamma\beta)}\frac{\epsilon_{\alpha\sigma\delta\gamma}x_\delta}{2\pi^2x^4}\\
\end{aligned}
\end{equation}
\begin{equation}
\begin{aligned}
F^{(+-)}_{\bar{q}Gq,2,\mu\nu\sigma\rho}(x)=&t_{\mu\nu\gamma\lambda\alpha\beta}\Bigg[\frac12\partial_{\beta}\left(\frac{\rho^4}{(x^2+\rho^2)^2x^2}\frac{x_{\gamma} x_\lambda}{x^2}\frac{x_\delta}{x^4}\right)-\frac{\rho^4}{(x^2+\rho^2)^2x^2}\frac{x_{\gamma} x_\lambda}{x^2}\partial_\beta\left(\frac{x_\delta}{x^4}\right)\\
&-\frac12\frac{\rho^4x_\rho x_\lambda x_\mu}{(x^2+\rho^2)^2x^8}\partial_\beta\ln K_D(x/\rho)\Bigg]K_D(x/\rho)\frac{g_{\sigma\delta}g_{\alpha\rho}-g_{\sigma\alpha}g_{\delta\rho}+g_{\sigma\rho}g_{\delta\alpha}}{2\pi^2}\\
=&\frac{1}{\pi^2}\left[-\frac{24\rho^4}{x^8\, (x^2 + \rho^2)^2}\left(K_D(x/\rho)-\frac16x\frac{d}{dx}K_D(x/\rho)\right)
+ \frac{16\rho^4}{x^6\, (x^2 + \rho^2)^3}K_D(x/\rho)\right]x_{[\rho}\epsilon_{\sigma]\mu \nu \lambda}x_\lambda
\end{aligned}
\end{equation}
\begin{equation}
    F^{(+-)}_{\bar{q}Gq,3,\mu\nu\alpha\beta}(x)=t_{\mu\nu\rho\lambda\alpha\beta}\frac{\rho^4}{(x^2+\rho^2)^2x^2}\frac{x_\rho x_\lambda}{x^2}\frac{1}{4\pi^2x^2}K_m(x/\rho)
\end{equation}
\begin{equation}
\begin{aligned}
   F^{(+-)}_{\bar{q}Gq,4,\mu\nu\sigma}(x)=& \frac{1}{2\pi^2}t_{\mu\nu\rho\lambda\sigma\beta}\Bigg[\frac12\partial_{\beta}\left(\frac{\rho^4}{(x^2+\rho^2)^2x^4}\frac{x_\rho x_\lambda}{x^2}\right)-\frac{\rho^4}{(x^2+\rho^2)^2x^2}\frac{x_\rho x_\lambda}{x^2}\partial_{\beta}\left(\frac{1}{x^2}\right)\\
   &-\frac12\frac{\rho^4}{(x^2+\rho^2)^2x^4}\frac{x_\rho x_\lambda}{x^2}\partial_\beta \ln K_m(x/\rho)\Bigg]K_m(x/\rho)\\
   =&\frac{1}{2\pi^2}\left[\frac{16\rho^4}
       {x^{6}\,(x^{2}+\rho^2)^2}\left(K_m(x/\rho)-\frac14x\frac{d}{dx}K_m(x/\rho)\right)-\frac{16\rho^4}
       {x^{4}\,(x^{2}+\rho^2)^3}K_m(x/\rho) \right]\epsilon_{\mu\nu\sigma\lambda}x_\lambda
\end{aligned}   
\end{equation}
\end{widetext}
with the help of the identity~\cite{Freese:2019bhb},


\[
\,\bar{\psi}\,\gamma^{[\mu} i\overleftrightarrow{\partial}^{\nu]} \psi
= \frac{1}{4} \,\epsilon^{\mu\nu\rho\sigma} \,\partial_{\rho}
\left( \bar{\psi}\,\gamma_{\sigma}\,\gamma_{5}\,\psi \right)
\]

\section{Molecular form factors}
\label{MOL-FF}
The molecular form-factors entering \eqref{eq:CFOp2} are
\begin{widetext}
\begin{align}
    \beta^{(+-)}_{\bar q Gq,1}(q)=&\frac{1}{q}\int_0^\infty dx\frac{16}{(x^2+1)^2x^2}\left(\frac{3J_3(qx)}{q^2x^2}-\frac{J_4(qx)}{qx}\right)K_D(x)\\
    \beta^{(+-)}_{\bar q Gq,2}(q)=&\frac{1}{q}\int_0^\infty dx\frac{16}{(x^2+1)^2x^2}\frac{J_3(qx)}{q^2x^2}K_D(x)
\end{align}

\begin{align}
    \beta^{(+-)}_{\bar q Gq,3}(q)=&\frac1q\int_0^\infty dx \Bigg[\frac{128}{x^4(x^2+1)^2}\left(K_D(x)-\frac14xK'_D(x)\right)\frac{J_2(qx)}{qx}\nonumber\\
    &\qquad\qquad-\frac{16}{(x^2+1)^2x^4}K_D(x)J_3(qx)\Bigg]\\
    \beta^{(+-)}_{\bar q Gq,4}(q)=&\frac1q\int_0^\infty dx \Bigg[\frac{64}{x^2(x^2+1)^2}\left(K_D(x)-\frac14xK'_D(x)\right)\frac{J_3(qx)}{q^2x^2}\nonumber\\
    &\qquad\qquad-\frac{8}{(x^2+1)^2x^2}K_D(x)\left(\frac{J_4(qx)}{qx}-\frac{2J_3(qx)}{q^2x^2}\right)\Bigg]\\
\beta^{(+-)}_{\bar q Gq,5}(q)=&\frac1q\int_0^\infty dx \Bigg[\frac{16}{x^2(x^2+1)^2}\left(K_m(x)-\frac12xK'_m(x)\right)\frac{J_2(qx)}{qx}-\frac{4}{(x^2+1)^2}K_m(x)J_3(qx)\Bigg]\\
\beta^{(+-)}_{\bar q Gq,6}(q)=&\frac1q\int_0^\infty dx \frac{16}{(x^2+1)^2}\frac{J_3(qx)}{q^2x^2}K_m(x)\\
\beta^{(+-)}_{\bar q Gq,7}(q)=&\frac1q\int_0^\infty dx \Bigg[\frac{16}{x^2(x^2+1)^2}\left(K_m(x)-\frac12xK'_m(x)\right)\frac{J_2(qx)}{qx}\Bigg]
\end{align}

\end{widetext}

At zero momentum transfer, their values are
\begin{align}
    \beta^{(+-)}_{\bar q Gq,1}(0)=&\frac2{15}\nonumber\\
    \beta^{(+-)}_{\bar q Gq,2}(0)=&0.0444
\end{align}

\begin{align}
    \beta^{(+-)}_{\bar q Gq,3}(0)=&3.048\nonumber\\
    \beta^{(+-)}_{\bar q Gq,4}(0)=&0.1460
\end{align}

\begin{align}
    \beta^{(+-)}_{\bar q Gq,5}(0)=&0.2331\nonumber\\
    \beta^{(+-)}_{\bar q Gq,6}(0)=&0.0935\nonumber\\   
    \beta^{(+-)}_{\bar q Gq,7}(0)=&0.2262
\end{align}

\begin{figure}
    \centering
    \includegraphics[width=\linewidth]{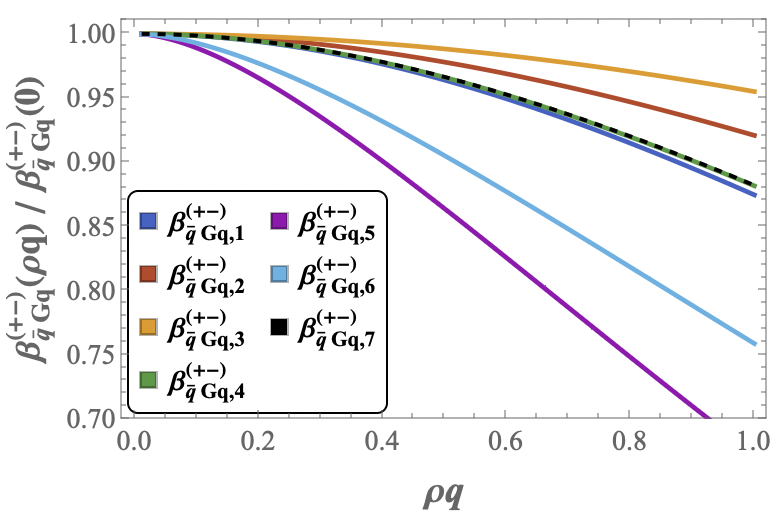}
    \caption{Emergent form factors $\beta^{(+-)}_{\bar{q}Gq}$ associated to the color Lorentz operator in a molecular (instanton-anti-instanton)  pair.}
    \label{fig:beta}
\end{figure}

\section{Off-forward Dirac structure}
For a spin-$1/2$ particle with mass $m$, symmetries such as Poincare covariance, parity, and hermicity allow for the determination of the off-forward tensor structures of any matrix element carrying Lorentz indices  $\{\gamma^\mu,\gamma^\mu\gamma^5,\bar{p}^\mu, q^\mu, g^{\mu\nu}, i\sigma^{\mu\nu},\epsilon^{\mu\nu\rho\lambda}\}$ alongside with the Dirac structrue $\{1,\slashed{\bar p}, \slashed{q}, [\slashed{\bar p},\slashed{q}]\}$.  The linear combinations of those can be used to establish the Lorentz structure of the baryon matrix elements and form factors. Those Lorentz structures can be further simplified using the identities
\begin{equation}
    \bar{u}_{s'}(p')\slashed{\bar{p}} u_s(p)= m_h \bar{u}_{s'}(p')u_s(p)
\end{equation}
\begin{equation}
    \bar{u}_{s'}(p')\slashed{q} u_s(p)= 0
\end{equation}
Between  on-shell baryon spinors $\bar{u}_s(p' )$ and $u_s(p' )$, not all 
allowed combinations are linearly independent. The following identities

\begin{equation}
    \bar{u}_{s'}(p')\gamma^\mu u_s(p)=\bar{u}_{s'}(p')\left(\frac{\bar p^\mu}{m_h} +\frac{i\sigma^{\mu\nu}q_\nu}{2m_h}\right)u_s(p)
\end{equation}

\begin{widetext}

\begin{equation}
\begin{aligned}
    \bar{u}_{s'}(p')\gamma^\mu\gamma^5 u_s(p)=\bar{u}_{s'}(p')\left(\frac{i\sigma^{\mu\nu}\gamma^5\bar{p}_\nu}{m_h} +\frac{q^\mu\gamma^5}{2m_h} \right)u_s(p)
\end{aligned}
\end{equation}

\begin{equation}
    \bar{u}_{s'}(p')\left(\gamma^\mu \bar{p}^\nu-\gamma^\nu \bar{p}^\mu\right)\bar{u}_s(p)=\frac{i}2\epsilon^{\mu\nu\rho\lambda}\bar{u}_{s'}(p')q_\rho \gamma_\lambda\gamma^5\bar{u}_s(p)
\end{equation}
\begin{equation}
    \bar{u}_{s'}(p')i\sigma^{\mu\nu} \bar{u}_s(p)=\bar{u}_{s'}(p')\left(\frac{\gamma^\mu q^\nu-\gamma^\nu q^\mu}{2m_h}-\frac{i}{m_h}\epsilon^{\mu\nu\rho\sigma}\bar{p}_\rho\gamma_\sigma\gamma^5\right) \bar{u}_s(p)
\end{equation}
\end{widetext}
can be used to simplify. Note that in the forward limit, 
\begin{equation}
\begin{aligned}
    &\bar{u}_s(p)\gamma^\mu u_s(p)=2p^\mu\\
    &\bar{u}_s(p)\gamma^\mu\gamma^5 u_s(p)=2m_hS^\mu\\
    &\bar{u}_s(p)\sigma^{\mu\nu} u_s(p)=2\epsilon^{\mu\nu\rho\lambda}S_\rho p_\lambda
\end{aligned}
\end{equation}
with $\epsilon^{0123}=1$. 
\red{Schouten identity of Levi-Civita symbol also simplifies the Lorentz decomposition of the form factors. Given abitrary Lorentz covariant 4-vector $\{a^\mu,b^\mu, c^\mu\}$, the identity reads}
\begin{equation}
\begin{aligned}
&a^\mu \epsilon^{\nu\sigma\rho\lambda} b_\rho c_\lambda - a^\nu \epsilon^{\mu\sigma\rho\lambda} b_\rho c_\lambda + a^\sigma \epsilon^{\mu\nu\rho\lambda} b_\rho c_\lambda\\
=&(a\cdot b) \epsilon^{\mu\nu\sigma\gamma} c_\gamma - (a \cdot c) \epsilon^{\mu\nu\sigma\gamma} b_\gamma
\end{aligned}
\end{equation}

Some useful identities of the Dirac $\sigma^{\mu\nu}$ are also useful in the process of simplifying the Lorentz structures, such as the relation between $\sigma^{\mu\nu}$ and $\sigma^{\mu\nu}\gamma^5$ through the Levi-Civita symbol
\begin{equation}
\frac i2\epsilon^{\mu\nu\rho\lambda}\sigma_{\rho\lambda}=\sigma_{\mu\nu}\gamma^5
\end{equation}
 and the identities involving three gamma matrices
\[
\sigma^{\mu\nu} \gamma^{\rho}
= i g^{\nu\rho} \gamma^{\mu}
- i g^{\mu\rho} \gamma^{\nu}
- \epsilon^{\mu\nu\rho\sigma} \gamma_{\sigma} \gamma_{5},
\]
\[
\gamma^{\rho} \sigma^{\mu\nu}
= i g^{\mu\rho} \gamma^{\nu}
- i g^{\nu\rho} \gamma^{\mu}
- \epsilon^{\mu\nu\rho\sigma} \gamma_{\sigma} \gamma_{5},
\]

\section{Hadronic form factors }
\label{App:had_FF}
In instanton vacuum, the presence of pseudoparticles modifies the point-like quark and gluon operators at low momentum transfer $Q^2$ characterized by $Q\lesssim 1/\rho$, \cite{Kim:2023pll,Liu:2024rdm,Balla:1997hf}. As a result, the color force operator in \eqref{eq:CFOpINS} and \eqref{eq:CFOp} can be mapped to various hadronic quark form factors: the scalar form factor $\sigma^q_h$, the energy-momentum tensor form factor $A^q_h$ and $D^q_h$ for the unpolarized part, and the axial form factors $G^q_A$, the pseudoscalar form factors $G^q_P$ and $\tilde{G}^q_P$, and the transversity form factors $A^q_{h,T}$ for the spin-dependent process. 

In this regime ($Q\lesssim 1/\rho$), the instantons and anti-instanton act collectively, inducing  effective glueball-quark or meson-quark interactions. The form factors are mostly described in the form of meson or glueball exchanges, characterized by hadronic parameters,  through their  masses or hadronic couplings \cite{Liu:2025ldh,Shuryak:2020ktq}.

In order to evaluate the color force form factor, here we present the details of the definition of each form factors, and its pertinent form in the ILM.

\subsection{Pion form factors}

To evaluate the $\Phi^q_{\pi,1}$ and $\Phi^q_{\pi,2}$ form factors, we need the quark scalar form factor $\sigma_\pi^q$ and energy-momentum tensor form factor $A_\pi^q$ and $D_\pi^q$. For pions, the quark scalar form factor is defined by
\begin{equation}
\begin{aligned}
    &\langle \pi'|m\bar\psi\psi|\pi\rangle=2m_\pi^2\sigma_\pi(Q^2)
\end{aligned}
\end{equation}
The traceless energy momentum tensor form factor for the pion is defined as
\begin{widetext}
\begin{equation}
\begin{aligned}
    & \langle \pi'|\bar{\psi}\left(\gamma_{(\mu} i\overleftrightarrow{\partial}_{\nu)}-\frac{1}{4}g_{\mu\nu}i\overleftrightarrow{\slashed{\partial}}\right)\psi|\pi\rangle=2A_\pi^{q}(Q^2)\left(\bar{p}_{\mu} \bar{p}_{\nu}-\frac14g_{\mu\nu}\bar{p}^2 \right)+\frac{1}{2}D_\pi^{q}(Q^2)\left(q_{\mu}q_{\nu}-\frac14g_{\mu\nu}q^2\right)
\end{aligned}
\end{equation}
These form factors are also related to the generalized form factors obtained by the second moment of unpolarized pion GPD.
\begin{equation}
    \int_{-1}^1 dx xH^{q/\pi}(x,\xi,t=-Q^2)=A^q_\pi(Q^2)+\xi^2 D^q_\pi(Q^2)
\end{equation}
\end{widetext}

As we are only interested in the twist-3 contribution, or the $\Phi^q_{\pi,1}$ form factor in this work, we only need to model the energy momentum tensor form factor $A_\pi^q$. It can be parameterized by a monopole form 
\begin{equation}
    A_\pi^{u+d}(Q^2)=\frac{0.481}{1+Q^2/1.262^2}
\end{equation}
with the lattice monopole fitting parameters from~\cite{Hackett:2023nkr},  assuming 
\begin{equation}
    A_\pi^{u-d}(Q^2)=0
\end{equation}
thanks to isospin symmetry. 


\subsection{Nucleon form factors}
To estimate  the nucleon $\Phi^q_{N,1}$ and $\Phi^q_{N,2}$ form factors, we need the quark scalar form factor $\sigma_N^q$  and the energy-momentum tensor form factors $A_N^q$, $J_N^q$, and $D_N^q$ both in the isoscalar and isovector channels. For the nucleon, the quark scalar form factor (sigma-term) is defined as
\begin{equation}
\begin{aligned}
    &\langle N'|m\bar\psi\psi|N\rangle=m_N\sigma_N(Q^2)\bar{u}_{s'}(p') u_s(p)
\end{aligned}
\end{equation}
The nucleon traceless energy momentum tensor form factor, is defined as
\begin{widetext}
\begin{equation}
\begin{aligned}
    & \langle N'|\bar{\psi}\left(\gamma_{(\mu} i\overleftrightarrow{\partial}_{\nu)}-\frac{1}{4}g_{\mu\nu}i\overleftrightarrow{\slashed{\partial}}\right)\psi|N\rangle=\\
    &\bar{u}_{s'}(p')\bigg[A_N^{q}(Q^2)\frac{1}{m_N}\left(\bar{p}^{\mu} \bar{p}^{\nu}-\frac14g_{\mu\nu}\bar{p}^2 \right)+J_N^{q}(Q^2)\frac{1}{2m_N}\left(2i\bar{p}^{(\mu}\sigma^{\nu)\alpha}q_\alpha-\frac14 g_{\mu\nu}q^2\right)\\
    &+D_N^{q}(Q^2)\frac{1}{4m_N}\left(q^{\mu}q^{\nu}-\frac14g^{\mu\nu}q^2\right)\bigg]u_{s}(p)
\end{aligned}
\end{equation}
where the angular momentum form factor is defined by $J_N^q=\frac12\left(A^q_N+B^q_N\right)$. Those form factors are also related to the generalized form factors derived from the unpolarized nucleon GPD $H^{q/N}$ and $E^{q/N}$ with skewness $2\xi=-q^+/\bar p^+ $ \cite{Diehl:2005jf}.
\begin{align}
    \int_{-1}^1 dx xH^{q/N}(x,\xi,t=-Q^2)=&A^q_N(Q^2)+\xi^2 D^q_N(Q^2) \\
    \int_{-1}^1 dx xE^{q/N}(x,\xi,t=-Q^2)=&B^q_N(Q^2)-\xi^2 D^q_N(Q^2)
\end{align}
\end{widetext}

To evaluate nucleon $\Phi^q_{N,4-8}$ form factors, we also need the quark axial form factor $G_A^q$, and the pseudoscalar form factors $\tilde{G}_P^q$ and $G_P^q$ both in isoscalar and isovector channels. The pseudoscalar form factor is defined  as
\begin{equation}
\begin{aligned}
    &\langle N'|m\bar\psi\gamma^5\psi|N\rangle=m_N\tilde{G}^q_P(Q^2)\bar{u}_{s'}(p') \gamma^5 u_s(p)
\end{aligned}
\end{equation}
The axial form factor and the induced pseudoscalar form factors, are defined as 
\begin{equation}
\begin{aligned}
    &\langle N'|\bar\psi\gamma_\mu\gamma^5\psi|N\rangle=\\
    &\bar{u}_{s'}(p') \left[\gamma_\mu\gamma^5 G^q_A(Q^2)+\frac{q_\mu\gamma^5}{2m_N}G^q_P(Q^2)\right] u_s(p)
\end{aligned}
\end{equation}
\red{However, these form factors $G^q_{A,P}$ and $\tilde{G}^q_{P}$ do not contribute to the twist-3 combination.}
To evaluate the color force form factors $\Phi^q_{N,1-3}$, $\Phi^q_{N,5}$, $\Phi^q_{N,6}$, and $\Phi^q_{N,8}$, \red{we also need the 7 tensor quark gravitational form factors $A^q_{T}$, $\tilde{A}^q_{T}$, $B^q_{T}$, $\tilde{B}^q_{T}$, $C^q_{T}$, $D^q_{T}$, and $\tilde{D}^q_{T}$ \cite{Bhoonah:2017olu}, while only the first 4 contribute to the twist-3 transversity gravitational form factors,} which correspond to the second moment of the transversity GPD is defined by 
\cite{Bhoonah:2017olu,Alexandrou:2024awx}
\begin{widetext}

\begin{equation}
\begin{aligned}
    &\langle N'|\bar\psi\sigma_{\mu\nu}\overleftrightarrow{\partial}_{\rho}\psi|N\rangle=\bar{u}_{s'}(p') \Bigg\{-i\sigma^{\mu\nu}\bar{p}^\rho  A_{T}^q-\frac{\bar{p}^\mu q^\nu-\bar{p}^\nu q^\mu}{m_N^2}\bar{p}^{\rho}\tilde{A}_{T}^q-\frac{\gamma^\mu q^\nu-\gamma^\nu q^\mu}{2m_N}\bar{p}^\rho B_{T}^q\\
    &-\frac{\gamma^\mu \bar{p}^\nu-\gamma^\nu \bar{p}^\mu}{m_N}q^{\rho}\tilde{B}_{T}^q-(g^{\mu\rho}q^{\nu}-g^{\nu\rho}q^{\mu})C_T^q-im_N\epsilon^{\mu\nu\rho\lambda}\gamma_\lambda\gamma^5D_T^q-i\epsilon^{\mu\nu\rho\lambda}q_\lambda\gamma^5\tilde{D}_T^q\Bigg\} u_s(p)
\end{aligned}
\end{equation}

Their relations to the nucleon transversity GPD in \cite{Kim:2025mol} are defined as 
\begin{align}
    \int_{-1}^1 dx xH_T^{q/N}(x,\xi,t=-Q^2)=&A^q_{T}(Q^2)\\
    \int_{-1}^1 dx xE_T^{q/N}(x,\xi,t=-Q^2)=&B^q_{T}(Q^2)\\
    \int_{-1}^1 dx x\tilde{H}_T^{q/N}(x,\xi,t=-Q^2)=&\tilde{A}^q_{T}(Q^2)\\
    \int_{-1}^1 dx x\tilde{E}_T^{q/N}(x,\xi,t=-Q^2)=&-2\xi\tilde{B}^q_{T}(Q^2)
\end{align}
\end{widetext}
\red{Using equation of motion \cite{Bhoonah:2017olu,Freese:2019bhb},}

\begin{equation}
\bar{\psi}\, i\sigma^{\lambda\mu}\gamma_5\, i\overleftrightarrow{\partial}_\mu \psi
= 2M\,\bar{\psi}\gamma^\lambda\gamma_5\psi
+ i\partial^\lambda(\bar{\psi}\gamma_5\psi)
\end{equation}
and
\begin{equation}
\epsilon_{\lambda\mu\nu\alpha}\bar{\psi}\, i\sigma^{\lambda\mu}\gamma_5\, i\overleftrightarrow{\partial}^{\nu} \psi
=2\,
\partial_\alpha(\bar{\psi}\psi).
\end{equation}
\red{those tensor quark gravitational form factors can be further constrained by the relations}  

\begin{equation}
\begin{aligned}
    \frac{2M}{m_N}G_A^q=&A_T^q-\frac{Q^2}{m_N^2}\tilde{A}_T^q-\frac{Q^2}{4m_N^2}B_T^q+\frac{Q^2}{2m_N^2}\tilde{B}_T^q\\
    &-2C_T^q-3D_T^q
\end{aligned}
\end{equation}

\begin{equation}
\begin{aligned}
    \frac{2M}{m_N}G_P^q-\frac{2m_N}m\tilde{G}^q_P=&-A_T^q+\frac{Q^2}{m_N^2}\tilde{A}_T^q-B_T^q\\
    &+2\tilde{B}_T^q+2C_T^q-6\tilde{D}_T^q
\end{aligned}
\end{equation}

\begin{equation}
\begin{aligned}
    -\frac{2m_N}{m}\sigma_N^q=&A_T^q+2\tilde{A}_T^q-\frac{Q^2}{m_N^2}\tilde{A}_T^q+B_T^q+4C_T^q
\end{aligned}
\end{equation}

As we are only interested in the twist-3 contribution in the Breit frame $q^+=0$, we will only focus on the $\Phi^q_{N,1}$, $\Phi^q_{N,3}$, $\Phi^q_{N,7}$, and $\Phi^q_{N,8}$ form factors in this work. In order to evaluate these form factors, we parameterize the energy momentum tensor form factors $A_N^q$ and $B_N^q$ by dipoles. Their scale dependence is also controlled by RG equation

\begin{equation}
    \begin{aligned}
\label{RG}
   A^{u-d}(Q^2,\mu)=\left(\frac{\alpha_s(\mu_0)}{\alpha_s(\mu)}\right)^{-\frac{8}{3}\frac{C_F}{\beta_0}}A^{u-d}(Q^2,\mu_0)     
    \end{aligned}
\end{equation}
and mixed equation in isoscalar sector

\begin{equation}
    \begin{aligned}
\label{DGLAP}
   \mu\frac{d}{d\mu}A^{u+d}(\mu)=&\frac{\alpha_s}{4\pi}\left[-\frac{16}{3}C_FA^{u+d}(\mu)+ \frac{8}{3}A^{g}(\mu)\right]\\
   \mu\frac{d}{d\mu}A^g(\mu)=&\frac{\alpha_s}{4\pi}\left[\frac{16}{3}C_FA^{u+d}(\mu)- \frac{8}{3}A^{g}(\mu)\right]\\   
    \end{aligned}
\end{equation}

At $\mu=2$ GeV, the lattice fitted dipole forms read
\begin{align}
    A_N^{u}(Q^2)=&\frac{0.4055}{(1 + Q^2/1.772^2)^2}\\
    A_N^{d}(Q^2)=&\frac{0.1385}{(1 + Q^2/1.555^2)^2}\\
    B_N^{u}(Q^2)=&\frac{0.171}{(1 + Q^2/1.535^2)^2}\\
    B_N^{d}(Q^2)=&\frac{-0.124}{(1 + Q^2/1.657^2)^2}
\end{align}
where the parameter for the isovector channel is obtained by fitting the lattice result of the second moment of GPD in~\cite{Bhattacharya:2023ays} with slightly heavier pion mass $m_\pi=260$ MeV, and the parameter for the isoscalar channel is obtained by fitting the lattice result of the gravitational form factor from~\cite{Yao:2024ixu}.

\red{For the transversity gravitational form factors, we also parameterize them by dipoles. Their scale dependence is also controlled by RG eqiation.}

\begin{equation}
    A^q_T(Q^2,\mu)=\left(\frac{\alpha_s(\mu_0)}{\alpha_s(\mu)}\right)^{-\frac{13}3\frac{C_F}{\beta_0}}A^q_T(Q^2,\mu_0)
\end{equation}
At $\mu=2$ GeV, the lattice fitted dipole forms read
\begin{align}
A^u_{T}(Q^2) &= \frac{0.268}{\left(1 + Q^2/2.312^2\right)^2} \\
A^d_{T}(Q^2) &= \frac{-0.052}{\left(1 + Q^2/2.448^2\right)^2}
\end{align}

\begin{align}
2\tilde{A}^{u+d}_{T}(t)+B^{u+d}_{T}(t) &=\frac{0.680}{\left(1+Q^2/1.736^2\right)^{2}}\\
2\tilde{A}^{u-d}_{T}(t)+B^{u-d}_{T}(t) &=\frac{0.267}{\left(1 + Q^2/1.395^2\right)^{2}}
\end{align}

\begin{align}
\label{eq:AT}
\tilde{A}^{u+d}_{T}(t) &= \frac{0.576}{\left(1 + Q^2/1.517^2\right)^{2}}\\
\tilde{A}^{u-d}_{T}(t) &= \frac{-0.0764}{\left(1 + Q^2/1.517^2\right)^{2}}
\end{align}
\red{where the parameters for $A_T^{u,d}$ is
obtained by fitting the lattice result from QCDSF/UKQCD collaboration in \cite{Gockeler:2005cj}, and the parameters
for $\tilde{A}_T^{u+d}$ and $B_T^{u+d}$ in isoscalar channel is obtained by fitting
the lattice result from QCDSF/UKQCD collaboration \cite{QCDSF:2006tkx} where they extrapolate the pion mass to $m_\pi=140$ MeV. The parameters
for $\tilde{A}_T^{u-d}$ and $B_T^{u-d}$ in isovector channel is obtained by fitting
the lattice result from Cyprus lattice group \cite{Alexandrou:2022dtc}. Here we assume that the dipole mass for $\tilde{A}^{u+d}_T$ in \eqref{eq:AT} is identical to that of $\tilde{A}^{u-d}_T$, due to the lack of available data constraining its momentum dependence. }


\bibliography{twist3,Ref3,hi4,ref,orbital}

\begin{thebibliography}{52}%
\makeatletter
\providecommand \@ifxundefined [1]{%
 \@ifx{#1\undefined}
}%
\providecommand \@ifnum [1]{%
 \ifnum #1\expandafter \@firstoftwo
 \else \expandafter \@secondoftwo
 \fi
}%
\providecommand \@ifx [1]{%
 \ifx #1\expandafter \@firstoftwo
 \else \expandafter \@secondoftwo
 \fi
}%
\providecommand \natexlab [1]{#1}%
\providecommand \enquote  [1]{``#1''}%
\providecommand \bibnamefont  [1]{#1}%
\providecommand \bibfnamefont [1]{#1}%
\providecommand \citenamefont [1]{#1}%
\providecommand \href@noop [0]{\@secondoftwo}%
\providecommand \href [0]{\begingroup \@sanitize@url \@href}%
\providecommand \@href[1]{\@@startlink{#1}\@@href}%
\providecommand \@@href[1]{\endgroup#1\@@endlink}%
\providecommand \@sanitize@url [0]{\catcode `\\12\catcode `\$12\catcode
  `\&12\catcode `\#12\catcode `\^12\catcode `\_12\catcode `\%12\relax}%
\providecommand \@@startlink[1]{}%
\providecommand \@@endlink[0]{}%
\providecommand \url  [0]{\begingroup\@sanitize@url \@url }%
\providecommand \@url [1]{\endgroup\@href {#1}{\urlprefix }}%
\providecommand \urlprefix  [0]{URL }%
\providecommand \Eprint [0]{\href }%
\providecommand \doibase [0]{http://dx.doi.org/}%
\providecommand \selectlanguage [0]{\@gobble}%
\providecommand \bibinfo  [0]{\@secondoftwo}%
\providecommand \bibfield  [0]{\@secondoftwo}%
\providecommand \translation [1]{[#1]}%
\providecommand \BibitemOpen [0]{}%
\providecommand \bibitemStop [0]{}%
\providecommand \bibitemNoStop [0]{.\EOS\space}%
\providecommand \EOS [0]{\spacefactor3000\relax}%
\providecommand \BibitemShut  [1]{\csname bibitem#1\endcsname}%
\let\auto@bib@innerbib\@empty
\bibitem [{\citenamefont {Shuryak}\ and\ \citenamefont
  {Vainshtein}(1982{\natexlab{a}})}]{Shuryak:1981kj}%
  \BibitemOpen
  \bibfield  {author} {\bibinfo {author} {\bibfnamefont {Edward~V.}\
  \bibnamefont {Shuryak}}\ and\ \bibinfo {author} {\bibfnamefont {A.~I.}\
  \bibnamefont {Vainshtein}},\ }\bibfield  {title} {\enquote {\bibinfo {title}
  {{Theory of Power Corrections to Deep Inelastic Scattering in Quantum
  Chromodynamics. 1. Q**2 Effects}},}\ }\href {\doibase
  10.1016/0550-3213(82)90355-8} {\bibfield  {journal} {\bibinfo  {journal}
  {Nucl. Phys. B}\ }\textbf {\bibinfo {volume} {199}},\ \bibinfo {pages}
  {451--481} (\bibinfo {year} {1982}{\natexlab{a}})}\BibitemShut {NoStop}%
\bibitem [{\citenamefont {Shuryak}\ and\ \citenamefont
  {Vainshtein}(1982{\natexlab{b}})}]{Shuryak:1981pi}%
  \BibitemOpen
  \bibfield  {author} {\bibinfo {author} {\bibfnamefont {Edward~V.}\
  \bibnamefont {Shuryak}}\ and\ \bibinfo {author} {\bibfnamefont {A.~I.}\
  \bibnamefont {Vainshtein}},\ }\bibfield  {title} {\enquote {\bibinfo {title}
  {{Theory of Power Corrections to Deep Inelastic Scattering in Quantum
  Chromodynamics. 2. Q**4 Effects: Polarized Target}},}\ }\href {\doibase
  10.1016/0550-3213(82)90377-7} {\bibfield  {journal} {\bibinfo  {journal}
  {Nucl. Phys. B}\ }\textbf {\bibinfo {volume} {201}},\ \bibinfo {pages} {141}
  (\bibinfo {year} {1982}{\natexlab{b}})}\BibitemShut {NoStop}%
\bibitem [{\citenamefont {Hatta}\ and\ \citenamefont
  {Schoenleber}(2024)}]{Hatta:2024otc}%
  \BibitemOpen
  \bibfield  {author} {\bibinfo {author} {\bibfnamefont {Yoshitaka}\
  \bibnamefont {Hatta}}\ and\ \bibinfo {author} {\bibfnamefont {Jakob}\
  \bibnamefont {Schoenleber}},\ }\bibfield  {title} {\enquote {\bibinfo {title}
  {{Twist analysis of the spin-orbit correlation in QCD}},}\ }\href@noop {} {\
  (\bibinfo {year} {2024})},\ \Eprint {http://arxiv.org/abs/2404.18872}
  {arXiv:2404.18872 [hep-ph]} \BibitemShut {NoStop}%
\bibitem [{\citenamefont {Burkardt}(2013)}]{Burkardt:2008ps}%
  \BibitemOpen
  \bibfield  {author} {\bibinfo {author} {\bibfnamefont {Matthias}\
  \bibnamefont {Burkardt}},\ }\bibfield  {title} {\enquote {\bibinfo {title}
  {{Transverse force on quarks in deep-inelastic scattering}},}\ }\href
  {\doibase 10.1103/PhysRevD.88.114502} {\bibfield  {journal} {\bibinfo
  {journal} {Phys. Rev. D}\ }\textbf {\bibinfo {volume} {88}},\ \bibinfo
  {pages} {114502} (\bibinfo {year} {2013})},\ \Eprint
  {http://arxiv.org/abs/0810.3589} {arXiv:0810.3589 [hep-ph]} \BibitemShut
  {NoStop}%
\bibitem [{\citenamefont {Shuryak}(1982)}]{Shuryak:1981ff}%
  \BibitemOpen
  \bibfield  {author} {\bibinfo {author} {\bibfnamefont {Edward~V.}\
  \bibnamefont {Shuryak}},\ }\bibfield  {title} {\enquote {\bibinfo {title}
  {{The Role of Instantons in Quantum Chromodynamics. 1. Physical Vacuum}},}\
  }\href {\doibase 10.1016/0550-3213(82)90478-3} {\bibfield  {journal}
  {\bibinfo  {journal} {Nucl. Phys. B}\ }\textbf {\bibinfo {volume} {203}},\
  \bibinfo {pages} {93} (\bibinfo {year} {1982})}\BibitemShut {NoStop}%
\bibitem [{\citenamefont {Diakonov}\ and\ \citenamefont
  {Petrov}(1986)}]{Diakonov:1985eg}%
  \BibitemOpen
  \bibfield  {author} {\bibinfo {author} {\bibfnamefont {Dmitri}\ \bibnamefont
  {Diakonov}}\ and\ \bibinfo {author} {\bibfnamefont {V.~Yu.}\ \bibnamefont
  {Petrov}},\ }\bibfield  {title} {\enquote {\bibinfo {title} {{A Theory of
  Light Quarks in the Instanton Vacuum}},}\ }\href {\doibase
  10.1016/0550-3213(86)90011-8} {\bibfield  {journal} {\bibinfo  {journal}
  {Nucl. Phys. B}\ }\textbf {\bibinfo {volume} {272}},\ \bibinfo {pages}
  {457--489} (\bibinfo {year} {1986})}\BibitemShut {NoStop}%
\bibitem [{\citenamefont {Sch\"afer}\ and\ \citenamefont
  {Shuryak}(1998)}]{Schafer:1996wv}%
  \BibitemOpen
  \bibfield  {author} {\bibinfo {author} {\bibfnamefont {Thomas}\ \bibnamefont
  {Sch\"afer}}\ and\ \bibinfo {author} {\bibfnamefont {Edward~V.}\ \bibnamefont
  {Shuryak}},\ }\bibfield  {title} {\enquote {\bibinfo {title} {{Instantons in
  QCD}},}\ }\href {\doibase 10.1103/RevModPhys.70.323} {\bibfield  {journal}
  {\bibinfo  {journal} {Rev. Mod. Phys.}\ }\textbf {\bibinfo {volume} {70}},\
  \bibinfo {pages} {323--426} (\bibinfo {year} {1998})},\ \Eprint
  {http://arxiv.org/abs/hep-ph/9610451} {arXiv:hep-ph/9610451} \BibitemShut
  {NoStop}%
\bibitem [{\citenamefont {Balla}\ \emph {et~al.}(1998)\citenamefont {Balla},
  \citenamefont {Polyakov},\ and\ \citenamefont {Weiss}}]{Balla:1997hf}%
  \BibitemOpen
  \bibfield  {author} {\bibinfo {author} {\bibfnamefont {J.}~\bibnamefont
  {Balla}}, \bibinfo {author} {\bibfnamefont {Maxim~V.}\ \bibnamefont
  {Polyakov}}, \ and\ \bibinfo {author} {\bibfnamefont {C.}~\bibnamefont
  {Weiss}},\ }\bibfield  {title} {\enquote {\bibinfo {title} {{Nucleon matrix
  elements of higher twist operators from the instanton vacuum}},}\ }\href
  {\doibase 10.1016/S0550-3213(98)00638-5} {\bibfield  {journal} {\bibinfo
  {journal} {Nucl. Phys. B}\ }\textbf {\bibinfo {volume} {510}},\ \bibinfo
  {pages} {327--364} (\bibinfo {year} {1998})},\ \Eprint
  {http://arxiv.org/abs/hep-ph/9707515} {arXiv:hep-ph/9707515} \BibitemShut
  {NoStop}%
\bibitem [{\citenamefont {Vladimirov}\ \emph {et~al.}(2025)\citenamefont
  {Vladimirov}, \citenamefont {Portela},\ and\ \citenamefont
  {Rodini}}]{Vladimirov:2025qrh}%
  \BibitemOpen
  \bibfield  {author} {\bibinfo {author} {\bibfnamefont {Alexey}\ \bibnamefont
  {Vladimirov}}, \bibinfo {author} {\bibfnamefont {Guillermo}\ \bibnamefont
  {Portela}}, \ and\ \bibinfo {author} {\bibfnamefont {Simone}\ \bibnamefont
  {Rodini}},\ }\bibfield  {title} {\enquote {\bibinfo {title} {{Determination
  of quark-gluon-quark interference within the proton}},}\ }\href@noop {} {\
  (\bibinfo {year} {2025})},\ \Eprint {http://arxiv.org/abs/2511.04294}
  {arXiv:2511.04294 [hep-ph]} \BibitemShut {NoStop}%
\bibitem [{\citenamefont {Crawford}\ \emph {et~al.}(2024)\citenamefont
  {Crawford}, \citenamefont {Can}, \citenamefont {Horsley}, \citenamefont
  {Rakow}, \citenamefont {Schierholz}, \citenamefont {St\"uben}, \citenamefont
  {Young},\ and\ \citenamefont {Zanotti}}]{Crawford:2024wzx}%
  \BibitemOpen
  \bibfield  {author} {\bibinfo {author} {\bibfnamefont {J.~A.}\ \bibnamefont
  {Crawford}}, \bibinfo {author} {\bibfnamefont {K.~U.}\ \bibnamefont {Can}},
  \bibinfo {author} {\bibfnamefont {R.}~\bibnamefont {Horsley}}, \bibinfo
  {author} {\bibfnamefont {P.~E.~L.}\ \bibnamefont {Rakow}}, \bibinfo {author}
  {\bibfnamefont {G.}~\bibnamefont {Schierholz}}, \bibinfo {author}
  {\bibfnamefont {H.}~\bibnamefont {St\"uben}}, \bibinfo {author}
  {\bibfnamefont {R.~D.}\ \bibnamefont {Young}}, \ and\ \bibinfo {author}
  {\bibfnamefont {J.~M.}\ \bibnamefont {Zanotti}},\ }\bibfield  {title}
  {\enquote {\bibinfo {title} {{Transverse force distributions in the proton
  from lattice QCD}},}\ }\href@noop {} {\  (\bibinfo {year} {2024})},\ \Eprint
  {http://arxiv.org/abs/2408.03621} {arXiv:2408.03621 [hep-lat]} \BibitemShut
  {NoStop}%
\bibitem [{\citenamefont {Athenodorou}\ \emph {et~al.}(2018)\citenamefont
  {Athenodorou}, \citenamefont {Boucaud}, \citenamefont {De~Soto},
  \citenamefont {Rodr\'\i{}guez-Quintero},\ and\ \citenamefont
  {Zafeiropoulos}}]{Athenodorou:2018jwu}%
  \BibitemOpen
  \bibfield  {author} {\bibinfo {author} {\bibfnamefont {A.}~\bibnamefont
  {Athenodorou}}, \bibinfo {author} {\bibfnamefont {Ph.}\ \bibnamefont
  {Boucaud}}, \bibinfo {author} {\bibfnamefont {F.}~\bibnamefont {De~Soto}},
  \bibinfo {author} {\bibfnamefont {J.}~\bibnamefont
  {Rodr\'\i{}guez-Quintero}}, \ and\ \bibinfo {author} {\bibfnamefont
  {S.}~\bibnamefont {Zafeiropoulos}},\ }\bibfield  {title} {\enquote {\bibinfo
  {title} {{Instanton liquid properties from lattice QCD}},}\ }\href {\doibase
  10.1007/JHEP02(2018)140} {\bibfield  {journal} {\bibinfo  {journal} {JHEP}\
  }\textbf {\bibinfo {volume} {02}},\ \bibinfo {pages} {140} (\bibinfo {year}
  {2018})},\ \Eprint {http://arxiv.org/abs/1801.10155} {arXiv:1801.10155
  [hep-lat]} \BibitemShut {NoStop}%
\bibitem [{\citenamefont {Shuryak}\ and\ \citenamefont
  {Zahed}(2023)}]{Shuryak:2021fsu}%
  \BibitemOpen
  \bibfield  {author} {\bibinfo {author} {\bibfnamefont {Edward}\ \bibnamefont
  {Shuryak}}\ and\ \bibinfo {author} {\bibfnamefont {Ismail}\ \bibnamefont
  {Zahed}},\ }\bibfield  {title} {\enquote {\bibinfo {title} {{Hadronic
  structure on the light front. I. Instanton effects and quark-antiquark
  effective potentials}},}\ }\href {\doibase 10.1103/PhysRevD.107.034023}
  {\bibfield  {journal} {\bibinfo  {journal} {Phys. Rev. D}\ }\textbf {\bibinfo
  {volume} {107}},\ \bibinfo {pages} {034023} (\bibinfo {year} {2023})},\
  \Eprint {http://arxiv.org/abs/2110.15927} {arXiv:2110.15927 [hep-ph]}
  \BibitemShut {NoStop}%
\bibitem [{\citenamefont {Shuryak}(1988)}]{Shuryak:1987tr}%
  \BibitemOpen
  \bibfield  {author} {\bibinfo {author} {\bibfnamefont {Edward~V.}\
  \bibnamefont {Shuryak}},\ }\bibfield  {title} {\enquote {\bibinfo {title}
  {{Toward the Quantitative Theory of the 'Instanton Liquid' 4. Tunneling in
  the Double Well Potential}},}\ }\href {\doibase 10.1016/0550-3213(88)90191-5}
  {\bibfield  {journal} {\bibinfo  {journal} {Nucl. Phys. B}\ }\textbf
  {\bibinfo {volume} {302}},\ \bibinfo {pages} {621--644} (\bibinfo {year}
  {1988})}\BibitemShut {NoStop}%
\bibitem [{\citenamefont {Balitsky}\ and\ \citenamefont
  {Yung}(1986)}]{Balitsky:1986qn}%
  \BibitemOpen
  \bibfield  {author} {\bibinfo {author} {\bibfnamefont {I.~I.}\ \bibnamefont
  {Balitsky}}\ and\ \bibinfo {author} {\bibfnamefont {A.~V.}\ \bibnamefont
  {Yung}},\ }\bibfield  {title} {\enquote {\bibinfo {title} {{Collective -
  Coordinate Method for Quasizero Modes}},}\ }\href {\doibase
  10.1016/0370-2693(86)91471-1} {\bibfield  {journal} {\bibinfo  {journal}
  {Phys. Lett. B}\ }\textbf {\bibinfo {volume} {168}},\ \bibinfo {pages}
  {113--119} (\bibinfo {year} {1986})}\BibitemShut {NoStop}%
\bibitem [{\citenamefont {Verbaarschot}(1991)}]{Verbaarschot:1991sq}%
  \BibitemOpen
  \bibfield  {author} {\bibinfo {author} {\bibfnamefont {J.~J.~M.}\
  \bibnamefont {Verbaarschot}},\ }\bibfield  {title} {\enquote {\bibinfo
  {title} {{Streamlines and conformal invariance in Yang-Mills theories}},}\
  }\href {\doibase 10.1016/0550-3213(91)90554-B} {\bibfield  {journal}
  {\bibinfo  {journal} {Nucl. Phys. B}\ }\textbf {\bibinfo {volume} {362}},\
  \bibinfo {pages} {33--53} (\bibinfo {year} {1991})},\ \bibinfo {note}
  {[Erratum: Nucl.Phys.B 386, 236--236 (1992)]}\BibitemShut {NoStop}%
\bibitem [{\citenamefont {Ilgenfritz}\ and\ \citenamefont
  {Shuryak}(1989)}]{Ilgenfritz:1988dh}%
  \BibitemOpen
  \bibfield  {author} {\bibinfo {author} {\bibfnamefont {Ernst-Michael}\
  \bibnamefont {Ilgenfritz}}\ and\ \bibinfo {author} {\bibfnamefont
  {Edward~V.}\ \bibnamefont {Shuryak}},\ }\bibfield  {title} {\enquote
  {\bibinfo {title} {{Chiral Symmetry Restoration at Finite Temperature in the
  Instanton Liquid}},}\ }\href {\doibase 10.1016/0550-3213(89)90617-2}
  {\bibfield  {journal} {\bibinfo  {journal} {Nucl. Phys. B}\ }\textbf
  {\bibinfo {volume} {319}},\ \bibinfo {pages} {511--520} (\bibinfo {year}
  {1989})}\BibitemShut {NoStop}%
\bibitem [{\citenamefont {Shuryak}\ and\ \citenamefont
  {Zahed}(2021)}]{Shuryak:2020ktq}%
  \BibitemOpen
  \bibfield  {author} {\bibinfo {author} {\bibfnamefont {Edward}\ \bibnamefont
  {Shuryak}}\ and\ \bibinfo {author} {\bibfnamefont {Ismail}\ \bibnamefont
  {Zahed}},\ }\bibfield  {title} {\enquote {\bibinfo {title} {{Nonperturbative
  quark-antiquark interactions in mesonic form factors}},}\ }\href {\doibase
  10.1103/PhysRevD.103.054028} {\bibfield  {journal} {\bibinfo  {journal}
  {Phys. Rev. D}\ }\textbf {\bibinfo {volume} {103}},\ \bibinfo {pages}
  {054028} (\bibinfo {year} {2021})},\ \Eprint
  {http://arxiv.org/abs/2008.06169} {arXiv:2008.06169 [hep-ph]} \BibitemShut
  {NoStop}%
\bibitem [{\citenamefont {Miesch}\ \emph {et~al.}(2024)\citenamefont {Miesch},
  \citenamefont {Shuryak},\ and\ \citenamefont {Zahed}}]{Miesch:2024fhv}%
  \BibitemOpen
  \bibfield  {author} {\bibinfo {author} {\bibfnamefont {Nicholas}\
  \bibnamefont {Miesch}}, \bibinfo {author} {\bibfnamefont {Edward}\
  \bibnamefont {Shuryak}}, \ and\ \bibinfo {author} {\bibfnamefont {Ismail}\
  \bibnamefont {Zahed}},\ }\bibfield  {title} {\enquote {\bibinfo {title}
  {{Bridging hadronic and vacuum structure by heavy quarkonia}},}\ }\href@noop
  {} {\  (\bibinfo {year} {2024})},\ \Eprint {http://arxiv.org/abs/2403.18700}
  {arXiv:2403.18700 [hep-ph]} \BibitemShut {NoStop}%
\bibitem [{\citenamefont {'t~Hooft}(1976)}]{tHooft:1976snw}%
  \BibitemOpen
  \bibfield  {author} {\bibinfo {author} {\bibfnamefont {Gerard}\ \bibnamefont
  {'t~Hooft}},\ }\bibfield  {title} {\enquote {\bibinfo {title} {{Computation
  of the Quantum Effects Due to a Four-Dimensional Pseudoparticle}},}\ }\href
  {\doibase 10.1103/PhysRevD.14.3432} {\bibfield  {journal} {\bibinfo
  {journal} {Phys. Rev. D}\ }\textbf {\bibinfo {volume} {14}},\ \bibinfo
  {pages} {3432--3450} (\bibinfo {year} {1976})},\ \bibinfo {note} {[Erratum:
  Phys.Rev.D 18, 2199 (1978)]}\BibitemShut {NoStop}%
\bibitem [{\citenamefont {Liu}\ \emph {et~al.}(2024)\citenamefont {Liu},
  \citenamefont {Shuryak},\ and\ \citenamefont {Zahed}}]{Liu:2024rdm}%
  \BibitemOpen
  \bibfield  {author} {\bibinfo {author} {\bibfnamefont {Wei-Yang}\
  \bibnamefont {Liu}}, \bibinfo {author} {\bibfnamefont {Edward}\ \bibnamefont
  {Shuryak}}, \ and\ \bibinfo {author} {\bibfnamefont {Ismail}\ \bibnamefont
  {Zahed}},\ }\bibfield  {title} {\enquote {\bibinfo {title} {{Glue in hadrons
  at medium resolution and the QCD instanton vacuum}},}\ }\href@noop {} {\
  (\bibinfo {year} {2024})},\ \Eprint {http://arxiv.org/abs/2404.03047}
  {arXiv:2404.03047 [hep-ph]} \BibitemShut {NoStop}%
\bibitem [{\citenamefont {Bhattacharya}\ and\ \citenamefont
  {Metz}(2022)}]{Bhattacharya:2021boh}%
  \BibitemOpen
  \bibfield  {author} {\bibinfo {author} {\bibfnamefont {Shohini}\ \bibnamefont
  {Bhattacharya}}\ and\ \bibinfo {author} {\bibfnamefont {Andreas}\
  \bibnamefont {Metz}},\ }\bibfield  {title} {\enquote {\bibinfo {title}
  {{Burkhardt-Cottingham-type sum rules for light-cone and quasi-PDFs}},}\
  }\href {\doibase 10.1103/PhysRevD.105.054027} {\bibfield  {journal} {\bibinfo
   {journal} {Phys. Rev. D}\ }\textbf {\bibinfo {volume} {105}},\ \bibinfo
  {pages} {054027} (\bibinfo {year} {2022})},\ \Eprint
  {http://arxiv.org/abs/2105.07282} {arXiv:2105.07282 [hep-ph]} \BibitemShut
  {NoStop}%
\bibitem [{\citenamefont {Shuryak}\ and\ \citenamefont
  {Vainshtein}(1981)}]{Shuryak:1981dg}%
  \BibitemOpen
  \bibfield  {author} {\bibinfo {author} {\bibfnamefont {Edward~V.}\
  \bibnamefont {Shuryak}}\ and\ \bibinfo {author} {\bibfnamefont {A.~I.}\
  \bibnamefont {Vainshtein}},\ }\bibfield  {title} {\enquote {\bibinfo {title}
  {{QCD POWER CORRECTIONS TO DEEP INELASTIC SCATTERING}},}\ }\href {\doibase
  10.1016/0370-2693(81)90042-3} {\bibfield  {journal} {\bibinfo  {journal}
  {Phys. Lett. B}\ }\textbf {\bibinfo {volume} {105}},\ \bibinfo {pages}
  {65--67} (\bibinfo {year} {1981})}\BibitemShut {NoStop}%
\bibitem [{\citenamefont {Jaffe}\ and\ \citenamefont
  {Soldate}(1981)}]{Jaffe:1981td}%
  \BibitemOpen
  \bibfield  {author} {\bibinfo {author} {\bibfnamefont {R.~L.}\ \bibnamefont
  {Jaffe}}\ and\ \bibinfo {author} {\bibfnamefont {M.}~\bibnamefont
  {Soldate}},\ }\bibfield  {title} {\enquote {\bibinfo {title} {{Twist Four in
  the QCD Analysis of Leptoproduction}},}\ }\href {\doibase
  10.1016/0370-2693(81)91206-5} {\bibfield  {journal} {\bibinfo  {journal}
  {Phys. Lett. B}\ }\textbf {\bibinfo {volume} {105}},\ \bibinfo {pages}
  {467--472} (\bibinfo {year} {1981})}\BibitemShut {NoStop}%
\bibitem [{\citenamefont {Jaffe}\ and\ \citenamefont
  {Soldate}(1982)}]{Jaffe:1982pm}%
  \BibitemOpen
  \bibfield  {author} {\bibinfo {author} {\bibfnamefont {R.~L.}\ \bibnamefont
  {Jaffe}}\ and\ \bibinfo {author} {\bibfnamefont {M.}~\bibnamefont
  {Soldate}},\ }\bibfield  {title} {\enquote {\bibinfo {title} {{Twist Four in
  Electroproduction: Canonical Operators and Coefficient Functions}},}\ }\href
  {\doibase 10.1103/PhysRevD.26.49} {\bibfield  {journal} {\bibinfo  {journal}
  {Phys. Rev. D}\ }\textbf {\bibinfo {volume} {26}},\ \bibinfo {pages} {49--68}
  (\bibinfo {year} {1982})}\BibitemShut {NoStop}%
\bibitem [{\citenamefont {Aslan}\ \emph {et~al.}(2019)\citenamefont {Aslan},
  \citenamefont {Burkardt},\ and\ \citenamefont {Schlegel}}]{Aslan:2019jis}%
  \BibitemOpen
  \bibfield  {author} {\bibinfo {author} {\bibfnamefont {Fatma~P.}\
  \bibnamefont {Aslan}}, \bibinfo {author} {\bibfnamefont {Matthias}\
  \bibnamefont {Burkardt}}, \ and\ \bibinfo {author} {\bibfnamefont {Marc}\
  \bibnamefont {Schlegel}},\ }\bibfield  {title} {\enquote {\bibinfo {title}
  {{Transverse Force Tomography}},}\ }\href {\doibase
  10.1103/PhysRevD.100.096021} {\bibfield  {journal} {\bibinfo  {journal}
  {Phys. Rev. D}\ }\textbf {\bibinfo {volume} {100}},\ \bibinfo {pages}
  {096021} (\bibinfo {year} {2019})},\ \Eprint
  {http://arxiv.org/abs/1904.03494} {arXiv:1904.03494 [hep-ph]} \BibitemShut
  {NoStop}%
\bibitem [{\citenamefont {Wandzura}\ and\ \citenamefont
  {Wilczek}(1977)}]{Wandzura:1977qf}%
  \BibitemOpen
  \bibfield  {author} {\bibinfo {author} {\bibfnamefont {S.}~\bibnamefont
  {Wandzura}}\ and\ \bibinfo {author} {\bibfnamefont {Frank}\ \bibnamefont
  {Wilczek}},\ }\bibfield  {title} {\enquote {\bibinfo {title} {{Sum Rules for
  Spin Dependent Electroproduction: Test of Relativistic Constituent
  Quarks}},}\ }\href {\doibase 10.1016/0370-2693(77)90700-6} {\bibfield
  {journal} {\bibinfo  {journal} {Phys. Lett. B}\ }\textbf {\bibinfo {volume}
  {72}},\ \bibinfo {pages} {195--198} (\bibinfo {year} {1977})}\BibitemShut
  {NoStop}%
\bibitem [{\citenamefont {Aslan}\ \emph {et~al.}(2020)\citenamefont {Aslan},
  \citenamefont {Burkardt},\ and\ \citenamefont {Schlegel}}]{Aslan:2020eqo}%
  \BibitemOpen
  \bibfield  {author} {\bibinfo {author} {\bibfnamefont {Fatma~P.}\
  \bibnamefont {Aslan}}, \bibinfo {author} {\bibfnamefont {Matthias}\
  \bibnamefont {Burkardt}}, \ and\ \bibinfo {author} {\bibfnamefont {Marc}\
  \bibnamefont {Schlegel}},\ }\bibfield  {title} {\enquote {\bibinfo {title}
  {{Transverse Force Tomography}},}\ }in\ \href {\doibase
  10.1142/9789811214950_0038} {\emph {\bibinfo {booktitle} {{Probing Nucleons
  and Nuclei in High Energy Collisions}: {Dedicated to the Physics of the
  Electron Ion Collider}}}}\ (\bibinfo {year} {2020})\ pp.\ \bibinfo {pages}
  {186--189},\ \Eprint {http://arxiv.org/abs/2001.05978} {arXiv:2001.05978
  [hep-ph]} \BibitemShut {NoStop}%
\bibitem [{\citenamefont {Ji}(1995)}]{Ji:1995mg}%
  \BibitemOpen
  \bibfield  {author} {\bibinfo {author} {\bibfnamefont {Xiang-Dong}\
  \bibnamefont {Ji}},\ }\bibfield  {title} {\enquote {\bibinfo {title}
  {{Physics of the G2 structure function of the nucleon}},}\ }in\ \href@noop {}
  {\emph {\bibinfo {booktitle} {{3rd Workshop on Deep Inelastic Scattering and
  QCD (DIS 95)}}}}\ (\bibinfo {year} {1995})\ pp.\ \bibinfo {pages}
  {435--438}\BibitemShut {NoStop}%
\bibitem [{\citenamefont {Liu}\ \emph {et~al.}(2023{\natexlab{a}})\citenamefont
  {Liu}, \citenamefont {Shuryak},\ and\ \citenamefont {Zahed}}]{Liu:2023yuj}%
  \BibitemOpen
  \bibfield  {author} {\bibinfo {author} {\bibfnamefont {Wei-Yang}\
  \bibnamefont {Liu}}, \bibinfo {author} {\bibfnamefont {Edward}\ \bibnamefont
  {Shuryak}}, \ and\ \bibinfo {author} {\bibfnamefont {Ismail}\ \bibnamefont
  {Zahed}},\ }\bibfield  {title} {\enquote {\bibinfo {title} {{Hadronic
  structure on the light-front. VII. Pions and kaons and their partonic
  distributions}},}\ }\href {\doibase 10.1103/PhysRevD.107.094024} {\bibfield
  {journal} {\bibinfo  {journal} {Phys. Rev. D}\ }\textbf {\bibinfo {volume}
  {107}},\ \bibinfo {pages} {094024} (\bibinfo {year} {2023}{\natexlab{a}})},\
  \Eprint {http://arxiv.org/abs/2302.03759} {arXiv:2302.03759 [hep-ph]}
  \BibitemShut {NoStop}%
\bibitem [{\citenamefont {Liu}\ \emph {et~al.}(2023{\natexlab{b}})\citenamefont
  {Liu}, \citenamefont {Shuryak},\ and\ \citenamefont {Zahed}}]{Liu:2023fpj}%
  \BibitemOpen
  \bibfield  {author} {\bibinfo {author} {\bibfnamefont {Wei-Yang}\
  \bibnamefont {Liu}}, \bibinfo {author} {\bibfnamefont {Edward}\ \bibnamefont
  {Shuryak}}, \ and\ \bibinfo {author} {\bibfnamefont {Ismail}\ \bibnamefont
  {Zahed}},\ }\bibfield  {title} {\enquote {\bibinfo {title} {{Hadronic
  structure on the light-front VIII. Light scalar and vector mesons}},}\
  }\href@noop {} {\  (\bibinfo {year} {2023}{\natexlab{b}})},\ \Eprint
  {http://arxiv.org/abs/2307.16302} {arXiv:2307.16302 [hep-ph]} \BibitemShut
  {NoStop}%
\bibitem [{\citenamefont {Gockeler}\ \emph
  {et~al.}(2005{\natexlab{a}})\citenamefont {Gockeler}, \citenamefont
  {Horsley}, \citenamefont {Pleiter}, \citenamefont {Rakow}, \citenamefont
  {Schafer}, \citenamefont {Schierholz}, \citenamefont {Stuben},\ and\
  \citenamefont {Zanotti}}]{Gockeler:2005vw}%
  \BibitemOpen
  \bibfield  {author} {\bibinfo {author} {\bibfnamefont {M.}~\bibnamefont
  {Gockeler}}, \bibinfo {author} {\bibfnamefont {R.}~\bibnamefont {Horsley}},
  \bibinfo {author} {\bibfnamefont {D.}~\bibnamefont {Pleiter}}, \bibinfo
  {author} {\bibfnamefont {Paul E.~L.}\ \bibnamefont {Rakow}}, \bibinfo
  {author} {\bibfnamefont {A.}~\bibnamefont {Schafer}}, \bibinfo {author}
  {\bibfnamefont {G.}~\bibnamefont {Schierholz}}, \bibinfo {author}
  {\bibfnamefont {H.}~\bibnamefont {Stuben}}, \ and\ \bibinfo {author}
  {\bibfnamefont {J.~M.}\ \bibnamefont {Zanotti}},\ }\bibfield  {title}
  {\enquote {\bibinfo {title} {{Investigation of the second moment of the
  nucleon's g(1) and g(2) structure functions in two-flavor lattice QCD}},}\
  }\href {\doibase 10.1103/PhysRevD.72.054507} {\bibfield  {journal} {\bibinfo
  {journal} {Phys. Rev. D}\ }\textbf {\bibinfo {volume} {72}},\ \bibinfo
  {pages} {054507} (\bibinfo {year} {2005}{\natexlab{a}})},\ \Eprint
  {http://arxiv.org/abs/hep-lat/0506017} {arXiv:hep-lat/0506017} \BibitemShut
  {NoStop}%
\bibitem [{\citenamefont {B{\"u}rger}\ \emph {et~al.}(2022)\citenamefont
  {B{\"u}rger}, \citenamefont {Wurm}, \citenamefont {L{\"o}ffler},
  \citenamefont {G{\"o}ckeler}, \citenamefont {Bali}, \citenamefont {Collins},
  \citenamefont {Sch{\"a}fer},\ and\ \citenamefont
  {Sternbeck}}]{Burger:2021knd}%
  \BibitemOpen
  \bibfield  {author} {\bibinfo {author} {\bibfnamefont {S.}~\bibnamefont
  {B{\"u}rger}}, \bibinfo {author} {\bibfnamefont {T.}~\bibnamefont {Wurm}},
  \bibinfo {author} {\bibfnamefont {M.}~\bibnamefont {L{\"o}ffler}}, \bibinfo
  {author} {\bibfnamefont {M.}~\bibnamefont {G{\"o}ckeler}}, \bibinfo {author}
  {\bibfnamefont {G.}~\bibnamefont {Bali}}, \bibinfo {author} {\bibfnamefont
  {S.}~\bibnamefont {Collins}}, \bibinfo {author} {\bibfnamefont
  {A.}~\bibnamefont {Sch{\"a}fer}}, \ and\ \bibinfo {author} {\bibfnamefont
  {A.}~\bibnamefont {Sternbeck}} (\bibinfo {collaboration} {RQCD}),\ }\bibfield
   {title} {\enquote {\bibinfo {title} {{Lattice results for the longitudinal
  spin structure and color forces on quarks in a nucleon}},}\ }\href {\doibase
  10.1103/PhysRevD.105.054504} {\bibfield  {journal} {\bibinfo  {journal}
  {Phys. Rev. D}\ }\textbf {\bibinfo {volume} {105}},\ \bibinfo {pages}
  {054504} (\bibinfo {year} {2022})},\ \Eprint
  {http://arxiv.org/abs/2111.08306} {arXiv:2111.08306 [hep-lat]} \BibitemShut
  {NoStop}%
\bibitem [{\citenamefont {Abe}\ \emph {et~al.}(1998)\citenamefont {Abe} \emph
  {et~al.}}]{E143:1998hbs}%
  \BibitemOpen
  \bibfield  {author} {\bibinfo {author} {\bibfnamefont {K.}~\bibnamefont
  {Abe}} \emph {et~al.} (\bibinfo {collaboration} {E143}),\ }\bibfield  {title}
  {\enquote {\bibinfo {title} {{Measurements of the proton and deuteron spin
  structure functions g(1) and g(2)}},}\ }\href {\doibase
  10.1103/PhysRevD.58.112003} {\bibfield  {journal} {\bibinfo  {journal} {Phys.
  Rev. D}\ }\textbf {\bibinfo {volume} {58}},\ \bibinfo {pages} {112003}
  (\bibinfo {year} {1998})},\ \Eprint {http://arxiv.org/abs/hep-ph/9802357}
  {arXiv:hep-ph/9802357} \BibitemShut {NoStop}%
\bibitem [{\citenamefont {Liu}(2025)}]{Liu:2025ldh}%
  \BibitemOpen
  \bibfield  {author} {\bibinfo {author} {\bibfnamefont {Wei-Yang}\
  \bibnamefont {Liu}},\ }\bibfield  {title} {\enquote {\bibinfo {title}
  {{Generic framework for non-perturbative QCD in light hadrons}},}\
  }\href@noop {} {\  (\bibinfo {year} {2025})},\ \Eprint
  {http://arxiv.org/abs/2501.07776} {arXiv:2501.07776 [hep-ph]} \BibitemShut
  {NoStop}%
\bibitem [{\citenamefont {Hatsuda}\ and\ \citenamefont
  {Kunihiro}(1994)}]{Hatsuda:1994pi}%
  \BibitemOpen
  \bibfield  {author} {\bibinfo {author} {\bibfnamefont {Tetsuo}\ \bibnamefont
  {Hatsuda}}\ and\ \bibinfo {author} {\bibfnamefont {Teiji}\ \bibnamefont
  {Kunihiro}},\ }\bibfield  {title} {\enquote {\bibinfo {title} {{QCD
  phenomenology based on a chiral effective Lagrangian}},}\ }\href {\doibase
  10.1016/0370-1573(94)90022-1} {\bibfield  {journal} {\bibinfo  {journal}
  {Phys. Rept.}\ }\textbf {\bibinfo {volume} {247}},\ \bibinfo {pages}
  {221--367} (\bibinfo {year} {1994})},\ \Eprint
  {http://arxiv.org/abs/hep-ph/9401310} {arXiv:hep-ph/9401310} \BibitemShut
  {NoStop}%
\bibitem [{\citenamefont {Liu}\ and\ \citenamefont
  {Zahed}(2025)}]{Liu:2025kuc}%
  \BibitemOpen
  \bibfield  {author} {\bibinfo {author} {\bibfnamefont {Wei-Yang}\
  \bibnamefont {Liu}}\ and\ \bibinfo {author} {\bibfnamefont {Ismail}\
  \bibnamefont {Zahed}},\ }\bibfield  {title} {\enquote {\bibinfo {title}
  {{Nucleon electric dipole form factor in the QCD instanton vacuum}},}\ }\href
  {\doibase 10.1103/d9tl-ycqf} {\bibfield  {journal} {\bibinfo  {journal}
  {Phys. Rev. D}\ }\textbf {\bibinfo {volume} {112}},\ \bibinfo {pages}
  {094048} (\bibinfo {year} {2025})},\ \Eprint
  {http://arxiv.org/abs/2501.11856} {arXiv:2501.11856 [hep-ph]} \BibitemShut
  {NoStop}%
\bibitem [{\citenamefont {Diakonov}(2003)}]{Diakonov:2002fq}%
  \BibitemOpen
  \bibfield  {author} {\bibinfo {author} {\bibfnamefont {Dmitri}\ \bibnamefont
  {Diakonov}},\ }\bibfield  {title} {\enquote {\bibinfo {title} {{Instantons at
  work}},}\ }\href {\doibase 10.1016/S0146-6410(03)90014-7} {\bibfield
  {journal} {\bibinfo  {journal} {Prog. Part. Nucl. Phys.}\ }\textbf {\bibinfo
  {volume} {51}},\ \bibinfo {pages} {173--222} (\bibinfo {year} {2003})},\
  \Eprint {http://arxiv.org/abs/hep-ph/0212026} {arXiv:hep-ph/0212026}
  \BibitemShut {NoStop}%
\bibitem [{\citenamefont {Shuryak}\ and\ \citenamefont
  {Zahed}(2026)}]{Shuryak:2026pqt}%
  \BibitemOpen
  \bibfield  {author} {\bibinfo {author} {\bibfnamefont {Edward}\ \bibnamefont
  {Shuryak}}\ and\ \bibinfo {author} {\bibfnamefont {Ismail}\ \bibnamefont
  {Zahed}},\ }\bibfield  {title} {\enquote {\bibinfo {title} {{The
  Hadron-Parton Bridge, From the QCD Vacuum to Partons}},}\ }\href@noop {} {\
  (\bibinfo {year} {2026})},\ \Eprint {http://arxiv.org/abs/2601.15085}
  {arXiv:2601.15085 [hep-ph]} \BibitemShut {NoStop}%
\bibitem [{\citenamefont {Alberico}\ \emph {et~al.}(2009)\citenamefont
  {Alberico}, \citenamefont {Bilenky}, \citenamefont {Giunti},\ and\
  \citenamefont {Graczyk}}]{Alberico:2008sz}%
  \BibitemOpen
  \bibfield  {author} {\bibinfo {author} {\bibfnamefont {W.~M.}\ \bibnamefont
  {Alberico}}, \bibinfo {author} {\bibfnamefont {S.~M.}\ \bibnamefont
  {Bilenky}}, \bibinfo {author} {\bibfnamefont {C.}~\bibnamefont {Giunti}}, \
  and\ \bibinfo {author} {\bibfnamefont {K.~M.}\ \bibnamefont {Graczyk}},\
  }\bibfield  {title} {\enquote {\bibinfo {title} {{Electromagnetic form
  factors of the nucleon: New Fit and analysis of uncertainties}},}\ }\href
  {\doibase 10.1103/PhysRevC.79.065204} {\bibfield  {journal} {\bibinfo
  {journal} {Phys. Rev. C}\ }\textbf {\bibinfo {volume} {79}},\ \bibinfo
  {pages} {065204} (\bibinfo {year} {2009})},\ \Eprint
  {http://arxiv.org/abs/0812.3539} {arXiv:0812.3539 [hep-ph]} \BibitemShut
  {NoStop}%
\bibitem [{\citenamefont {Diakonov}\ \emph {et~al.}()\citenamefont {Diakonov},
  \citenamefont {Jaenicke},\ and\ \citenamefont {Polyakov}}]{Diakonov:1991}%
  \BibitemOpen
  \bibfield  {author} {\bibinfo {author} {\bibfnamefont {D.}~\bibnamefont
  {Diakonov}}, \bibinfo {author} {\bibfnamefont {J.}~\bibnamefont {Jaenicke}},
  \ and\ \bibinfo {author} {\bibfnamefont {M.}~\bibnamefont {Polyakov}},\
  }\href@noop {} {\enquote {\bibinfo {title} {Gluon exchange corrections to the
  nucleon mass in the chiral theory},}\ }\bibinfo {note} {Preprint LNPI-1738
  (1991), unpublished}\BibitemShut {NoStop}%
\bibitem [{\citenamefont {Freese}\ and\ \citenamefont
  {Clo{\"e}t}(2019)}]{Freese:2019bhb}%
  \BibitemOpen
  \bibfield  {author} {\bibinfo {author} {\bibfnamefont {Adam}\ \bibnamefont
  {Freese}}\ and\ \bibinfo {author} {\bibfnamefont {Ian~C.}\ \bibnamefont
  {Clo{\"e}t}},\ }\bibfield  {title} {\enquote {\bibinfo {title}
  {{Gravitational form factors of light mesons}},}\ }\href {\doibase
  10.1103/PhysRevC.100.015201} {\bibfield  {journal} {\bibinfo  {journal}
  {Phys. Rev. C}\ }\textbf {\bibinfo {volume} {100}},\ \bibinfo {pages}
  {015201} (\bibinfo {year} {2019})},\ \bibinfo {note} {[Erratum: Phys.Rev.C
  105, 059901 (2022)]},\ \Eprint {http://arxiv.org/abs/1903.09222}
  {arXiv:1903.09222 [nucl-th]} \BibitemShut {NoStop}%
\bibitem [{\citenamefont {Kim}\ and\ \citenamefont
  {Weiss}(2024)}]{Kim:2023pll}%
  \BibitemOpen
  \bibfield  {author} {\bibinfo {author} {\bibfnamefont {June-Young}\
  \bibnamefont {Kim}}\ and\ \bibinfo {author} {\bibfnamefont {Christian}\
  \bibnamefont {Weiss}},\ }\bibfield  {title} {\enquote {\bibinfo {title}
  {{Instanton effects in twist-3 generalized parton distributions}},}\ }\href
  {\doibase 10.1016/j.physletb.2023.138387} {\bibfield  {journal} {\bibinfo
  {journal} {Phys. Lett. B}\ }\textbf {\bibinfo {volume} {848}},\ \bibinfo
  {pages} {138387} (\bibinfo {year} {2024})},\ \Eprint
  {http://arxiv.org/abs/2310.16890} {arXiv:2310.16890 [hep-ph]} \BibitemShut
  {NoStop}%
\bibitem [{\citenamefont {Hackett}\ \emph {et~al.}(2023)\citenamefont
  {Hackett}, \citenamefont {Oare}, \citenamefont {Pefkou},\ and\ \citenamefont
  {Shanahan}}]{Hackett:2023nkr}%
  \BibitemOpen
  \bibfield  {author} {\bibinfo {author} {\bibfnamefont {Daniel~C.}\
  \bibnamefont {Hackett}}, \bibinfo {author} {\bibfnamefont {Patrick~R.}\
  \bibnamefont {Oare}}, \bibinfo {author} {\bibfnamefont {Dimitra~A.}\
  \bibnamefont {Pefkou}}, \ and\ \bibinfo {author} {\bibfnamefont {Phiala~E.}\
  \bibnamefont {Shanahan}},\ }\bibfield  {title} {\enquote {\bibinfo {title}
  {{Gravitational form factors of the pion from lattice QCD}},}\ }\href@noop {}
  {\  (\bibinfo {year} {2023})},\ \Eprint {http://arxiv.org/abs/2307.11707}
  {arXiv:2307.11707 [hep-lat]} \BibitemShut {NoStop}%
\bibitem [{\citenamefont {Diehl}\ and\ \citenamefont
  {Hagler}(2005)}]{Diehl:2005jf}%
  \BibitemOpen
  \bibfield  {author} {\bibinfo {author} {\bibfnamefont {M.}~\bibnamefont
  {Diehl}}\ and\ \bibinfo {author} {\bibfnamefont {Ph.}\ \bibnamefont
  {Hagler}},\ }\bibfield  {title} {\enquote {\bibinfo {title} {{Spin densities
  in the transverse plane and generalized transversity distributions}},}\
  }\href {\doibase 10.1140/epjc/s2005-02342-6} {\bibfield  {journal} {\bibinfo
  {journal} {Eur. Phys. J. C}\ }\textbf {\bibinfo {volume} {44}},\ \bibinfo
  {pages} {87--101} (\bibinfo {year} {2005})},\ \Eprint
  {http://arxiv.org/abs/hep-ph/0504175} {arXiv:hep-ph/0504175} \BibitemShut
  {NoStop}%
\bibitem [{\citenamefont {Bhoonah}\ and\ \citenamefont
  {Lorc{\'e}}(2017)}]{Bhoonah:2017olu}%
  \BibitemOpen
  \bibfield  {author} {\bibinfo {author} {\bibfnamefont {Amit}\ \bibnamefont
  {Bhoonah}}\ and\ \bibinfo {author} {\bibfnamefont {C{\'e}dric}\ \bibnamefont
  {Lorc{\'e}}},\ }\bibfield  {title} {\enquote {\bibinfo {title} {{Quark
  transverse spin{\textendash}orbit correlations}},}\ }\href {\doibase
  10.1016/j.physletb.2017.10.003} {\bibfield  {journal} {\bibinfo  {journal}
  {Phys. Lett. B}\ }\textbf {\bibinfo {volume} {774}},\ \bibinfo {pages}
  {435--440} (\bibinfo {year} {2017})},\ \Eprint
  {http://arxiv.org/abs/1703.08322} {arXiv:1703.08322 [hep-ph]} \BibitemShut
  {NoStop}%
\bibitem [{\citenamefont {Alexandrou}(2024)}]{Alexandrou:2024awx}%
  \BibitemOpen
  \bibfield  {author} {\bibinfo {author} {\bibfnamefont {Constantia}\
  \bibnamefont {Alexandrou}},\ }\bibfield  {title} {\enquote {\bibinfo {title}
  {{Nucleon Transversity from lattice QCD}},}\ }\href {\doibase
  10.22323/1.477.0002} {\bibfield  {journal} {\bibinfo  {journal} {PoS}\
  }\textbf {\bibinfo {volume} {Transversity2024}},\ \bibinfo {pages} {002}
  (\bibinfo {year} {2024})},\ \Eprint {http://arxiv.org/abs/2408.14370}
  {arXiv:2408.14370 [hep-lat]} \BibitemShut {NoStop}%
\bibitem [{\citenamefont {Kim}(2025)}]{Kim:2025mol}%
  \BibitemOpen
  \bibfield  {author} {\bibinfo {author} {\bibfnamefont {June-Young}\
  \bibnamefont {Kim}},\ }\bibfield  {title} {\enquote {\bibinfo {title}
  {{Chiral-odd generalized parton distributions in the large-$N_{c}$ limit of
  QCD: Next-to-leading-order contributions}},}\ }\href@noop {} {\  (\bibinfo
  {year} {2025})},\ \Eprint {http://arxiv.org/abs/2506.21013} {arXiv:2506.21013
  [hep-ph]} \BibitemShut {NoStop}%
\bibitem [{\citenamefont {Bhattacharya}\ \emph {et~al.}(2023)\citenamefont
  {Bhattacharya}, \citenamefont {Cichy}, \citenamefont {Constantinou},
  \citenamefont {Gao}, \citenamefont {Metz}, \citenamefont {Miller},
  \citenamefont {Mukherjee}, \citenamefont {Petreczky}, \citenamefont
  {Steffens},\ and\ \citenamefont {Zhao}}]{Bhattacharya:2023ays}%
  \BibitemOpen
  \bibfield  {author} {\bibinfo {author} {\bibfnamefont {Shohini}\ \bibnamefont
  {Bhattacharya}}, \bibinfo {author} {\bibfnamefont {Krzysztof}\ \bibnamefont
  {Cichy}}, \bibinfo {author} {\bibfnamefont {Martha}\ \bibnamefont
  {Constantinou}}, \bibinfo {author} {\bibfnamefont {Xiang}\ \bibnamefont
  {Gao}}, \bibinfo {author} {\bibfnamefont {Andreas}\ \bibnamefont {Metz}},
  \bibinfo {author} {\bibfnamefont {Joshua}\ \bibnamefont {Miller}}, \bibinfo
  {author} {\bibfnamefont {Swagato}\ \bibnamefont {Mukherjee}}, \bibinfo
  {author} {\bibfnamefont {Peter}\ \bibnamefont {Petreczky}}, \bibinfo {author}
  {\bibfnamefont {Fernanda}\ \bibnamefont {Steffens}}, \ and\ \bibinfo {author}
  {\bibfnamefont {Yong}\ \bibnamefont {Zhao}},\ }\bibfield  {title} {\enquote
  {\bibinfo {title} {{Moments of proton GPDs from the OPE of nonlocal quark
  bilinears up to NNLO}},}\ }\href {\doibase 10.1103/PhysRevD.108.014507}
  {\bibfield  {journal} {\bibinfo  {journal} {Phys. Rev. D}\ }\textbf {\bibinfo
  {volume} {108}},\ \bibinfo {pages} {014507} (\bibinfo {year} {2023})},\
  \Eprint {http://arxiv.org/abs/2305.11117} {arXiv:2305.11117 [hep-lat]}
  \BibitemShut {NoStop}%
\bibitem [{\citenamefont {Yao}\ \emph {et~al.}(2025)\citenamefont {Yao},
  \citenamefont {Xu}, \citenamefont {Binosi}, \citenamefont {Cui},
  \citenamefont {Ding}, \citenamefont {Raya}, \citenamefont {Roberts},
  \citenamefont {Rodr{\'\i}guez-Quintero},\ and\ \citenamefont
  {Schmidt}}]{Yao:2024ixu}%
  \BibitemOpen
  \bibfield  {author} {\bibinfo {author} {\bibfnamefont {Z.~Q.}\ \bibnamefont
  {Yao}}, \bibinfo {author} {\bibfnamefont {Y.~Z.}\ \bibnamefont {Xu}},
  \bibinfo {author} {\bibfnamefont {D.}~\bibnamefont {Binosi}}, \bibinfo
  {author} {\bibfnamefont {Z.~F.}\ \bibnamefont {Cui}}, \bibinfo {author}
  {\bibfnamefont {M.}~\bibnamefont {Ding}}, \bibinfo {author} {\bibfnamefont
  {K.}~\bibnamefont {Raya}}, \bibinfo {author} {\bibfnamefont {C.~D.}\
  \bibnamefont {Roberts}}, \bibinfo {author} {\bibfnamefont {J.}~\bibnamefont
  {Rodr{\'\i}guez-Quintero}}, \ and\ \bibinfo {author} {\bibfnamefont {S.~M.}\
  \bibnamefont {Schmidt}},\ }\bibfield  {title} {\enquote {\bibinfo {title}
  {{Nucleon gravitational form factors}},}\ }\href {\doibase
  10.1140/epja/s10050-025-01557-x} {\bibfield  {journal} {\bibinfo  {journal}
  {Eur. Phys. J. A}\ }\textbf {\bibinfo {volume} {61}},\ \bibinfo {pages} {92}
  (\bibinfo {year} {2025})},\ \Eprint {http://arxiv.org/abs/2409.15547}
  {arXiv:2409.15547 [hep-ph]} \BibitemShut {NoStop}%
\bibitem [{\citenamefont {Gockeler}\ \emph
  {et~al.}(2005{\natexlab{b}})\citenamefont {Gockeler}, \citenamefont {Hagler},
  \citenamefont {Horsley}, \citenamefont {Pleiter}, \citenamefont {Rakow},
  \citenamefont {Schafer}, \citenamefont {Schierholz},\ and\ \citenamefont
  {Zanotti}}]{Gockeler:2005cj}%
  \BibitemOpen
  \bibfield  {author} {\bibinfo {author} {\bibfnamefont {M.}~\bibnamefont
  {Gockeler}}, \bibinfo {author} {\bibfnamefont {Ph.}\ \bibnamefont {Hagler}},
  \bibinfo {author} {\bibfnamefont {R.}~\bibnamefont {Horsley}}, \bibinfo
  {author} {\bibfnamefont {D.}~\bibnamefont {Pleiter}}, \bibinfo {author}
  {\bibfnamefont {Paul E.~L.}\ \bibnamefont {Rakow}}, \bibinfo {author}
  {\bibfnamefont {A.}~\bibnamefont {Schafer}}, \bibinfo {author} {\bibfnamefont
  {G.}~\bibnamefont {Schierholz}}, \ and\ \bibinfo {author} {\bibfnamefont
  {J.~M.}\ \bibnamefont {Zanotti}} (\bibinfo {collaboration} {QCDSF, UKQCD}),\
  }\bibfield  {title} {\enquote {\bibinfo {title} {{Quark helicity flip
  generalized parton distributions from two-flavor lattice QCD}},}\ }\href
  {\doibase 10.1016/j.physletb.2005.09.002} {\bibfield  {journal} {\bibinfo
  {journal} {Phys. Lett. B}\ }\textbf {\bibinfo {volume} {627}},\ \bibinfo
  {pages} {113--123} (\bibinfo {year} {2005}{\natexlab{b}})},\ \Eprint
  {http://arxiv.org/abs/hep-lat/0507001} {arXiv:hep-lat/0507001} \BibitemShut
  {NoStop}%
\bibitem [{\citenamefont {G{\"o}ckeler}\ \emph {et~al.}(2007)\citenamefont
  {G{\"o}ckeler}, \citenamefont {H{\"a}gler}, \citenamefont {Horsley},
  \citenamefont {Nakamura}, \citenamefont {Pleiter}, \citenamefont {Rakow},
  \citenamefont {Sch{\"a}fer}, \citenamefont {Schierholz}, \citenamefont
  {St{\"u}ben},\ and\ \citenamefont {Zanotti}}]{QCDSF:2006tkx}%
  \BibitemOpen
  \bibfield  {author} {\bibinfo {author} {\bibfnamefont {M.}~\bibnamefont
  {G{\"o}ckeler}}, \bibinfo {author} {\bibfnamefont {Ph.}\ \bibnamefont
  {H{\"a}gler}}, \bibinfo {author} {\bibfnamefont {R.}~\bibnamefont {Horsley}},
  \bibinfo {author} {\bibfnamefont {Y.}~\bibnamefont {Nakamura}}, \bibinfo
  {author} {\bibfnamefont {D.}~\bibnamefont {Pleiter}}, \bibinfo {author}
  {\bibfnamefont {P.~E.~L.}\ \bibnamefont {Rakow}}, \bibinfo {author}
  {\bibfnamefont {A.}~\bibnamefont {Sch{\"a}fer}}, \bibinfo {author}
  {\bibfnamefont {G.}~\bibnamefont {Schierholz}}, \bibinfo {author}
  {\bibfnamefont {H.}~\bibnamefont {St{\"u}ben}}, \ and\ \bibinfo {author}
  {\bibfnamefont {J.~M.}\ \bibnamefont {Zanotti}} (\bibinfo {collaboration}
  {QCDSF, UKQCD}),\ }\bibfield  {title} {\enquote {\bibinfo {title}
  {{Transverse spin structure of the nucleon from lattice QCD simulations}},}\
  }\href {\doibase 10.1103/PhysRevLett.98.222001} {\bibfield  {journal}
  {\bibinfo  {journal} {Phys. Rev. Lett.}\ }\textbf {\bibinfo {volume} {98}},\
  \bibinfo {pages} {222001} (\bibinfo {year} {2007})},\ \Eprint
  {http://arxiv.org/abs/hep-lat/0612032} {arXiv:hep-lat/0612032} \BibitemShut
  {NoStop}%
\bibitem [{\citenamefont {Alexandrou}\ \emph {et~al.}(2023)\citenamefont
  {Alexandrou} \emph {et~al.}}]{Alexandrou:2022dtc}%
  \BibitemOpen
  \bibfield  {author} {\bibinfo {author} {\bibfnamefont {C.}~\bibnamefont
  {Alexandrou}} \emph {et~al.},\ }\bibfield  {title} {\enquote {\bibinfo
  {title} {{Moments of the nucleon transverse quark spin densities using
  lattice QCD}},}\ }\href {\doibase 10.1103/PhysRevD.107.054504} {\bibfield
  {journal} {\bibinfo  {journal} {Phys. Rev. D}\ }\textbf {\bibinfo {volume}
  {107}},\ \bibinfo {pages} {054504} (\bibinfo {year} {2023})},\ \Eprint
  {http://arxiv.org/abs/2202.09871} {arXiv:2202.09871 [hep-lat]} \BibitemShut
  {NoStop}%
\end{thebibliography}%
\end{document}